# The Complexity of Circumscription in Description Logic

**Piero A. Bonatti**                                             BONATTI@NA.INFN.IT
*Section of Computer Science, Department of Physics*
*University of Naples Federico II, Italy*

**Carsten Lutz**                                        CLU@INFORMATIK.UNI-BREMEN.DE
*Department of Mathematics and Computer Science*
*University of Bremen, Germany*

**Frank Wolter**                                          WOLTER@LIVERPOOL.AC.UK
*Department of Computer Science*
*University of Liverpool, UK*

## Abstract

As fragments of first-order logic, Description logics (DLs) do not provide nonmonotonic features such as defeasible inheritance and default rules. Since many applications would benefit from the availability of such features, several families of nonmonotonic DLs have been developed that are mostly based on default logic and autoepistemic logic. In this paper, we consider circumscription as an interesting alternative approach to nonmonotonic DLs that, in particular, supports defeasible inheritance in a natural way. We study DLs extended with circumscription under different language restrictions and under different constraints on the sets of minimized, fixed, and varying predicates, and pinpoint the exact computational complexity of reasoning for DLs ranging from $\mathcal{ALC}$ to $\mathcal{ALCIO}$ and $\mathcal{ALCQO}$. When the minimized and fixed predicates include only concept names but no role names, then reasoning is complete for $\mathrm{NExp}^{\mathrm{NP}}$. It becomes complete for $\mathrm{NP}^{\mathrm{NExp}}$ when the number of minimized and fixed predicates is bounded by a constant. If roles can be minimized or fixed, then complexity ranges from $\mathrm{NExp}^{\mathrm{NP}}$ to undecidability.

## 1. Introduction

Early knowledge representation (KR) formalisms such as semantic networks and frames included a wealth of features in order to provide as much expressive power as possible (Quillian, 1968; Minsky, 1975). In particular, such formalisms usually admitted a structured representation of classes and objects similar to modern description logics (DLs), but also mechanisms for defeasible inheritance, default rules, and other features that are nowadays studied in the area of nonmonotonic logics (NMLs). When KR theory developed further, the all-embracing approaches were largely given up in favour of more specialized ones due to unfavourable computational properties and problems with the semantics. This process has caused DLs and NMLs to develop into two independent subfields. Consequently, modern description logics such as $\mathcal{SHIQ}$ lack the expressive power to represent defeasible inheritance and other nonmonotonic features (Horrocks, Sattler, & Tobies, 2000).

Despite (or due to) this development, there has been a continuous interest in the (re)integration of nonmonotonic features into description logics. In recent years, the advent of several new applications of DLs have increased this interest even further. We briefly discuss two of them. First, DLs have become a popular tool for the formalization of biomed-ical ontologies such as GALEN (Rector & Horrocks, 1997) and SNOMED (Cote, Rothwell, Palotay, Beckett, & Brochu, 1993). As argued for example by Rector (2004) and Stevens et al. (2005), it is important for such ontologies to represent exceptions of the form "in humans, the heart is usually located on the left-hand side of the body; in humans with situs inversus, the heart is located on the right-hand side of the body". Modelling such situations requires *defeasible inheritance*, i.e., properties transfer to all instances of a class by default, but can be explicitly overridden in special cases (McCarthy, 1986; Horty, 1994; Brewka, 1994; Baader & Hollunder, 1995b). The second application is the use of DLs as security policy languages (Uszok, Bradshaw, Johnson, Jeffers, Tate, Dalton, & Aitken, 2004; Kagal, Finin, & Joshi, 2003; Tonti, Bradshaw, Jeffers, Montanari, Suri, & Uszok, 2003). In formalizing access control policies, one must deal with the situation where a given request is neither explicitly allowed nor explicitly denied. Then, a default decision has to be taken such as in *open* and *closed* policies, where authorizations are respectively granted or denied by default (Bonatti & Samarati, 2003). Moreover, policies are often formulated incrementally, i.e., start with general authorizations for large classes of subjects, objects, and actions, and then progressively refine them by introducing exceptions for specific subclasses. This approach is clearly an incarnation of defeasible inheritance.

While the above applications illustrate that integrating nonmonotonic features into DLs is worthwhile, the actual engineering of a computationally well-behaved nonmonotonic DL that provides sufficient expressive power turns out to be a non-trivial task. In particular, combinations of DLs and nonmonotonic logics typically involve subtle interactions between the two component logics and this easily leads to undecidability. It appears that there is no one optimal way to circumnavigate these difficulties, and thus many different combinations of DLs and nonmonotonic logics have been proposed in the literature, each with individual strengths and limitations (we provide a survey in Section 7). However, there is a striking gap: almost all existing approaches are based on default logic and autoepistemic logic, while circumscription has received very little attention in connection with DLs, and the computational properties of DLs with circumscription have been almost completely unknown. This is all the more surprising since circumscription is known to be one of the weakest forms of nonmonotonic reasoning—see the work by Janhunen (1999) for one of the most recent surveys, and the paper by Bonatti and Eiter (1996) for an expressiveness analysis in terms of queries. Therefore, it is a natural idea to use circumscription for defining a computationally well-behaved, yet expressive DL with nonmonotonic features.

In this paper, we study circumscription (McCarthy, 1980) as an alternative approach to defining nonmonotonic DLs. In particular, we define a family of DLs with circumscription that enable a natural modelling of defeasible inheritance. Our general approach is to generalize standard DL knowledge bases to *circumscribed knowledge bases (cKBs)* which, additionally to a TBox for representing terminological knowledge and an ABox for representing knowledge about individuals, are equipped with a *circumscription pattern*. This pattern lists predicates (i.e., concept and role names) to be *minimized* in the sense that, in all admitted models of the cKB, the extension of the listed predicates has to be minimal w.r.t. set inclusion. Following McCarthy (1986), the minimized predicates can be used as "abnormality predicates" that identify instances which are not typical for their class. Circumscription patterns can require other predicates to be fixed during minimization, or allow them to vary freely (McCarthy, 1986). A main feature of the DLs of our family is that they come with a built-in mechanism for defeasible inheritance: by default, the properties of a class ("humans" in the first example above) transfer to each subclass ("humans with situs inversus"), but exceptions can be specified based on a priority mechanism. It is well-known that defeasible inheritance with priority cannot be modularly encoded with pure default or autoepistemic logic (Horty, 1994), and workarounds such as the explicit listing of exceptions lead to serious maintainability problems. Circumscription lends itself naturally to priorities, based on circumscription patterns that can express preferences between minimized predicates in terms of a partial ordering. As argued by Baader and Hollunder (1995b), such an approach is well-suited to ensure a smooth interplay between defeasible inheritance and DL subsumption, and thus we prefer it over traditional prioritized circumscription.

To achieve decidability, nonmonotonic DLs usually have to adopt suitable restrictions on the expressive power of the DL component, on non-monotonic features, or their interaction. In the case of default logic and autoepistemic logic, a typical restriction concerns the different treatment of individuals that are explicitly denoted by a constant, and those that are not. This goes back to reasoning in first-order default logic (Reiter, 1980) and autoepistemic logic (Moore, 1985), which also involve tricky technical issues related to the denotation of individuals. To make reasoning decidable in DLs based on default logic, default rules are applied only to the individuals denoted by constants that occur explicitly in the knowledge base (Baader & Hollunder, 1995a), but not to unnamed individuals. As a consequence, named and unnamed individuals are not treated uniformly. In approaches based on autoepistemic logic (Donini, Lenzerini, Nardi, Nutt, & Schaerf, 1998; Donini, Nardi, & Rosati, 1997, 2002), an alternative solution is to restrict the domain to a fixed, denumerable set of constants. This approach overcomes the different treatment of named and unnamed individuals since all individuals are named. The flipside is that ad-hoc encodings are required when the domain is finite or the unique name assumption is not enforced, i.e., when different constants are allowed to denote the same individual. In this respect, DLs with circumscription pose no difficulty at all, and named individuals are treated in exactly the same way as unnamed ones without any assumptions on the domain. At the same time, we are able to base our nonmonotonic DLs on rather expressive DL components such as $\mathcal{ALCIO}$ and $\mathcal{ALCQO}$ without losing decidability. However, we cannot do without any restrictions either: we only allow to fix and minimize concept names during circumscription and require that all role names vary.

The main contribution of this paper is a detailed analysis of the computational properties of reasoning with cKBs. We show that, in the expressive DLs $\mathcal{ALCIO}$ and $\mathcal{ALCQO}$, instance checking, satisfiability and subsumption are decidable for *concept-circumscribed* KBs in which only concept names (and no role names) are minimized and fixed. More precisely, we prove that these reasoning problems are $\mathrm{NExp}^{\mathrm{NP}}$-complete, where the lower bound applies already to concept-circumscribed KBs in $\mathcal{ALC}$ with empty preference relation and without fixed concept names and (1) empty TBox or (2) empty ABox and acyclic TBox. In addition, we show that if a constant bound is imposed on the number of minimized and fixed concept names, then the complexity drops to $\mathrm{NP}^{\mathrm{NExp}}$.

The situation is completely different when role names are minimized or fixed. First, the complexity of reasoning with cKBs formulated in $\mathcal{ALC}$ with a single fixed role name, empty TBox, empty preference relation, and no minimized role names turns out to be outside the analytic hierarchy, and thus very highly undecidable. This result is shown by a reduction of satisfiability in monadic second-order logic (MSO) with binary predicates over arbitrary (i.e., not necessarily tree-shaped) structures. The reduction does not apply to cKBs in which role names can be minimized, but not fixed. Surprisingly, we find that in this case reasoning with empty TBoxes becomes decidable (and again $\mathrm{NExP^{NP}}$-complete) for DLs between $\mathcal{ALC}$ and $\mathcal{ALCQO}$, and only for $\mathcal{ALCI}$ and its extensions it is undecidable. For all these logics, however, adding acyclic TBoxes leads to undecidability. The reader can find a table summarising the complexity results in Section 7.

| Name | Syntax | Semantics |
|------|--------|-----------|
| inverse role | $r^-$ | $(r^{\mathcal{I}})^{\smile} = \{(d,e) \mid (e,d) \in r^{\mathcal{I}}\}$ |
| nominal | $\{a\}$ | $\{a^{\mathcal{I}}\}$ |
| negation | $\neg C$ | $\Delta^{\mathcal{I}} \setminus C^{\mathcal{I}}$ |
| conjunction | $C \sqcap D$ | $C^{\mathcal{I}} \cap D^{\mathcal{I}}$ |
| disjunction | $C \sqcup D$ | $C^{\mathcal{I}} \cup D^{\mathcal{I}}$ |
| at-least restriction | $(\geqslant n\, r\, C)$ | $\{d \in \Delta^{\mathcal{I}} \mid \#\{e \in C^{\mathcal{I}} \mid (d,e) \in r^{\mathcal{I}}\} \geq n\}$ |
| at-most restriction | $(\leqslant n\, r\, C)$ | $\{d \in \Delta^{\mathcal{I}} \mid \#\{e \in C^{\mathcal{I}} \mid (d,e) \in r^{\mathcal{I}}\} \leq n\}$ |

Figure 1: Syntax and semantics of $\mathcal{ALCQIO}$.

It is interesting to note that our results are somewhat unusual from the perspective of NMLs. First, the *arity* of predicates has an impact on decidability: fixing concept names (unary predicates) does not impair decidability, whereas fixing a single role name (binary predicate) leads to a strong undecidability result. Second, the *number* of predicates that are minimized or fixed (bounded vs. unbounded) affects the computational complexity of reasoning. Although (as we note in passing) a similar effect can be observed in propositional logic with circumscription, this has, to the best of our knowledge, never been explicitly noted.

The paper is organized as follows. In the next section we introduce syntax, semantics, and reasoning problems for circumscribed KBs, and provide some examples. Section 3 provides some basic results such as a polynomial simulation of fixed concepts by means of minimized concepts, a polynomial reduction of reasoning with general TBoxes to reasoning with acyclic TBoxes, and a polynomial reduction of the simultaneous satisfiability of multiple cKBs to standard satisfiability. Then, Section 4 proves the decidability and complexity results for concept-circumscribed knowledge bases. Fixed and minimized roles are considered in Sections 5 and 6, respectively. Section 7 discusses related work, and Section 8 concludes the paper by summarizing the main results and pointing out some interesting directions for further research. To improve readability, many proof details are deferred to the appendix. This paper is an extended version of the article by Bonatti, Lutz, and Wolter (2006).

## 2. Description Logics and Circumscription

In DLs, *concepts* are inductively defined with the help of a set of *constructors*, starting with a set $\mathsf{N_C}$ of *concept names*, a set $\mathsf{N_R}$ of *role names*, and (possibly) a set $\mathsf{N_I}$ of *individual* *names* (all countably infinite). We use the term *predicates* to refer to elements of $N_C \cup N_R$. The concepts of the expressive DL $\mathcal{ALCQIO}$ are formed using the constructors shown in Figure 1.

There, the inverse role constructor is a role constructor, whereas the remaining six constructors are concept constructors. In Figure 1 and throughout this paper, we use $\#S$ to denote the cardinality of a set $S$, $a$ and $b$ to denote individual names, $r$ and $s$ to denote roles (i.e., role names and inverses thereof), $A, B$ to denote concept names, and $C, D$ to denote (possibly complex) concepts. As usual, we use $\top$ as abbreviation for an arbitrary (but fixed) propositional tautology, $\bot$ for $\neg\top$, $\rightarrow$ and $\leftrightarrow$ for the usual Boolean abbreviations, $\exists r.C$ (*existential restriction*) for $(\geqslant 1\ r\ C)$, and $\forall r.C$ (*universal restriction*) for $(\leqslant 0\ r\ \neg C)$.

In this paper, we will not be concerned with $\mathcal{ALCQIO}$ itself, but with several of its fragments.[1] The basic such fragment allows only for negation, conjunction, disjunction, and universal and existential restrictions, and is called $\mathcal{ALC}$. The availability of additional constructors is indicated by concatenation of a corresponding letter: $\mathcal{Q}$ stands for number restrictions, $\mathcal{I}$ stands for inverse roles, and $\mathcal{O}$ for nominals. This explains the name $\mathcal{ALCQIO}$, and also allows us to refer to fragments such as $\mathcal{ALCIO}$, $\mathcal{ALCQO}$, and $\mathcal{ALCQI}$.

The semantics of $\mathcal{ALCQIO}$-concepts is defined in terms of an *interpretation* $\mathcal{I} = (\Delta^{\mathcal{I}}, \cdot^{\mathcal{I}})$. The *domain* $\Delta^{\mathcal{I}}$ is a non-empty set of individuals and the *interpretation function* $\cdot^{\mathcal{I}}$ maps each concept name $A \in N_C$ to a subset $A^{\mathcal{I}}$ of $\Delta^{\mathcal{I}}$, each role name $r \in N_R$ to a binary relation $r^{\mathcal{I}}$ on $\Delta^{\mathcal{I}}$, and each individual name $a \in N_I$ to an individual $a^{\mathcal{I}} \in \Delta^{\mathcal{I}}$. The extension of $\cdot^{\mathcal{I}}$ to inverse roles and arbitrary concepts is inductively defined as shown in the third column of Figure 1. An interpretation $\mathcal{I}$ is called a *model* of a concept $C$ if $C^{\mathcal{I}} \neq \emptyset$. If $\mathcal{I}$ is a model of $C$, we also say that $C$ is *satisfied* by $\mathcal{I}$.

A *(general) TBox* is a finite set of *concept implications (CIs)* $C \sqsubseteq D$ where $C$ and $D$ are concepts. As usual, we use $C \doteq D$ as an abbreviation for $C \sqsubseteq D$ and $D \sqsubseteq C$. An *ABox* is a finite set of *concept assertions* $C(a)$ and *role assertions* $r(a, b)$, where $a, b$ are individual names, $r$ is a role name, and $C$ is a concept. An interpretation $\mathcal{I}$ *satisfies* (i) a CI $C \sqsubseteq D$ if $C^{\mathcal{I}} \subseteq D^{\mathcal{I}}$, (ii) an assertion $C(a)$ if $a^{\mathcal{I}} \in C^{\mathcal{I}}$, and (iii) an assertion $r(a, b)$ if $(a^{\mathcal{I}}, b^{\mathcal{I}}) \in r^{\mathcal{I}}$. Then, $\mathcal{I}$ is a *model* of a TBox $\mathcal{T}$ if it satisfies all implications in $\mathcal{T}$, and a *model* of an ABox $\mathcal{A}$ if it satisfies all assertions in $\mathcal{A}$.

An important class of TBoxes are *acyclic* TBoxes: call a TBox $\mathcal{T}$ acyclic if it is a set of *concept definitions* $A \doteq C$, where $A$ is a concept name and the following two conditions hold:

- no concept name occurs more than once on the left hand side of a definition in $\mathcal{T}$;

- the relation $\prec_{\mathcal{T}}$, defined by setting $A \prec_{\mathcal{T}} B$ iff $A \doteq C \in \mathcal{T}$ and $B$ occurs in $C$, is well-founded.

---

1. The reason that we do not consider $\mathcal{ALCQIO}$ in this paper is that it does not have the finite model property; i.e., there are satisfiable concepts that are not satisfiable in finite models. Our proofs of the complexity upper bounds assume the finite model property and, therefore, do not work for $\mathcal{ALCQIO}$. Investigating circumscription for description logics without the finite model property remains an interesting open problem.

## 2.1 Circumscription, Varying Predicates, and Partial Priority Ordering

Circumscription is a logical approach suitable for modelling what *normally* or *typically* holds, and thus admits the modeling of defeasible inheritance (McCarthy, 1986; Lifschitz, 1993). The idea is to define, in a standard first-order language, both domain knowledge and so-called *abnormality predicates* that identify instances of a class that violate the normal or typical properties of that class. To capture the intuition that abnormality is exceptional, inference is not based on the set of all models of the resulting theory as in classical logic, but rather restricted to those models where the extension of the abnormality predicates is *minimal* with respect to set inclusion. Intuitively, this means that reasoning is based only on models that are "as normal as possible". Since such models are classical models of the given knowledge base, all classical first-order inferences are valid in circumscription (but additional inferences may become possible).

Since description logics are fragments of first-order logic, circumscription can be readily applied. Using $\mathcal{ALC}$ syntax, we can assert that mammals normally inhabitate land, and that whales do not live on land:

$$\texttt{Mammal} \quad \sqsubseteq \quad \exists\texttt{habitat.Land} \sqcup \texttt{Ab}_{\texttt{Mammal}}$$
$$\texttt{Whale} \quad \sqsubseteq \quad \texttt{Mammal} \sqcap \neg\exists\texttt{habitat.Land}$$

The upper inclusion states that any mammal not inhabitating land is an abnormal mammal, thus satisfying the abnormality predicate $\texttt{Ab}_{\texttt{Mammal}}$. When applying circumscription to the above TBox, we should thus only consider models where the extension of $\texttt{Ab}_{\texttt{Mammal}}$ is minimal. However, there is more than one way of defining such preferred models because each non-minimized predicate can be treated in two different ways during minimization: we may either fix its extension or let it vary freely.

Intuitively, fixed predicates retain their classical semantics while varying predicates may be affected by minimization. As a concrete example, consider once more the above TBox and assume that all non-minimized predicates are fixed. Then we can derive the following subsumptions:

$$\begin{aligned} \texttt{Whale} \quad &\sqsubseteq \quad \texttt{Ab}_{\texttt{Mammal}} \\ \texttt{Ab}_{\texttt{Mammal}} \quad &\doteq \quad \texttt{Mammal} \sqcap \neg\exists\texttt{habitat.Land}. \end{aligned} \qquad (\dagger)$$

Here, $\texttt{Whale} \sqsubseteq \texttt{Ab}_{\texttt{Mammal}}$ and $\texttt{Ab}_{\texttt{Mammal}} \sqsupseteq \texttt{Mammal} \sqcap \neg\exists\texttt{habitat.Land}$ are classical consequences of the TBox. The minimization of $\texttt{Ab}_{\texttt{Mammal}}$ adds only the inclusion $\texttt{Ab}_{\texttt{Mammal}} \sqsubseteq \texttt{Mammal} \sqcap \neg\exists\texttt{habitat.Land}.$

To further analyze fixed predicates, suppose that we explicitly introduce a concrete mammal that is not a whale by adding an ABox assertion

$$\texttt{Mammal} \sqcap \neg\texttt{Whale(flipper)}$$

We might expect that we can derive $\exists\texttt{habitat.Land(flipper)}$, but actually that is not the case. To see this, observe that there is a classical model of the knowledge base that falsifies $\exists\texttt{habitat.Land(flipper)}$; the extension of the fixed predicates $\texttt{habitat}$ and $\texttt{Land}$ are not affected by minimization, so $\exists\texttt{habitat.Land(flipper)}$ must still be false after minimization. The same argument can be applied to the negation of $\exists\texttt{habitat.Land(flipper)}$, which is thus also not derivable. What we have just seen is that if a sentence uses only fixed predicates, then it is a consequence of a circumscribed knowledge base if, and only if, it is a classical consequence of the knowledge base.

Now assume that we let the role `habitat` and the concept name `Land` vary freely, and fix only `Mammal` and `Whale`. In view of the concept inclusion for `Mammal` in our original TBox, this setup may be interpreted as expressing that it is *very* unlikely for a mammal not to live on land: we are willing to modify the extension of `habitat` and `Land` in order to avoid abnormality. We obtain an additional consequence, namely:

$$\texttt{Whale} \ \dot{=} \ \texttt{Ab}_{\texttt{Mammal}}. \tag{‡}$$

To see that this is indeed a consequence note that, during minimization, we can (i) make `Land` non-empty and (ii) for any mammal $m$ that is not a whale, ensure that $m$ is not an $\texttt{Ab}_{\texttt{Mammal}}$ by linking it via `habitat` to the generated instance of `Land`.[2] Intuitively, the equality (‡) can be seen as reflecting the unlikeliness of being abnormal: a mammal is only abnormal if there is a reason, and the only reason that we have captured in our knowledge base is being a whale.

Let us now return to the assertion `Mammal ⊓ ¬Whale(flipper)`. By applying classical reasoning to (†) and (‡), we derive `Whale ⊒ Mammal ⊓ ¬∃habitat.Land` (i.e., whales are the only mammals that do not live on land). Thus we can now derive the expected conclusion `∃habitat.Land(flipper)`. In summary, by turning `habitat` and `Land` into varying predicates, we have obtained a more natural modelling in which the `habitat` attribute of mammals can be forced to its default value.

Driving our example further, we might now consider whales abnormal to such a degree that we do not believe they exist unless there is evidence that they do. Then we should, additionally, let `Whale` vary freely. The result is that (†) and (‡) can still be derived, and additionally we obtain the consequence

$$\texttt{Whale} \dot{=} \texttt{Ab}_{\texttt{Mammal}} \dot{=} \bot.$$

We can then use an ABox to add evidence that whales exist, e.g. through the assertion `Whale(mobydick)`. As expected, the result of this change is that

$$\texttt{Whale} \dot{=} \texttt{Ab}_{\texttt{Mammal}} \dot{=} \{\texttt{mobydick}\}.$$

Evidence for the existence of another, anonymous whale could be generated by adding the ABox assertion `Male(mobydick)` and the TBox statement

$$\texttt{Whale} \sqsubseteq \exists\texttt{mother}.(\texttt{Whale} \sqcap \neg\texttt{Male})$$

with `mother` and `Male` varying freely. This knowledge base classically entails that there exist two whales, satisfying `Male` and `¬Male`, respectively. The former is denoted by `mobydick`, while the latter is not denoted by any ABox individual (which corresponds to a first-order constant). After minimization, `Whale` contains exactly those two individuals.

In general, the appropriate combination of fixed and varying predicates depends on the application. Therefore, we adhere to standard circumscription and give users the freedom to choose which predicates are minimized, fixed, and varying.

---

2. Indeed, this is the only reason to let `Land` vary: to ensure that it can be made non-empty during minimization.

As another example, consider the sentences: *"In humans, the heart is usually located on the left-hand side of the body; in humans with situs inversus, the heart is located on the right-hand side of the body".* They can be axiomatized as follows:

$$\texttt{Human} \sqsubseteq \exists\texttt{has\_heart}.\exists\texttt{has\_position}.\{\texttt{Left}\} \sqcup \texttt{Ab}_{\texttt{Human}}$$
$$\texttt{Situs\_Inversus} \sqsubseteq \exists\texttt{has\_heart}.\exists\texttt{has\_position}.\{\texttt{Right}\}$$
$$\exists\texttt{has\_heart}.\exists\texttt{has\_position}.\{\texttt{Left}\} \sqcap \exists\texttt{has\_heart}.\exists\texttt{has\_position}.\{\texttt{Right}\} \sqsubseteq \bot.$$

The predicate $\texttt{Ab}_{\texttt{Human}}$ represents abnormal humans and should be minimized. If humans with situs inversus are to be restricted to those individuals that are explicitly declared to have this property, then by analogy with the previous example the roles specifying heart position and the class of exceptional individuals $\texttt{Situs\_Inversus}$ should be allowed to vary while $\texttt{Human}$ can be fixed and retain its classical semantics. As a result and in the absence of any further axioms, $\texttt{Ab}_{\texttt{Human}}$ and $\texttt{Situs\_Inversus}$ are empty in all minimized models. The additional axiom $\exists\texttt{has\_friend}.\texttt{Situs\_Inversus}(\texttt{John})$ turns $\texttt{Ab}_{\texttt{Human}}$ and $\texttt{Situs\_Inversus}$ into a singleton set containing an anonymous individual (though in some models, it may be John himself). As an example for a nonclassical consequence, consider:

$$\texttt{Human} \sqcap \neg\texttt{Situs\_Inversus} \sqsubseteq \exists\texttt{has\_heart}.\exists\texttt{has\_position}.\{\texttt{Left}\},$$

that is, all humans have the default heart position with the only exception of those that are explicitly declared to have situs inversus.

It has been extensively argued (McCarthy, 1986; Horty, 1994; Brewka, 1994; Baader & Hollunder, 1995b) that there is an interplay between subsumption and abnormality predicates that should be addressed in nonmonotonic DLs. Consider, for example, the following TBox:

$$
\begin{aligned}
\texttt{User} &\sqsubseteq \neg\exists\texttt{hasAccessTo}.\{\texttt{ConfidentialFile}\} \sqcup \texttt{Ab}_{\texttt{User}} \\
\texttt{Staff} &\sqsubseteq \texttt{User} \\
\texttt{Staff} &\sqsubseteq \exists\texttt{hasAccessTo}.\{\texttt{ConfidentialFile}\} \sqcup \texttt{Ab}_{\texttt{Staff}} \\
\texttt{BlacklistedStaff} &\sqsubseteq \texttt{Staff} \sqcap \neg\exists\texttt{hasAccessTo}.\{\texttt{ConfidentialFile}\}
\end{aligned}
$$

To get models that are "as normal as possible", as a first attempt we could minimize the two abnormality predicates $\texttt{Ab}_{\texttt{User}}$ and $\texttt{Ab}_{\texttt{Staff}}$ in parallel. Assume that $\texttt{hasAccessTo}$ is varying, and $\texttt{User}$, $\texttt{Staff}$, and $\texttt{BlacklistedStaff}$ are fixed. Then, the result of parallel minimization is that staff members may or may not have access to confidential files, with equal preference. In the first case, they are abnormal users, and in the second case, they are abnormal staff. However, one may argue that the first option should be preferred: since $\texttt{Staff} \sqsubseteq \texttt{User}$ (but not the other way round), the normality information for staff is more *specific* than the normality information for users and should have higher priority. Such effects are well-known also from the propositional/first-order case and indeed, circumscription has soon after its introduction been extended with priorities to address issues of specificity (McCarthy, 1986).

In our formalism, users can specify priorities between minimized predicates. Typically, these priorities will reflect the subsumption hierarchy (as computed w.r.t. the class of *all* models). Since the subsumption hierarchy is in general a partial order, the priorities between minimized predicates may form a partial order, too. This approach is analogous to partially ordering priorities on default rules, as proposed by Brewka (1994). It is more general than standard prioritized circumscription, which assumes a total ordering (McCarthy, 1986; Lifschitz, 1985), and a special case of *nested circumscription* (Lifschitz, 1995).

## 2.2 Circumscribed Knowledge Bases

To define DLs with circumscription, we start by introducing *circumscription patterns*. They describe how individual predicates are treated during minimization.

**Definition 1 (Circumscription pattern, $<_{\mathsf{CP}}$)** *A circumscription pattern is a tuple* $\mathsf{CP}$ *of the form* $(\prec, M, F, V)$, *where* $\prec$ *is a strict partial order over* $M$, *and* $M$, $F$, *and* $V$ *are mutually disjoint subsets of* $\mathsf{N_C} \cup \mathsf{N_R}$, *the* minimized, fixed, *and* varying *predicates, respectively. By* $\preceq$, *we denote the reflexive closure of* $\prec$. *Define a preference relation* $<_{\mathsf{CP}}$ *on interpretations by setting* $\mathcal{I} <_{\mathsf{CP}} \mathcal{J}$ *iff the following conditions hold:*

1. $\Delta^{\mathcal{I}} = \Delta^{\mathcal{J}}$ *and, for all* $a \in \mathsf{N_I}$, $a^{\mathcal{I}} = a^{\mathcal{J}}$,

2. *for all* $p \in F$, $p^{\mathcal{I}} = p^{\mathcal{J}}$,

3. *for all* $p \in M$, *if* $p^{\mathcal{I}} \not\subseteq p^{\mathcal{J}}$ *then there exists* $q \in M$, $q \prec p$, *such that* $q^{\mathcal{I}} \subset q^{\mathcal{J}}$,

4. *there exists* $p \in M$ *such that* $p^{\mathcal{I}} \subset p^{\mathcal{J}}$ *and for all* $q \in M$ *such that* $q \prec p$, $q^{\mathcal{I}} = q^{\mathcal{J}}$.

*When* $M \cup F \subseteq \mathsf{N_C}$ *(i.e., the minimized and fixed predicates are all concepts) we call* $(\prec, M, F, V)$ *a concept circumscription pattern.* $\triangle$

We use the term *concept circumscription* if only concept circumscription patterns are admitted. Based on circumscription patterns, we can define circumscribed DL knowledge bases and their models.

**Definition 2 (Circumscribed KB)** *A circumscribed knowledge base (cKB) is an expression* $\mathsf{Circ_{CP}}(\mathcal{T}, \mathcal{A})$, *where* $\mathcal{T}$ *is a TBox,* $\mathcal{A}$ *an ABox, and* $\mathsf{CP} = (\prec, M, F, V)$ *a circumscription pattern such that* $M, F, V$ *partition the predicates used in* $\mathcal{T}$ *and* $\mathcal{A}$. *An interpretation* $\mathcal{I}$ *is a* model *of* $\mathsf{Circ_{CP}}(\mathcal{T}, \mathcal{A})$ *if it is a model of* $\mathcal{T}$ *and* $\mathcal{A}$ *and there exists no model* $\mathcal{I}'$ *of* $\mathcal{T}$ *and* $\mathcal{A}$ *such that* $\mathcal{I}' <_{\mathsf{CP}} \mathcal{I}$.

*A cKB* $\mathsf{Circ_{CP}}(\mathcal{T}, \mathcal{A})$ *is called a* concept-circumscribed KB *if* $\mathsf{CP}$ *is a concept circumscription pattern.* $\triangle$

Note that partially ordered circumscription becomes standard parallel circumscription if the empty relation is used for $\prec$.

The main reasoning tasks for (non-circumscribed) KBs are satisfiability of concepts w.r.t. KBs, subsumption w.r.t. KBs, and instance checking w.r.t. KBs. These reasoning tasks are fundamental for circumscribed KBs as well. We now provide precise definitions of these tasks. Throughout this and the following section, $\mathcal{DL}$ denotes the set of DLs introduced in the previous section; i.e., $\mathcal{ALC}$, $\mathcal{ALCI}$, $\mathcal{ALCO}$, $\mathcal{ALCQ}$, $\mathcal{ALCQI}$, $\mathcal{ALCIO}$, $\mathcal{ALCQO}$, and $\mathcal{ALCQIO}$.

**Definition 3 (Reasoning tasks)**

- *A concept $C$ is* satisfiable *w.r.t. a cKB $\mathsf{Circ}_{\mathsf{CP}}(\mathcal{T}, \mathcal{A})$ if some model $\mathcal{I}$ of $\mathsf{Circ}_{\mathsf{CP}}(\mathcal{T}, \mathcal{A})$ satisfies $C^{\mathcal{I}} \neq \emptyset$. Let $\mathcal{L} \in \mathcal{DL}$. The* satisfiability problem *w.r.t. cKBs in $\mathcal{L}$ is defined as follows: given a concept $C$ in $\mathcal{L}$ and a cKB $\mathsf{Circ}_{\mathsf{CP}}(\mathcal{T}, \mathcal{A})$ in $\mathcal{L}$, decide whether $C$ is satisfiable w.r.t. $\mathsf{Circ}_{\mathsf{CP}}(\mathcal{T}, \mathcal{A})$.*

- *A concept $C$ is* subsumed by *a concept $D$ w.r.t. a cKB $\mathsf{Circ}_{\mathsf{CP}}(\mathcal{T}, \mathcal{A})$, in symbols $\mathsf{Circ}_{\mathsf{CP}}(\mathcal{T}, \mathcal{A}) \models C \sqsubseteq D$, if $C^{\mathcal{I}} \subseteq D^{\mathcal{I}}$ for all models $\mathcal{I}$ of $\mathsf{Circ}_{\mathsf{CP}}(\mathcal{T}, \mathcal{A})$. Let $\mathcal{L} \in \mathcal{DL}$. The* subsumption problem *w.r.t. cKBs in $\mathcal{L}$ is defined as follows: given concepts $C$ and $D$ in $\mathcal{L}$ and a cKB $\mathsf{Circ}_{\mathsf{CP}}(\mathcal{T}, \mathcal{A})$ in $\mathcal{L}$, decide whether $\mathsf{Circ}_{\mathsf{CP}}(\mathcal{T}, \mathcal{A}) \models C \sqsubseteq D$.*

- *An individual name $a$ is an* instance of *a concept $C$ w.r.t. a cKB $\mathsf{Circ}_{\mathsf{CP}}(\mathcal{T}, \mathcal{A})$, in symbols $\mathsf{Circ}_{\mathsf{CP}}(\mathcal{T}, \mathcal{A}) \models C(a)$, if $a^{\mathcal{I}} \in C^{\mathcal{I}}$ for all models $\mathcal{I}$ of $\mathsf{Circ}_{\mathsf{CP}}(\mathcal{T}, \mathcal{A})$. Let $\mathcal{L} \in \mathcal{DL}$. The* instance problem *w.r.t. cKBs in $\mathcal{L}$ is defined as follows: given a concept $C$ in $\mathcal{L}$, an individual name $a$, and a cKB $\mathsf{Circ}_{\mathsf{CP}}(\mathcal{T}, \mathcal{A})$ in $\mathcal{L}$, decide whether $\mathsf{Circ}_{\mathsf{CP}}(\mathcal{T}, \mathcal{A}) \models C(a)$.*

$\triangle$

These reasoning problems can be polynomially reduced to one another: first, $C$ is satisfiable w.r.t. $\mathsf{Circ}_{\mathsf{CP}}(\mathcal{T}, \mathcal{A})$ iff $\mathsf{Circ}_{\mathsf{CP}}(\mathcal{T}, \mathcal{A}) \not\models C \sqsubseteq \bot$, and $\mathsf{Circ}_{\mathsf{CP}}(\mathcal{T}, \mathcal{A}) \models C \sqsubseteq D$ iff $C \sqcap \neg D$ is not satisfiable w.r.t. $\mathsf{Circ}_{\mathsf{CP}}(\mathcal{T}, \mathcal{A})$. And second, $C$ is satisfiable w.r.t. $\mathsf{Circ}_{\mathsf{CP}}(\mathcal{T}, \mathcal{A})$ iff $\mathsf{Circ}_{\mathsf{CP}}(\mathcal{T}, \mathcal{A}) \not\models \neg C(a)$, where $a$ is an individual name not appearing in $\mathcal{T}$ and $\mathcal{A}$; conversely, we have $\mathsf{Circ}_{\mathsf{CP}}(\mathcal{T}, \mathcal{A}) \models C(a)$ iff $A \sqcap \neg C$ is not satisfiable w.r.t. $\mathsf{Circ}_{\mathsf{CP}'}(\mathcal{T}, \mathcal{A} \cup \{A(a)\})$, where $A$ is a concept name not occurring in $\mathcal{T}$ and $\mathcal{A}$, and $\mathsf{CP}'$ is obtained from $\mathsf{CP}$ by adding $A$ to $M$ (and leaving $\prec$ as it is). In this paper, we use satisfiability w.r.t. cKBs as the basic reasoning problem.

## 3. Basic Reductions

We present three basic reductions between reasoning problems for circumscribed knowledge bases that are interesting in their own right and, additionally, will be useful for establishing the main results of this paper later on. More precisely, we replay a well-known reduction of fixed predicates to minimized predicates in the context of DLs, reduce reasoning w.r.t. cKBs with general TBoxes to reasoning w.r.t. cKBs with acyclic TBoxes, and show that, under certain conditions, simultaneous satisfiability w.r.t. a collection of cKBs is reducible to satisfiability w.r.t. a single cKB.

### 3.1 Fixed and minimized concepts

In circumscription, it is folklore that fixed predicates can be simulated in terms of minimized predicates, see e.g. de Kleer (1989). In the case of DLs, the same simulation is possible for concept names. To see this, let $C_0$ be a concept and $\mathsf{Circ}_{\mathsf{CP}}(\mathcal{T}, \mathcal{A})$ a circumscribed KB with $\mathsf{CP} = (\prec, M, F, V)$ and $F_0 = \{A_1, \dots, A_k\} = F \cap \mathsf{N_C}$. Define a new pattern $\mathsf{CP}' = (\prec, M', F \setminus F_0, V)$ with

- $M' = M \cup \{A_1, \dots, A_k, A_1', \dots, A_k'\}$, where $A_1', \dots, A_k'$ are concept names that do not occur in $C_0$, $M$, $F$, $V$, $\mathcal{T}$, and $\mathcal{A}$;

- $\mathcal{T}' = \mathcal{T} \cup \{A_i' \doteq \neg A_i \mid 1 \leq i \leq k\}$.

It is not difficult to see that $C_0$ is satisfiable w.r.t. $\mathsf{Circ}_{\mathsf{CP}}(\mathcal{T}, \mathcal{A})$ iff it is satisfiable w.r.t. $\mathsf{Circ}_{\mathsf{CP}'}(\mathcal{T}', \mathcal{A})$. Thus, we get the following result.

**Lemma 4** *Let $\mathcal{L} \in \mathcal{DL}$. Then satisfiability w.r.t. (concept-)circumscribed KBs in $\mathcal{L}$ can be polynomially reduced to satisfiability w.r.t. (concept-)circumscribed KBs in $\mathcal{L}$ that have no fixed concept names.*

In contrast to concept names, fixed role names cannot be reduced to minimized role names since Boolean operators on roles are not available in standard DLs such as $\mathcal{ALCQIO}$. A proof is given in Section 6, where we show that, in some cases, reasoning with minimized role names is decidable, whereas the corresponding reasoning task for cKBs with fixed role names is undecidable.

The reduction above clearly relies on TBoxes. However, in this paper we will sometimes work with circumscribed KBs in which the TBox is empty. The following lemma, proved in the appendix, shows that for cKBs in $\mathcal{ALC}$ without fixed role names and with empty TBox, one can simulate fixed concept names using minimized concept names without introducing a TBox. The proof, which may be viewed as a much more careful version of the proof of Lemma 4, can be adapted to yield an analogous result for the other logics in $\mathcal{DL}$.

**Lemma 5** *In $\mathcal{ALC}$, satisfiability w.r.t. (concept-)circumscribed KBs with empty TBox and without fixed roles can be polynomially reduced to satisfiability w.r.t. (concept-)circumscribed KBs with empty TBox and without fixed predicates.*

### 3.2 Acyclic and General TBoxes

For many DLs, satisfiability w.r.t. (non-circumscribed) KBs with general TBoxes is harder than satisfiability w.r.t. (non-circumscribed) KBs with acyclic TBoxes. In the case of $\mathcal{ALC}$, $\mathcal{ALCI}$, $\mathcal{ALCQ}$, and $\mathcal{ALCQO}$, the latter problem is PSPACE-complete (Baader, McGuiness, Nardi, & Patel-Schneider, 2003; Baader, Milicic, Lutz, Sattler, & Wolter, 2005b; Y. Ding & Wu, 2007) while the former is EXPTIME-complete (Baader et al., 2003). The only DLs considered in this paper for which satisfiability is EXPTIME-hard already with acyclic TBoxes are $\mathcal{ALCIO}$ and its extensions (Areces, Blackburn, & Marx, 2000). We show that, for circumscribed KBs, there is no difference in computational complexity between acyclic and general TBoxes.

Let $C_0$ be a concept and $\mathsf{Circ}_{\mathsf{CP}}(\mathcal{T}, \mathcal{A})$ a cKB with $\mathsf{CP} = (\prec, M, F, V)$. We may assume without loss of generality that $\mathcal{T} = \{\top \doteq C\}$ for some concept $C$. (To see this, observe that axioms $C \sqsubseteq D$ are equivalent to $\top \doteq \neg C \sqcup D$.) Define

- an acyclic TBox $\mathcal{T}' = \{A \doteq C, B \doteq \exists u.\neg A, A' \doteq \neg A, B' \doteq \neg B\}$, where $A, B, A', B', u$ are new concept and role names not occurring in $\mathcal{T}$, $\mathcal{A}$, $M$, $F$, $V$, and $C_0$.

- a circumscription pattern $\mathsf{CP}' = (\prec, M', F, V')$, where $M' = M \cup \{A', B'\}$ and $V' = V \cup \{A, B, u\}$.

We will ad $B'$ conjunctively to $C_0$ and thus be interested in models of $\mathsf{Circ}_{\mathsf{CP}'}(\mathcal{T}', \mathcal{A})$ where $(B')^{\mathcal{I}} \neq \emptyset$. In such models, we have $A^{\mathcal{I}} = \Delta^{\mathcal{I}}$ (and thus $C^{\mathcal{I}} = \Delta^{\mathcal{I}}$) since, otherwise, we can turn each instance $d$ of $B'$ into an instance of $\neg B'$ by making $d$ an instance of $B$ and linking it via the role $u$ to an instance of $\neg A$, thus obtaining a more preferred model w.r.t. $<_{\mathsf{CP}'}$. This is the basis of the proof of the following lemma, given in the appendix.

**Lemma 6** $C_0$ *is satisfiable w.r.t.* $\mathsf{Circ}_{\mathsf{CP}}(\mathcal{T}, \mathcal{A})$ *iff* $C_0 \sqcap B'$ *is satisfiable w.r.t.* $\mathsf{Circ}_{\mathsf{CP}'}(\mathcal{T}', \mathcal{A})$.

Thus, we have obtained the following result.

**Proposition 7** *Let* $\mathcal{L} \in \mathcal{DL}$. *Satisfiability w.r.t. (concept-)circumscribed KBs in* $\mathcal{L}$ *can be polynomially reduced to satisfiability w.r.t. (concept-)circumscribed KBs in* $\mathcal{L}$ *with acyclic TBoxes and without changing the ABox.*

This shows that satisfiability w.r.t. cKBs with acyclic TBoxes is of the same complexity as satisfiability w.r.t. cKBs with general TBoxes. In many cases considered in this paper, the same is even true for cKBs with empty TBoxes, c.f. Section 4. However, we also identify cases where cKBs with non-empty TBoxes have higher complexity (see Theorems 24 and 28), and thus a general reduction as the one underlying Proposition 7 cannot exist for the case of empty TBoxes.

### 3.3 Simultaneous Satisfiability

In applications, it is often necessary to merge TBoxes, ABoxes, and whole knowledge bases by taking their union. We show that, under certain conditions, reasoning w.r.t. the union of several circumscribed KBs can be reduced to reasoning w.r.t. the component cKBs. A concept $C$ is *simultaneously satisfiable w.r.t. cKBs* $\mathsf{Circ}_{\mathsf{CP}_1}(\mathcal{T}_1, \mathcal{A}_1), \ldots, \mathsf{Circ}_{\mathsf{CP}_k}(\mathcal{T}_k, \mathcal{A}_k)$ if there exists an interpretation $\mathcal{I}$ that is a model of all the cKBs and satisfies $C^{\mathcal{I}} \neq \emptyset$. The following lemma says that simultaneous satisfiability can be polynomially reduced to satisfiability w.r.t. a single cKB if there are no two cKBs that share a role name.

The proof idea for the case $k = 2$ is as follows. Given $\mathsf{Circ}_{\mathsf{CP}_1}(\mathcal{T}_1, \mathcal{A}_1)$ and $\mathsf{Circ}_{\mathsf{CP}_2}(\mathcal{T}_2, \mathcal{A}_2)$, we first take the union of these two cKBs, replacing in $\mathsf{Circ}_{\mathsf{CP}_2}(\mathcal{T}_2, \mathcal{A}_2)$ each concept name $A$ that is also used in $\mathsf{Circ}_{\mathsf{CP}_1}(\mathcal{T}_1, \mathcal{A}_1)$ with a fresh concept name $A'$. We then introduce an additional concept name $P$ (for 'problem') and make sure that $P$ is satisfied by each ABox individual whenever there is a point in the model where the interpretation of $A$ and $A'$ disagrees. We then look for a model where $P$ is not satisfied in the ABox. Intuitively, the additional concept name $P$ satisfies the purpose of 'decoupling' $A$ and $A'$, which is important e.g. in the case where $A/A'$ is minimized both in $\mathsf{Circ}_{\mathsf{CP}_1}(\mathcal{T}_1, \mathcal{A}_1)$ and $\mathsf{Circ}_{\mathsf{CP}_2}(\mathcal{T}_2, \mathcal{A}_2)$. Details are given in the appendix.

**Lemma 8** *For all* $\mathcal{L} \in \mathcal{DL}$, *simultaneous satisfiability w.r.t. (concept-)circumscribed KBs* $\mathsf{Circ}_{\mathsf{CP}_1}(\mathcal{T}_1, \mathcal{A}_1), \ldots \mathsf{Circ}_{\mathsf{CP}_k}(\mathcal{T}_k, \mathcal{A}_k)$, *such that* $\mathsf{Circ}_{\mathsf{CP}_i}(\mathcal{T}_i, \mathcal{A}_i)$ *and* $\mathsf{Circ}_{\mathsf{CP}_j}(\mathcal{T}_j, \mathcal{A}_j)$ *share no role names for* $1 \leq i < j \leq k$, *can be reduced in polynomial time to satisfiability w.r.t. single (concept-)circumscribed KBs.*

## 4. The Complexity of Reasoning in Concept-Circumscribed KBs

The main contributions of this paper consist in (i) showing that, in many cases, reasoning with circumscribed knowledge bases is decidable; and (ii) performing a detailed analysis of the computational complexity of these decidable cases. In this section, we show that satisfiability w.r.t. concept-circumscribed KBs is $\text{NExp}^{\text{NP}}$-complete for the DL $\mathcal{ALC}$ and its extensions $\mathcal{ALCIO}$ and $\mathcal{ALCQO}$. We also show that it is $\text{NP}^{\text{NExP}}$-complete if the number of fixed and minimized concept names is bounded by a constant. We first present proofs of the upper bounds and then establish matching lower bounds.

## 4.1 Upper Bounds

We start with the general case in which there is no bound on the number of fixed and minimized predicates.

### 4.1.1 THE GENERAL CASE

We prepare the upper bound proof by showing that if a concept is satisfiable w.r.t. a concept-circumscribed KB, then it is satisfiable in a model of bounded size. We use $|C|$ to denote the length of the concept $C$, i.e., the number of (occurrences of) symbols needed to write $C$. The *size* $|\mathcal{T}|$ of a TBox $\mathcal{T}$ is $\sum_{C \sqsubseteq D \in \mathcal{T}} |C| + |D|$, and the *size* $|\mathcal{A}|$ of an ABox $\mathcal{A}$ is the sum of the sizes of all assertions in $\mathcal{A}$, where the size of each role assertion is 1 and the size of concept assertions $C(a)$ is $|C|$.

**Lemma 9** *Let $C_0$ be a concept, $\text{Circ}_{\text{CP}}(\mathcal{T}, \mathcal{A})$ a concept-circumscribed KB, and $n := |C_0| + |\mathcal{T}| + |\mathcal{A}|$. If $C_0$ is satisfiable w.r.t. $\text{Circ}_{\text{CP}}(\mathcal{T}, \mathcal{A})$, then the following holds:*

*(i) If $\mathcal{T}$, $\mathcal{A}$ and $C_0$ are formulated in $\mathcal{ALCIO}$, then $C_0$ is satisfied in a model $\mathcal{I}$ of $\text{Circ}_{\text{CP}}(\mathcal{T}, \mathcal{A})$ with $\#\Delta^{\mathcal{I}} \leq 2^{2n}$.*

*(ii) If $\mathcal{T}$, $\mathcal{A}$ and $C_0$ are formulated in $\mathcal{ALCQO}$ and $m$ is the maximal parameter occurring in a number restriction in $\mathcal{T}$, $\mathcal{A}$, or $C_0$, then $C_0$ is satisfied in a model $\mathcal{I}$ of $\text{Circ}_{\text{CP}}(\mathcal{T}, \mathcal{A})$ with $\#\Delta^{\mathcal{I}} \leq 2^{2n} \times (m + 1) \times n$.*

**Proof.** Let $\text{CP}$, $\mathcal{T}$, $\mathcal{A}$, and $C_0$ be as in Lemma 9. We may assume that $\mathcal{A} = \emptyset$ as every assertion $C(a)$ can be expressed as an implication $\{a\} \sqsubseteq C$, and every assertion $r(a, b)$ can be expressed as $\{a\} \sqsubseteq \exists r.\{b\}$. Denote by $\text{cl}(C, \mathcal{T})$ the smallest set of concepts that contains all subconcepts of $C$, all subconcepts of concepts appearing in $\mathcal{T}$, and is closed under single negations (i.e., if $D \in \text{cl}(C, \mathcal{T})$ and $D$ does not start with $\neg$, then $\neg D \in \text{cl}(C, \mathcal{T})$).

Let $\mathcal{I}$ be a common model of $C_0$ and $\text{Circ}_{\text{CP}}(\mathcal{T}, \mathcal{A})$, and let $d_0 \in C_0^{\mathcal{I}}$. Define an equivalence relation "$\sim$" on $\Delta^{\mathcal{I}}$ by setting $d \sim d'$ iff

$$\{C \in \text{cl}(C_0, \mathcal{T}) \mid d \in C^{\mathcal{I}}\} = \{C \in \text{cl}(C_0, \mathcal{T}) \mid d' \in C^{\mathcal{I}}\}.$$

We use $[d]$ to denote the equivalence class of $d \in \Delta^{\mathcal{I}}$ w.r.t. the "$\sim$" relation. Pick from each equivalence class $[d]$ exactly one member and denote the resulting subset of $\Delta^{\mathcal{I}}$ by $\Delta'$.

We first prove Point (i). Thus, assume that $\mathcal{T}$ and $C_0$ are formulated in $\mathcal{ALCIO}$. We define a new interpretation $\mathcal{J}$ as follows:

$$
\begin{aligned}
\Delta^{\mathcal{J}} &:= \Delta' \\
A^{\mathcal{J}} &:= \{d \in \Delta' \mid d \in A^{\mathcal{I}}\} \\
r^{\mathcal{J}} &:= \{(d_1, d_2) \in \Delta' \times \Delta' \mid \exists d_1' \in [d_1], d_2' \in [d_2] : (d_1', d_2') \in r^{\mathcal{I}}\} \\
a^{\mathcal{J}} &:= d \in \Delta' \text{ if } a^{\mathcal{I}} \in [d].
\end{aligned}
$$

The following claim is easily proved using induction on the structure of $C$.

**Claim**: For all $C \in \mathsf{cl}(C_0, \mathcal{T})$ and all $d \in \Delta^{\mathcal{I}}$, we have $d \in C^{\mathcal{I}}$ iff $d' \in C^{\mathcal{J}}$ for the element $d' \in [d]$ of $\Delta^{\mathcal{J}}$.

Thus, $\mathcal{J}$ is a model of $\mathcal{T}$ satisfying $C_0$. To show that $\mathcal{J}$ is a model of $\mathsf{Circ}_{\mathsf{CP}}(\mathcal{T}, \mathcal{A})$, it thus remains to show that there is no model $\mathcal{J}'$ of $\mathcal{T}$ with $\mathcal{J}' <_{\mathsf{CP}} \mathcal{J}$. Assume to the contrary that there is such a $\mathcal{J}'$. We define an interpretation $\mathcal{I}'$ as follows:

$$
\begin{aligned}
\Delta^{\mathcal{I}'} &:= \Delta^{\mathcal{I}} \\
A^{\mathcal{I}'} &:= \bigcup_{d \in A^{\mathcal{J}'}} [d] \\
r^{\mathcal{I}'} &:= \bigcup_{(d_1, d_2) \in r^{\mathcal{J}'}} [d_1] \times [d_2] \\
a^{\mathcal{I}'} &:= a^{\mathcal{I}}.
\end{aligned}
$$

It is a matter of routine to show the following:

**Claim**: For all concepts $C \in \mathsf{cl}(C_0, \mathcal{T})$ and all $d \in \Delta^{\mathcal{I}}$, we have $d \in C^{\mathcal{I}'}$ iff $d' \in C^{\mathcal{J}'}$ for the element $d' \in [d]$ from $\Delta^{\mathcal{J}}$.

It follows that $\mathcal{I}'$ is a model of $\mathcal{T}$. Observe that $A^{\mathcal{I}} \odot A^{\mathcal{I}'}$ iff $A^{\mathcal{J}} \odot A^{\mathcal{J}'}$ for each concept name $A$ and $\odot \in \{\supseteq, \subseteq\}$. Therefore – and since $\mathsf{CP}$ is a concept circumscription pattern – $\mathcal{I}' <_{\mathsf{CP}} \mathcal{I}$ follows from $\mathcal{J}' <_{\mathsf{CP}} \mathcal{J}$. We have derived a contradiction and conclude that $\mathcal{J}$ is a model of $\mathsf{Circ}_{\mathsf{CP}}(\mathcal{T}, \mathcal{A})$. Thus we are done since the size of $\mathcal{J}$ is bounded by $2^{2n}$.

Now for Point (ii). Pick, for each $d \in \Delta'$ and each concept $(\geqslant k \, r \, C) \in \mathsf{cl}(C_0, \mathcal{T})$ such that $d \in (\geqslant k \, r \, C)^{\mathcal{I}}$, $k$ elements from $\{d' \mid d' \in C^{\mathcal{I}}, (d, d') \in r^{\mathcal{I}}\}$. Also pick, for each concept $(\leqslant k \, r \, C) \in \mathsf{cl}(C_0, \mathcal{T})$ such that $d \in (\neg(\leqslant k \, r \, C))^{\mathcal{I}}$, $k + 1$ elements from $\{d' \mid d' \in C^{\mathcal{I}}, (d, d') \in r^{\mathcal{I}}\}$. Denote by $\Delta''$ the collection of the elements picked. Take for each $d \in \Delta'' \setminus \Delta'$ an element $d^s \in \Delta'$ such that $d \sim d^s$ and define an interpretation $\mathcal{J}$ by

$$
\begin{aligned}
\Delta^{\mathcal{J}} &:= \Delta' \cup \Delta'' \\
A^{\mathcal{J}} &:= \{d \in \Delta' \cup \Delta'' \mid d \in A^{\mathcal{I}}\} \\
r^{\mathcal{J}} &:= \{(d_1, d_2) \in \Delta' \times (\Delta' \cup \Delta'') \mid (d_1, d_2) \in r^{\mathcal{I}}\} \\
&\quad \cup \{(d_1, d_2) \in (\Delta'' \setminus \Delta') \times (\Delta' \cup \Delta'') \mid (d_1^s, d_2) \in r^{\mathcal{I}}\} \\
a^{\mathcal{J}} &:= d \text{ if } a^{\mathcal{I}} \in [d].
\end{aligned}
$$

The following claim is easily proved.

**Claim**: For all $C \in \mathsf{cl}(C_0, \mathcal{T})$, we have the following:

(i) for all $d, d' \in \Delta^{\mathcal{J}}$, if $d \sim d'$, then $d \in C^{\mathcal{J}}$ iff $d' \in C^{\mathcal{J}}$;

(ii) for all $d \in \Delta^{\mathcal{I}}$, we have $d \in C^{\mathcal{I}}$ iff $d' \in C^{\mathcal{J}}$ for some element $d' \in [d]$ of $\Delta^{\mathcal{J}}$.

Thus, $\mathcal{J}$ is a model of $\mathcal{T}$ satisfying $C_0$. To show that $\mathcal{J}$ is a model of $\mathsf{Circ}_{\mathsf{CP}}(\mathcal{T}, \mathcal{A})$, it thus remains to show that there is no model $\mathcal{J}'$ of $\mathcal{T}$ with $\mathcal{J}' <_{\mathsf{CP}} \mathcal{J}$. Assume to the contrary that there is such a $\mathcal{J}'$. We define an interpretation $\mathcal{I}'$. To this end, take for each $d \in \Delta^{\mathcal{I}} \setminus \Delta^{\mathcal{J}}$ the $d^p \in \Delta'$ such that $d \sim d^p$. Now define $\mathcal{I}'$ as follows

$$
\begin{aligned}
\Delta^{\mathcal{I}'} &:= \Delta^{\mathcal{I}} \\
A^{\mathcal{I}'} &:= A^{\mathcal{J}'} \cup \{d \in \Delta^{\mathcal{I}} \setminus \Delta^{\mathcal{J}} \mid d^p \in A^{\mathcal{J}'}\} \\
r^{\mathcal{I}'} &:= r^{\mathcal{J}'} \cup \{(d_1, d_2) \in (\Delta^{\mathcal{I}} \setminus \Delta^{\mathcal{J}}) \times \Delta^{\mathcal{I}} \mid (d_1^p, d_2) \in r^{\mathcal{J}'}\} \\
a^{\mathcal{I}'} &:= a^{\mathcal{I}}.
\end{aligned}
$$

Again, it is a matter of routine to show:

**Claim**: For all concepts $C \in \mathsf{cl}(C_0, \mathcal{T})$ and all $d \in \Delta^{\mathcal{I}}$, we have $d \in C^{\mathcal{I}'} \cap \Delta^{\mathcal{J}}$ iff $d \in C^{\mathcal{J}'}$ and $d \in C^{\mathcal{I}'} \cap (\Delta^{\mathcal{I}} \setminus \Delta^{\mathcal{J}})$ iff $d^p \in C^{\mathcal{J}'}$ for the element $d^p \in [d]$ from $\Delta'$.

It follows that $\mathcal{I}'$ is a model for $\mathcal{T}$. Observe that $A^{\mathcal{I}} \odot A^{\mathcal{I}'}$ iff $A^{\mathcal{J}} \odot A^{\mathcal{J}'}$ for each concept name $A$ and $\odot \in \{\supseteq, \subseteq\}$. Therefore – and since CP is a concept circumscription pattern – $\mathcal{I}' <_{\mathsf{CP}} \mathcal{I}$ follows from $\mathcal{J}' <_{\mathsf{CP}} \mathcal{J}$. We have derived a contradiction and conclude that $\mathcal{J}$ is a model of $\mathsf{Circ}_{\mathsf{CP}}(\mathcal{T}, \mathcal{A})$. Thus we are done since the size of $\mathcal{J}$ is clearly bounded by $2^{2n} \times (m+1) \times n$. ❏

It is interesting to note that the proof of Lemma 9 does not go through if role names are minimized or fixed. This problem cannot be overcome, as proved by the undecidability results presented in Sections 5 and 6.

Using the bounded model property just established, we can now prove decidability of reasoning with concept-circumscribed KBs formulated in $\mathcal{ALCIO}$ and $\mathcal{ALCQO}$. More precisely, Lemma 9 suggests a non-deterministic decision procedure for satisfiability w.r.t. concept circumscription patterns: simply guess an interpretation of bounded size and then check whether it is a model. It turns out that this procedure shows containment of satisfiability in the complexity class $\mathrm{NExp}^{\mathrm{NP}}$, which contains those problems that can be solved by a non-deterministic exponentially time-bounded Turing machine that has access to an NP oracle. It is known that $\mathrm{NExp} \subseteq \mathrm{NExp}^{\mathrm{NP}} \subseteq \mathrm{ExpSpace}$.

**Theorem 10** *In $\mathcal{ALCIO}$ and $\mathcal{ALCQO}$, it is in $\mathrm{NExp}^{\mathrm{NP}}$ to decide whether a concept is satisfiable w.r.t. a concept-circumscribed KB $\mathsf{Circ}_{\mathsf{CP}}(\mathcal{T}, \mathcal{A})$.*

**Proof.** It is not hard to see that there exists an NP algorithm that takes as input a cKB $\mathsf{Circ}_{\mathsf{CP}}(\mathcal{T}, \mathcal{A})$ and a finite interpretation $\mathcal{I}$, and checks whether $\mathcal{I}$ is *not* a model of $\mathsf{Circ}_{\mathsf{CP}}(\mathcal{T}, \mathcal{A})$: the algorithm first verifies in polynomial time whether $\mathcal{I}$ is a model of $\mathcal{T}$ and $\mathcal{A}$, answering "yes" if this is not the case. Otherwise, the algorithm guesses an interpretation $\mathcal{J}$ that has the same domain as $\mathcal{I}$ and interprets all individual names in the same way, and then checks whether (i) $\mathcal{J}$ is a model of $\mathcal{T}$ and $\mathcal{A}$, and (ii) $\mathcal{J} <_{\mathsf{CP}} \mathcal{I}$. It answers "yes" if both checks succeed, and "no" otherwise. Clearly, checking whether $\mathcal{J} <_{\mathsf{CP}} \mathcal{I}$ can be done in time polynomial w.r.t. the size of $\mathcal{J}$ and $\mathcal{I}$.

This NP algorithm may now be used as an oracle in a NExp-algorithm for deciding satisfiability of a concept $C_0$ w.r.t. a cKB $\mathsf{Circ}_{\mathsf{CP}}(\mathcal{T}, \mathcal{A})$: by Lemma 9, it suffices to guess an interpretation of size $2^{4k}$ with $k = |C_0| + |\mathcal{T}| + |\mathcal{A}|$,[3] and then use the NP algorithm to check whether $\mathcal{I}$ is a model of $\mathsf{Circ}_{\mathsf{CP}}(\mathcal{T}, \mathcal{A})$. This proves that concept satisfiability is in $\mathrm{NExp}^{\mathrm{NP}}$. ❏

---

3. The bound $2^{4k}$ clearly dominates the two bounds given in Parts (i) and (ii) of Lemma 9.

By the reductions given in Section 2, Theorem 10 yields co-NExp$^{\text{NP}}$ upper bounds for subsumption and the instance problem. We will show in Section 4.2 that these upper bounds are tight.

### 4.1.2 Bounded Number of Minimized and Fixed Predicates

Since NExp$^{\text{NP}}$ is a rather large complexity class, it is a natural question whether we can impose restrictions on concept circumscription such that reasoning becomes simpler. In the following, we identify such a case by considering concept-circumscribed KBs in which the number of minimized and fixed concept names is bounded by some constant. In this case, the complexity of satisfiability w.r.t. concept-circumscribed KBs drops to NP$^{\text{NExp}}$. For readers uninitiated to oracle complexity classes, we recall that NExp $\subseteq$ NP$^{\text{NExp}}$ $\subseteq$ NExp$^{\text{NP}}$, and that NP$^{\text{NExp}}$ is believed to be much less powerful than NExp$^{\text{NP}}$, see for example the work by Eiter et al. (2004).

To prove the NP$^{\text{NExp}}$ upper bound, we first introduce counting formulas as a common generalization of TBoxes and ABoxes.

**Definition 11 (Counting Formula)** *A counting formula $\phi$ is a Boolean combination of concept implications, ABox assertions $C(a)$, and cardinality assertions $(C = n)$ where $C$ is a concept and $n$ a non-negative integer. We use $\wedge$, $\vee$, $\neg$ and $\rightarrow$ to denote the Boolean operators of counting formulas. An interpretation $\mathcal{I}$ satisfies a cardinality assertion $(C = n)$ if $\#C^{\mathcal{I}} = n$. The satisfaction relation $\mathcal{I} \models \phi$ between models $\mathcal{I}$ and counting formulas $\phi$ is defined in the obvious way.* △

In the following, we assume that the integers occurring in cardinality assertions are coded in binary. The NP$^{\text{NExp}}$ algorithm to be devised will use an algorithm for satisfiability of (non-circumscribed) counting formulas as an oracle. Therefore, we should first determine the computational complexity of the latter. It follows from results by Tobies (2000) that, in $\mathcal{ALC}$, satisfiability of counting formulas is NExp-hard. A matching upper bound for the DLs $\mathcal{ALCIO}$ and $\mathcal{ALCQO}$ is obtained from the facts that (i) there is a polynomial translation of counting formulas formulated in these languages into C2, the two-variable fragment of first-order logic extended with counting quantifiers (Grädel, Otto, & Rosen, 1997; Pacholski, Szwast, & Tendera, 2000), and (ii) satisfiability in C2 is in NExp even if the numbers in counting quantifiers are coded in binary (Pratt-Hartmann, 2005).

**Theorem 12 (Tobies, Pratt)** *In $\mathcal{ALC}$, $\mathcal{ALCIO}$ and $\mathcal{ALCQO}$, satisfiability of counting formulas is NExp-complete even if numbers in number restrictions are coded in binary.*

We now establish the improved upper bound.

**Theorem 13** *Let $c$ be a constant. In $\mathcal{ALCIO}$ and $\mathcal{ALCQO}$, it is in NP$^{\text{NExp}}$ to decide satisfiability w.r.t. concept-circumscribed KBs $\mathsf{Circ}_{\mathsf{CP}}(\mathcal{T}, \mathcal{A})$, where $\mathsf{CP} = (\prec, M, F, V)$ is such that $\#M \leq c$ and $\#F \leq c$.*

**Proof.** Assume that we want to decide satisfiability of the concept $C_0$ w.r.t. the cKB $\mathsf{Circ}_{\mathsf{CP}}(\mathcal{T}, \mathcal{A})$, where $\mathsf{CP} = (\prec, M, F, V)$ with $\#M \leq c$ and $\#F \leq c$. By Lemma 4, we may assume that $F = \emptyset$ (we may have to increase the constant $c$ appropriately). We may assume without loss of generality that the cardinality of $M$ is exactly $c$. Thus, let $M = \{A_0, \ldots, A_c\}$. By Lemma 9, $C_0$ is satisfiable w.r.t. $\mathsf{Circ}_{\mathsf{CP}}(\mathcal{T}, \mathcal{A})$ iff there exists a model of $C_0$ and $\mathsf{Circ}_{\mathsf{CP}}(\mathcal{T}, \mathcal{A})$ of size $2^{4k}$, with $k = |C_0| + |\mathcal{T}| + |\mathcal{A}|$. Consider, for all $S \subseteq M$, the concept

$$C_S := \bigsqcap_{A \in S} A \sqcap \bigsqcap_{A \in \{A_1, \ldots, A_c\} \setminus S} \neg A.$$

As $c$ is constant, the number $2^c$ of such concepts is constant as well. Clearly, the sets $C_S^{\mathcal{I}}$, $S \subseteq M$, form a partition of the domain $\Delta^{\mathcal{I}}$ of any model $\mathcal{I}$. Introduce, for each concept name $B$ and role name $r$ in $\mathcal{T} \cup \mathcal{A}$, a fresh concept name $B'$ and a fresh role name $r'$, respectively. For a concept $C$, denote by $C'$ the result of replacing in $C$ each concept name $B$ and role name $r$ with $B'$ and $r'$, respectively. The primed versions $\mathcal{A}'$ and $\mathcal{T}'$ of $\mathcal{A}$ and $\mathcal{T}$ are defined analogously. Denote by $N$ the set of individual names in $\mathcal{T} \cup \mathcal{A} \cup \{C_0\}$.

The $\mathrm{NExp}$-oracle we are going to use in our algorithm checks whether a counting formula $\phi$ is satisfiable or not. Now, the $\mathrm{NP}^{\mathrm{NExp}}$-algorithm is as follows (we use $C \sqsubseteq D$ as an abbreviation for the counting formula $(C \sqsubseteq D) \wedge \neg(D \sqsubseteq C)$):

1. Guess

   - a sequence $(n_S \mid S \subseteq M)$ of numbers $n_S \leq 2^{4k}$ coded in binary;

   - for each individual name $a \in N$, exactly one set $S_a \subseteq M$;

   - a subset $E$ of $N \times N$.

2. By calling the oracle, check whether the counting formula $\phi_1$ is satisfiable, where $\phi_1$ is the conjunction over

   - $\mathcal{T} \cup \mathcal{A} \cup \{\neg(C_0 = 0)\}$;

   - $(C_S = n_S)$, for all $S \subseteq M$;

   - $C_{S_a}(a)$, for each $a \in N$;

   - $\{(\{a\} \sqsubseteq \{b\}) \mid (a, b) \in E\} \cup \{\neg(\{a\} \sqsubseteq \{b\}) \mid (a, b) \in N - E\}$.

3. By calling the oracle, check whether the counting formula $\phi_2$ is satisfiable, where $\phi_2$ is the conjunction over

   - $\mathcal{T}' \cup \mathcal{A}'$;

   - $(C_S = n_S)$, for all $S \subseteq M$ (note that we use the unprimed versions);

   - $C_{S_a}(a)$, for each individual name $a \in N$ (we use the unprimed versions);

   - $\{(\{a\} \sqsubseteq \{b\}) \mid (a, b) \in E\} \cup \{\neg(\{a\} \sqsubseteq \{b\}) \mid (a, b) \in N - E\}$;

   - for all $A \in M$,

     $$\neg(A' \sqsubseteq A) \rightarrow \bigvee_{B \in M, B \prec A} (B' \sqsubseteq B);$$

733

- and, finally,

$$\bigvee_{A \in M} \left( (A' \sqsubset A) \wedge \bigwedge_{B \in M, B \prec A} (B \doteq B') \right).$$

4. The algorithm states that $C_0$ is satisfiable in a model of $\mathsf{Circ}_{\mathsf{CP}}(\mathcal{T}, \mathcal{A})$ if, and only if, $\phi_1$ is satisfiable and $\phi_2$ is not satisfiable.

Using the fact that $c$ is fixed, is is not hard to verify that this is a $\mathrm{NP}^{\mathrm{NEXP}}$-algorithm. It remains to show correctness and completeness.

Suppose that there exists a model of $\mathsf{Circ}_{\mathsf{CP}}(\mathcal{T}, \mathcal{A})$ satisfying $C_0$. Then there is such a model $\mathcal{I}$ of size bounded by $2^{4k}$. Let the algorithm guess

- the numbers $n_S = \#C_S^{\mathcal{I}}$, $S \subseteq M$,

- the sets $S_a$ such that $a^{\mathcal{I}} \in C_{S_a}^{\mathcal{I}}$,

- the set $E = \{(a, b), (b, a) \mid a^{\mathcal{I}} = b^{\mathcal{I}}, a, b \in N\}$.

Clearly, $\phi_1$ is satisfied in $\mathcal{I}$. It remains to show that $\phi_2$ is unsatisfiable. But suppose there exists a model $\mathcal{J}$ satisfying $\phi_2$. By the definitions of $\phi_1$ and $\phi_2$, we may assume that

- $\Delta^{\mathcal{I}} = \Delta^{\mathcal{J}}$;

- $A^{\mathcal{I}} = A^{\mathcal{J}}$ for all $A \in M$;

- $a^{\mathcal{I}} = a^{\mathcal{J}}$ for all individual names $a$.

Moreover, as no unprimed role names occur in $\phi_2$ and the only unprimed concept names in $\phi_2$ are those in $M$, we may assume that the interpretation of all unprimed concept and role names in $\mathcal{I}$ and $\mathcal{J}$ coincide. Thus, $\mathcal{J}$ is a model of $\mathsf{Circ}_{\mathsf{CP}}(\mathcal{T}, \mathcal{A})$ satisfying $C_0$. But now define a model $\mathcal{J}'$ with domain $\Delta^{\mathcal{J}}$ by setting

- $a^{\mathcal{J}'} = a^{\mathcal{J}}$, for all individual names $a$;

- $r^{\mathcal{J}'} = (r')^{\mathcal{J}}$, for all role names $r$;

- $A^{\mathcal{J}'} = (A')^{\mathcal{J}}$, for all concept names $A$.

Then, by the conjunct under Item 1 of the definition of $\phi_2$, $\mathcal{J}'$ is a model for $\mathcal{A} \cup \mathcal{T}$. By Items 5 and 6 of the definition of $\phi_2$, $\mathcal{J}' <_{\mathsf{CP}} \mathcal{J}$, and we have derived a contradiction.

Conversely, suppose the algorithm says that there exists a model of $\mathsf{Circ}_{\mathsf{CP}}(\mathcal{T}, \mathcal{A})$ satisfying $C_0$. Then take a model $\mathcal{I}$ for $\phi_1$. By the conjunct under Item 1 of $\phi_1$, $\mathcal{I}$ is a model for $\mathcal{T} \cup \mathcal{A}$ satisfying $C_0$. It follows from the unsatisfiability of $\phi_2$ that $\mathcal{I}$ is a model for $\mathsf{Circ}_{\mathsf{CP}}(\mathcal{T}, \mathcal{A})$. ❏

As a corollary, we obtain co-$\mathrm{NP}^{\mathrm{NExp}}$ upper bounds for subsumption and the instance problem. A similar drop of complexity occurs in propositional logic, where satisfiability w.r.t. circumscribed theories is complete for $\mathrm{NP}^{\mathrm{NP}}$ and it is not difficult to see that bounding the minimized and fixed predicates allows us to find a $\mathrm{P}^{\mathrm{NP}}$ algorithm.

## 4.2 Lower Bounds

We prove lower complexity bounds for reasoning with concept-circumscribed KBs that match the upper bounds given in Section 4.1.

### 4.2.1 THE GENERAL CASE

As in Section 4.1, we start with the general case in which the number of fixed and minimized predicates is not bounded. Our aim is to establish two $\mathrm{NExp}^{\mathrm{NP}}$-lower bounds that both match the upper bound established in Theorem 10. The first bound is for satisfiability w.r.t. concept-circumscribed KBs that are formulated in $\mathcal{ALC}$ and have an empty TBox, but a non-empty ABox. The second bound is also for satisfiability w.r.t. concept-circumscribed KBs formulated in $\mathcal{ALC}$, but assumes an acyclic TBox and empty ABox. Both reductions work already in the case of an empty preference relation, and without any fixed predicates. Note that considering satisfiability of a concept $C$ w.r.t. a concept-circumscribed KB $\mathsf{Circ}_{\mathsf{CP}}(\mathcal{T}, \mathcal{A})$ with *both* $\mathcal{T}$ and $\mathcal{A}$ empty is not very interesting: it can be seen that $C$ is satisfiable w.r.t. $\mathsf{Circ}_{\mathsf{CP}}(\mathcal{T}, \mathcal{A})$ iff $C_{\perp}$ is satisfiable (without reference to any KB), where $C_{\perp}$ is the concept obtained from $C$ by replacing all minimized concept names with $\perp$.

The proof of our first result is by reduction of a succinct version of the problem co-CERT3COL, which is $\mathrm{NExp}^{\mathrm{NP}}$-complete (Eiter, Gottlob, & Mannila, 1997), to satisfiability w.r.t. concept-circumscribed KBs with empty TBox. Let us first introduce the regular (non-succinct) version of co-CERT3COL:

*Instance of size $n$*: an undirected graph $G$ on the vertices $\{0, 1, \ldots, n-1\}$ such that every edge is labelled with a disjunction of two literals over the Boolean variables $\{V_{i,j} \mid i, j < n\}$.

*Yes-Instance of size $n$*: an instance $G$ of size $n$ such that, for some truth value assignment $t$ to the Boolean variables, the graph $t(G)$ obtained from $G$ by including only those edges whose label evaluates to true under $t$ is not 3-colorable.

As shown by Stewart (1991), co-CERT3COL is complete for $\mathrm{NP}^{\mathrm{NP}}$. To obtain a problem complete for $\mathrm{NExp}^{\mathrm{NP}}$, Eiter et al. use the complexity upgrade technique: by encoding the input in a succinct form using Boolean circuits, the complexity is raised by one exponential to $\mathrm{NExp}^{\mathrm{NP}}$ (Eiter et al., 1997). More precisely, the succinct version co-CERT3COL$_S$ of co-CERT3COL is obtained by representing the input graph $G$ with nodes $\{0, \ldots, 2^n - 1\}$ as $4n + 3$ Boolean circuits with $2n$ inputs (and one output) each. The Boolean circuits are named $c_E$, $c_S^{(1)}$, $c_S^{(2)}$, and $c_j^{(i)}$, with $i \in \{1, 2, 3, 4\}$ and $j < n$. For all circuits, the $2n$ inputs are the bits of the binary representation of two nodes of the graph. The purpose of the circuits is as follows:

- circuit $c_E$ outputs 1 if there is an edge between the two input nodes, and 0 otherwise;

- if there is an edge between the input nodes, circuit $c_S^{(1)}$ outputs 1 if the first literal in the disjunction labelling this edge is positive, and 0 otherwise; circuit $c_S^{(2)}$ does the same for the second literal; if there is no edge between the input nodes, the output is arbitrary;

- if there is an edge between the input nodes, the circuits $c_j^{(i)}$ compute the labelling $V_{k_1,k_2} \vee V_{k_3,k_4}$ of this edge between the input nodes by generating the numbers $k_1, \ldots, k_4$: circuit $c_j^{(i)}$ outputs the $j$-th bit of $k_i$; if there is no edge between the input nodes, the output is arbitrary.

We reduce co-CERT3COL$_S$ to satisfiability w.r.t. concept-circumscribed KBs that are formulated in $\mathcal{ALC}$ and whose TBox and preference relation are empty. It then remains to apply Lemma 5 to eliminate fixed concept names (we note that the construction in the proof of the lemma leaves the preference relation untouched). Let

$$G = (n, c_E, c_S^{(1)}, c_S^{(2)}, \{c_j^{(i)}\}_{i \in \{1,..,4\}, j < n})$$

be the (succinct representation of the) input graph with $2^n$ nodes. We construct an ABox $\mathcal{A}_G = \{C_0 \sqcap \mathsf{Root}(a_0)\}$, a circumscription pattern $\mathsf{CP}_G$, and a concept $C_G$ such that $G$ is a yes-instance of co-CERT3COL$_S$ iff $C_G$ is satisfiable w.r.t. $\mathsf{Circ}_{\mathsf{CP}_G}(\emptyset, \mathcal{A}_G)$.

The concept $C_0$ used in $\mathcal{A}_G$ is a conjunction whose presentation is split into two parts. Intuitively, the purpose of the first group of conjuncts is to fix a truth assignment $t$ for the variables $\{V_{i,j} \mid i, j < n\}$, and to construct (an isomorphic image of) the graph $t(G)$ obtained from $G$ by including only those edges whose label evaluates to true under $t$. Then, the purpose of the second group is to make sure that $t(G)$ is not 3-colorable.

When formulating $C_0$, we use several binary counters for counting modulo $2^n$ (the number of nodes in the input graph). The main counters $X$ and $Y$ use concept names $X_0, \ldots, X_{n-1}$ and $Y_0, \ldots, Y_{n-1}$ as their bits, respectively. Additionally, we introduce concept names $K_0^{(i)}, \ldots, K_{n-1}^{(i)}$, $i \in \{1, 2, 3, 4\}$, that serve as four additional counters $K^{(1)}, \ldots, K^{(4)}$. The first group of conjuncts of $C_0$ can be found in Figure 2, where the following abbreviations are used:

- $\forall r^i.C$ denotes the $n$-fold nesting $\forall r. \cdots . \forall r.C$;

- $\forall r.(K^{(i)} = X)$ is an abbreviation for $\bigsqcap_{j<n} \left( (X_j \rightarrow \forall r.K_j^{(i)}) \sqcap (\neg X_j \rightarrow \forall r.\neg K_j^{(i)}) \right)$ and similarly for $\forall r.(K^{(i)} = Y)$;

- the abbreviations $W_c$, $c$ a Boolean circuit, are explained later on.

The intuition behind Figure 2 is as follows. Lines (1) to (5) build up a binary tree of depth $2n$ whose edges are labeled with the role name $r$. The $2^{2n}$ leaves of the tree are instances of the concept name $\mathsf{Leaf}$, and they are labeled with all possible values of the counters $X$ and $Y$. Since we will minimize $\mathsf{Leaf}$ via the circumscription pattern $\mathsf{CP}_G$, this concept name denotes *precisely* the leaves of the tree. Due to the use of the counters $X$ and $Y$, the leaves are all distinct.

The leaves of the tree just established satisfy a number of purposes. To start with, each leaf with counter values $X = i$ and $Y = j$ corresponds to the variable $V_{i,j}$ of co-3CERTCOL$_S$ and determines a truth value for this variable via truth/falsity of the concept

$$\forall r^i.(\exists r.X_i \sqcap \exists r.\neg X_i) \qquad \text{for } i < n \tag{1}$$

$$\forall r^j.((X_i \to \forall r.X_i) \sqcap (\neg X_i \to \forall r.\neg X_i)) \qquad \text{for } i < n, j < 2n \tag{2}$$

$$\forall r^{n+i}.(\exists r.Y_i \sqcap \exists r.\neg Y_i) \qquad i < n \tag{3}$$

$$\forall r^{n+j}.((Y_i \to \forall r.Y_i) \sqcap (\neg Y_i \to \forall r.\neg Y_i)) \qquad \text{for } i < j < n \tag{4}$$

$$\forall r^{2n}.\mathsf{Leaf} \tag{5}$$

$$\forall r^{2n}.(W_{c_E} \sqcap W_{c_S^{(1)}} \sqcap W_{c_S^{(2)}}) \tag{6}$$

$$\forall r^{2n}.(W_{c_j^{(1)}} \sqcap \cdots \sqcap W_{c_j^{(4)}}) \qquad \text{for } j < n \tag{7}$$

$$\forall r^{2n}.(\exists \mathsf{var1}.\mathsf{LeafFix} \sqcap \forall \mathsf{var1}.(K^{(1)}{=}X) \sqcap \forall \mathsf{var1}.(K^{(2)}{=}Y)) \tag{8}$$

$$\forall r^{2n}.(\exists \mathsf{var2}.\mathsf{LeafFix} \sqcap \forall \mathsf{var2}.(K^{(3)}{=}X) \sqcap \forall \mathsf{var2}.(K^{(4)}{=}Y)) \tag{9}$$

$$P \leftarrow \exists r^{2n}.\exists \mathsf{var1}.\neg \mathsf{Leaf} \tag{10}$$

$$P \leftarrow \exists r^{2n}.\exists \mathsf{var2}.\neg \mathsf{Leaf} \tag{11}$$

$$\forall r^{2n}.(S_1 \to (\mathsf{Tr}_1 \leftrightarrow \forall \mathsf{var1}.\mathsf{Tr})) \tag{12}$$

$$\forall r^{2n}.(\neg S_1 \to (\neg \mathsf{Tr}_1 \leftrightarrow \forall \mathsf{var1}.\mathsf{Tr})) \tag{13}$$

$$\forall r^{2n}.(S_2 \to (\mathsf{Tr}_2 \leftrightarrow \forall \mathsf{var2}.\mathsf{Tr})) \tag{14}$$

$$\forall r^{2n}.(\neg S_2 \to (\neg \mathsf{Tr}_2 \leftrightarrow \forall \mathsf{var1}.\mathsf{Tr})) \tag{15}$$

$$\forall r^{2n}.(\mathsf{Elim} \leftrightarrow (\neg E \sqcup \neg(\mathsf{Tr}_1 \sqcup \mathsf{Tr}_2))) \tag{16}$$

Figure 2: The first group of conjuncts of $C_0$.

name $\mathsf{Tr}$. Thus, the leaves jointly describe a truth assignment $t$ for the instance $G$ of co-3CERTCOL$_S$. A second purpose of the leaves is to represent the potential edges of $G$: additionally to representing a variable, a leaf with $X = i$ and $Y = j$ corresponds to the potential edge between the nodes $i$ and $j$. To explain this more properly, we must first discuss the abbreviations $W_c$ used in Lines (6) and (7) of Figure 2.

Each concept $W_c$, $c$ a Boolean circuit with $2n$ inputs, is the result of converting $c$ into a concept that uses only the constructors $\neg, \sqcap, \sqcup$ such that the following condition is satisfied: if a $d \in \Delta^{\mathcal{I}}$ is an instance of $W_c$, the output of $c$ upon input $b_0, \ldots, b_{2n-1}$ is $b$, and the truth value of the concept names $X_0, \ldots, X_{n-1}, Y_0, \ldots, Y_{n-1}$ at $d$ is described by $b_0, \ldots, b_{2n-1}$, then the truth value of some concept name $\mathsf{Out}$ at $d$ is described by $b$. By introducing one auxiliary concept name for every inner gate of $c$, the translation can be done such that the size of $W_c$ is linear in the size of $c$. The following concept names are used as output:

- $W_{c_E}$ uses the concept name $E$ as output;

- $W_{c_S^{(i)}}$ uses the concept name $S_i$ as output, for $i \in \{1, 2\}$;

- $W_{c_j^{(i)}}$ uses the concept name $K_j^{(i)}$ as output, for $i \in \{1, \ldots, 4\}$ and $j < n$.

$$\forall r^{2n}.(\exists\mathsf{col1}.\mathsf{LeafFix} \sqcap \exists\mathsf{col2}.\mathsf{LeafFix}) \tag{17}$$

$$\forall r^{2n}.(\forall\mathsf{col1}.(X = X) \sqcap \forall\mathsf{col1}.(Y = 0)) \tag{18}$$

$$\forall r^{2n}.(\forall\mathsf{col2}.(Y = X) \sqcap \forall\mathsf{col2}.(Y = 0)) \tag{19}$$

$$P \leftarrow \exists r^{2n}.\exists\mathsf{col1}.\neg\mathsf{Leaf} \tag{20}$$

$$P \leftarrow \exists r^{2n}.\exists\mathsf{col2}.\neg\mathsf{Leaf} \tag{21}$$

$$\forall r^{2n}.((Y = 0) \rightarrow (R \sqcup B \sqcup G)) \tag{22}$$

$$\forall r^{2n}.((Y = 0) \rightarrow (\neg(R \sqcap B) \sqcap \neg(R \sqcap G) \sqcap \neg(B \sqcap G))) \tag{23}$$

$$\forall r^{2n}.((\neg\mathsf{Elim} \sqcap \exists\mathsf{col1}.R \sqcap \exists\mathsf{col2}.R) \rightarrow \mathsf{Clash}) \tag{24}$$

$$\forall r^{2n}.((\neg\mathsf{Elim} \sqcap \exists\mathsf{col1}.G \sqcap \exists\mathsf{col2}.G) \rightarrow \mathsf{Clash}) \tag{25}$$

$$\forall r^{2n}.((\neg\mathsf{Elim} \sqcap \exists\mathsf{col1}.B \sqcap \exists\mathsf{col2}.B) \rightarrow \mathsf{Clash}) \tag{26}$$

Figure 3: The second group of conjuncts of $C_0$.

Lines (6) and (7) ensure that these concepts are propagated to all leaves. Our next aim is to ensure that each leaf that represents a potential edge $(i, j)$ is connected via the role var1 to the leaf that represents the variable in the first disjunct of the label of $(i, j)$, and analogously for the role var2 and the variable in the second disjunct of the edge label. If we replaced the concept name LeafFix with Leaf in Lines (8) and (9), then these lines would apparently encode these properties. However, we have to be careful as the mentioned replacement would interact with the minimization of Leaf. To fix this problem, we resort to a trick: we use the concept name LeafFix instead of Leaf. In this way, we may or may not reach an instance of Leaf. If we do not, we force the concept name $P$ to be true at the root of the tree in Lines (10) and (11). We will use $C_G$ to rule out models in which $P$ is true. Finally, we fix LeafFix via $\mathsf{CP}_G$ to eliminate interaction with the minimization of Leaf.

The remaining Lines (12) to (16) ensure that a leaf is an instance of Elim iff the potential edge that it represents is not present in the graph $t(G)$ induced by the truth assignment $t$ described by the leaves.

The second group of conjuncts of $C_0$ can be found in Figure 3. Here, $(Y = 0)$ stands for the concept $(\neg Y_0 \sqcap \cdots \sqcap \neg Y_{n-1})$. As already mentioned, the purpose of these conjuncts is to ensure that the graph $t(G)$ described by the leaves does not have a 3-coloring. The strategy for ensuring this is as follows: we use the $2^n$ leaves with $Y = 0$ to store the colors of the nodes, i.e., the leaf with $X = i$ and $Y = 0$ stores the color of node $i$. By Lines (22) and (23), there is a unique coloring. Then, Lines (17) to (21) ensure that each leaf (viewed as an edge) is connected via the role col1 to the leaf that stores the color of the first node of the edge, and analogously for the role col2 and the second node of the edge. LeafFix and $P$ have the same role as before. Lines (24) to (26) guarantee that the concept name Clash identifies problems in the coloring: a leaf is in Clash if it represents an edge that exists in $G$, is not dropped in $t(G)$, and where both endpoints have the same color. The idea is that Clash will be minimized while $R$, $G$, and $B$ vary. When some additional concept names are fixed, this corresponds to a universal quantification over all possible colorings.

Set $C_G = \mathsf{Root} \sqcap \neg P \sqcap \exists r^{2n}.\mathsf{Clash}$, and recall that $\mathcal{A}_G = \{C_0 \sqcap \mathsf{Root}(a_0)\}$. The following lemma is proved in the appendix.

**Lemma 14** $G$ *is a yes-instance of co-3CERTCOL$_S$ iff $C_G$ is satisfiable with respect to* $\mathsf{Circ}_{\mathsf{CP}_G}(\emptyset, \mathcal{A}_G)$, *where* $\mathsf{CP}_G = (\prec, M, F, V)$ *with* $\prec = \emptyset$, $M = \{\mathsf{Root}, \mathsf{Leaf}, \mathsf{Clash}\}$,

$$F = \{\mathsf{LeafFix}, \mathsf{Tr}, X_0, \ldots, X_{n-1}, Y_0, \ldots, Y_{n-1}, \},$$

*and $V$ the set of all remaining predicates in $\mathcal{A}_G$.*

Since the size of $\mathcal{A}_G$ is polynomial in $n$, we get the following result by applying Lemma 5.

**Theorem 15** *In $\mathcal{ALC}$, satisfiability w.r.t. concept-circumscribed KBs is* $\mathrm{NExp}^{\mathrm{NP}}$-*hard even if the TBox and the preference relation are empty and there are no fixed predicates.*

It is now rather straightforward to establish the announced second $\mathrm{NExp}^{\mathrm{NP}}$ lower bound by a reduction of satisfiability w.r.t. concept-circumscribed KBs in the special case formulated in Theorem 15. Details are given in the appendix.

**Corollary 16** *In $\mathcal{ALC}$, satisfiability w.r.t. concept-circumscribed KBs is* $\mathrm{NExp}^{\mathrm{NP}}$-*hard even if the TBox is acyclic, the ABox and preference relations are empty, and there are no fixed predicates.*

Corresponding lower bounds for subsumption and the instance problems follow from the reduction given in Section 2.

### 4.2.2 Bounded Number of Minimized and Fixed Predicates

We now establish a matching lower bound for Theorem 13 by showing that, in $\mathcal{ALC}$, satisfiability w.r.t. concept-circumscribed KBs is $\mathrm{NP}^{\mathrm{NExp}}$-hard even if only a constant number of predicates is allowed to be minimized and fixed. In contrast to the previous section, we ignore the case of empty TBoxes and directly establish the lower bound for the case of non-empty TBoxes and empty ABoxes. This allows us to demonstrate the usefulness of Lemma 8 for separating different parts of a lower bound proof: in the main reduction from the previous section, the two parts of the reduction shown in Figure 2 and 3 are not truly independent, which forced us to implement the technical trick that involves the concept names $\mathsf{LeafFix}$ and $P$. When using Lemma 8, in contrast, we achieve a true separation of concerns. In general, though, we conjecture that the lower bound proved in this section can also be established for the case of empty TBoxes by adapting the mentioned technical trick. We leave this as a problem to the interested reader.

Recall that a (non-deterministic) $k$-*tape Turing machine* is described by a tuple

$$(Q, \Sigma, q_0, \Delta, q_{\mathsf{acc}}, q_{\mathsf{rej}}),$$

with $Q$ a finite set of states, $\Sigma$ a finite alphabet, $q_0 \in Q$ a starting state,

$$\Delta \subseteq Q \times \Sigma^k \times Q \times \Sigma^k \times \{L, R\}^k$$

a transition relation, and $q_{\mathsf{acc}}, q_{\mathsf{rej}} \in Q$ the accepting and rejecting states. For our purposes, an *oracle Turing machine* is a 2-tape Turing machine $\mathcal{M}$ that, additionally, is equipped with

- a 1-tape Turing machine $\mathcal{M}'$ (the *oracle*) whose alphabet is identical to that of $\mathcal{M}$,

- a query state $q_?$, and

- two answer states $q_{\text{yes}}$ and $q_{\text{no}}$.

The second tape of $\mathcal{M}$ is called the *oracle tape*. When $\mathcal{M}$ enters $q_?$, the oracle determines the next state of $\mathcal{M}$: if the content of the oracle tape is contained in the language accepted by the oracle, the next state is $q_{yes}$. Otherwise, it is $q_{no}$. During this transition, the head is not moved and no symbols are written. The state $q_?$ cannot occur as the left-most component of a tuple in $\mathcal{M}$'s transition relation.

Let $\mathcal{M} = (Q, \Sigma, q_0, \Delta, q_{\text{acc}}, q_{\text{rej}}, \mathcal{M}', q_?, q_{\text{yes}}, q_{\text{no}})$ be an oracle Turing machine such that the following holds:

- $\mathcal{M}$ solves an $\text{NP}^{\text{NExp}}$-complete problem;

- the time consumption of $\mathcal{M}$ is bounded by a polynomial $p$ (where oracle calls contribute with a single clock tick);

- the time consumption of $\mathcal{M}' = (Q', \Sigma, q_0', \Delta', q_{\text{acc}}', q_{\text{rej}}')$ is bounded by $2^{q(n)}$, with $q$ a polynomial.

We assume without loss of generality that $\mathcal{M}$ and $\mathcal{M}'$ never attempt to move left from the left-most position of the tape (neither right from the right-most position). Our $\text{NP}^{\text{NExp}}$-hardness proof uses a reduction of the word problem of $\mathcal{M}$. Thus, let $w \in \Sigma^*$ be an input for $\mathcal{M}$ of length $n$, and let $m = p(n)$ and $m' = q(p(n))$. We construct three TBoxes $\mathcal{T}_w$, $\mathcal{T}_w'$, and $\mathcal{T}_w''$, circumscription patterns $\text{CP}$, $\text{CP}'$, and $\text{CP}''$, and a concept $C_w$ such that $\mathcal{M}$ accepts $w$ iff $C_w$ is simultaneously satisfiable w.r.t. $\text{Circ}_{\text{CP}}(\mathcal{T}_w, \emptyset)$, $\text{Circ}_{\text{CP}'}(\mathcal{T}_w', \emptyset)$, and $\text{Circ}_{\text{CP}''}(\mathcal{T}_w'', \emptyset)$. Then, Lemma 8 yields a reduction to (non-simultaneous) satisfiability w.r.t. concept-circumscribed cKBs. Intuitively, the purpose of the first TBox $\mathcal{T}_w$ is to impose a basic structure on the domain, while $\mathcal{T}_w'$ describes computations of $\mathcal{M}$, and $\mathcal{T}_w''$ describes computations of $\mathcal{M}'$. We use general TBoxes rather than acyclic ones since, by Lemma 6, this can be done without loss of generality.

The TBox $\mathcal{T}_w$ is shown in Figure 4. As in the previous reduction, we use concept names $X_0, \ldots, X_{m'-1}$ and $Y_0, \ldots, Y_{m'-1}$ to implement two binary counters for counting modulo $2^{m'}$. We also use the same abbreviations as in the previous reduction. Additionally, $\forall r.(X{+}{+})$ states that the value of the counter $X_0, \ldots, X_{m'-1}$ is incremented when going to $r$-successors, i.e.,

$$\bigsqcap_{k=0..m'-1} \big( \bigsqcap_{j=0..k-1} X_j \big) \to \big( (X_k \to \forall r.\neg X_k) \sqcap (\neg X_k \to \forall r.X_k) \big)$$

$$\bigsqcap_{k=0..m'-1} \big( \bigsqcup_{j=0..k-1} \neg X_j \big) \to \big( (X_k \to \forall r.X_k) \sqcap (\neg X_k \to \forall r.\neg X_k) \big)$$

The purpose of Lines (27) to (30) is to ensure that, for each possible value $(i, j)$ of the counters $X$ and $Y$, there is at least one instance of $\text{NExp}$ that satisfies $(X = i)$ and $(Y = j)$. We minimize $\text{NExp}$, and thus enforce that $\text{NExp}$ has *exactly* $2^{m'} \times 2^{m'}$ elements. These elements are interconnected via the roles $r$ (for "right") and $u$ (for "up") to form a $2^{m'} \times 2^{m'}$-grid. Later on, we use this grid to encode computations of the oracle machine $\mathcal{M}'$. Observe

$$\top \sqsubseteq \exists\mathsf{aux}.\mathsf{NExp} \tag{27}$$

$$\mathsf{NExp} \sqsubseteq (\neg(X = 2^{m'} - 1) \to \exists r.\mathsf{NExp}) \sqcap (\neg(Y = 2^{m'} - 1) \to \exists u.\mathsf{NExp}) \tag{28}$$

$$\mathsf{NExp} \sqsubseteq \forall r.(Y{=}Y) \sqcap \forall r.(X{+}{+}) \tag{29}$$

$$\mathsf{NExp} \sqsubseteq \forall u.(X{=}X) \sqcap \forall u.(Y{+}{+}) \tag{30}$$

$$\top \sqsubseteq \bigsqcap_{i<m} \exists\mathsf{aux}.(\mathsf{Result} \sqcap R_i) \tag{31}$$

$$\mathsf{Result} \sqsubseteq \bigsqcap_{i<j<m} \neg(R_i \sqcap R_j) \tag{32}$$

$$\top \sqsubseteq \exists\mathsf{aux}.\mathsf{NP} \tag{33}$$

Figure 4: The TBox $\mathcal{T}_w$.

that, since we are working with simultaneous satisfiability, the minimization of $\mathsf{NExp}$ does not interact with anything that we are going to put into the TBoxes $\mathcal{T}'_w$ and $\mathcal{T}''_w$.

We also minimize the concept name $\mathsf{Result}$, and thus Lines (31) and (32) guarantee that there are exactly $m$ instances of $\mathsf{Result}$, identified by the concept names $R_0, \dots, R_{m-1}$. If $\mathcal{M}$ makes a call to the oracle in the $i$-th step, then the result of this call will be stored in the (unique) instance of $\mathsf{Result} \sqcap R_i$, i.e., this instance will satisfy the concept name $\mathsf{Rej}$ if $\mathcal{M}'$ rejected the input and falsify it otherwise. Finally, we also minimize $\mathsf{NP}$, and thus Line (33) guarantees that there is exactly one instance of $\mathsf{NP}$. This instance will be used to represent the computation of $\mathcal{M}$. Summing up, the circumscription pattern for $\mathcal{T}_w$ is

$$\mathsf{CP} := (\emptyset, \{\mathsf{NExp}, \mathsf{Result}, \mathsf{NP}\}, \emptyset, V),$$

with $V$ containing all remaining predicates used in $\mathcal{T}_w$.

The purpose of $\mathcal{T}'_w$ is to describe computations of $\mathcal{M}$. We use the following concept names:

- For all $a \in \Sigma$, $i, j < m$, and $k \in \{1, 2\}$, we introduce a concept name $S_a^{i,j,k}$. Intuitively, $S_a^{i,j,k}$ expresses that $a$ is the symbol in the $j$-th cell of the $k$-th tape in the $i$-th step of $\mathcal{M}$'s computation. We start our numbering of tape cells and steps with 0.

- For all $q \in Q$ and $i < m$, $Q_q^i$ is a concept name expressing that $\mathcal{M}$ is in state $q$ in the $i$-th step of the computation.

- For all $q \in Q$, $i, j < m$, and $k \in \{1, 2\}$, $H_j^{i,k}$ is a concept name expressing that the $k$-th head of $\mathcal{M}$ is on cell $j$ in the $i$-th step of the computation.

- For all $q \in Q$, $a \in \Sigma$, $i, j < m$, $k \in \{1, 2\}$, and $M \in \{L, R\}$, concept names $A_q^i$, $A_a^{i,j,k}$, $A_M^{i,k}$ that serve as markers. More precisely, $A_q^i$ means that, at time point $i$, $\mathcal{M}$ has executed a transition that switches to state $q$. Similarly, $A_a^{i,j,k}$ describes the symbol written in that transition on tape $k$, and $A_M^{i,k}$ describes the move on tape $k$.

The details of $\mathcal{T}'_w$ are given in Figure 5. One copy of the concept inclusions in this figure are needed for every $i, j, j' < m$ and every $k \in \{1, 2\}$. We assume that $w = a_0 \cdots a_{n-1}$ and

$$\text{NP} \sqsubseteq Q_{q_0}^0 \sqcap H_0^{0,1} \sqcap S_{a_0}^{0,0,1} \sqcap \cdots S_{a_{n-1}}^{0,n-1,1} \sqcap S_B^{0,n,1} \sqcap \cdots \sqcap S_B^{0,m-1,1} \tag{34}$$

$$\text{NP} \sqsubseteq H_0^{0,2} \sqcap S_B^{0,0,2} \sqcap \cdots \sqcap S_B^{0,m-1,2} \tag{35}$$

$$\text{NP} \sqsubseteq \bigsqcap_{a,b \in \Sigma, q \in Q \setminus \{q_?\}} \big( (S_a^{i,j,1} \sqcap S_b^{i,j',2} \sqcap Q_q^i \sqcap H_j^{i,1} \sqcap H_{j'}^{i,2}) \to \tag{36}$$

$$\bigsqcup_{(q,a,b,q',a',b',M,M') \in \Delta} A_q^i \sqcap A_a^{i,j,1} \sqcap A_b^{i,j',2} \sqcap A_M^{i,1} \sqcap A_{M'}^{i,2} \big) \big) \tag{37}$$

$$\text{NP} \sqsubseteq \bigsqcap_{q \in Q} A_q^i \to Q_q^{i+1} \quad \text{if } i < m-1 \tag{38}$$

$$\text{NP} \sqsubseteq \bigsqcap_{a \in \Sigma} A_a^{i,j,k} \to S_a^{i+1,j,k} \quad \text{if } i < m-1 \tag{39}$$

$$\text{NP} \sqsubseteq A_L^{i,k} \to H_{j-1}^{i+1,k} \quad \text{if } i < m-1 \text{ and } j > 0 \tag{40}$$

$$\text{NP} \sqsubseteq A_R^{i,k} \to H_{j+1}^{i+1,k} \quad \text{if } i < m-1 \text{ and } j < m-1 \tag{41}$$

$$\text{NP} \sqsubseteq \bigsqcap_{a \in \Sigma} \big( (S_a^{i,j,k} \sqcap H_{j'}^{i,k}) \to S_a^{i+1,j,k} \big) \quad \text{if } i < m-1 \text{ and } j \neq j' \tag{42}$$

$$\text{NP} \sqsubseteq \bigsqcap_{a,b \in \Sigma, a \neq b} \neg (S_a^{i,j,k} \sqcap S_b^{i,j,k}) \sqcap \bigsqcap_{q,q' \in Q, q \neq q'} \neg (Q_q^i \sqcap Q_{q'}^i) \tag{43}$$

$$\text{NP} \sqsubseteq \neg (H_j^{i,k} \sqcap H_{j'}^{i,k}) \quad \text{if } j \neq j' \tag{44}$$

$$\text{NP} \sqsubseteq \exists \mathsf{res}_i.(\mathsf{Result} \sqcap R_i) \sqcap \forall \mathsf{res}_i.(\mathsf{Result} \sqcap R_i) \tag{45}$$

$$\text{NP} \sqsubseteq (Q_{q_?}^i \sqcap \exists \mathsf{res}_i.\mathsf{Rej}) \to Q_{q_{no}}^{i+1} \quad \text{if } i < m-1 \tag{46}$$

$$\text{NP} \sqsubseteq (Q_{q_?}^i \sqcap \exists \mathsf{res}_i.\neg \mathsf{Rej}) \to Q_{q_{yes}}^{i+1} \quad \text{if } i < m-1 \tag{47}$$

$$\text{NP} \sqsubseteq (Q_{q_?}^i \sqcap H_j^{i,k}) \to H_j^{i+1,k} \quad \text{if } i < m-1 \tag{48}$$

$$\text{NP} \sqsubseteq \bigsqcap_{a \in \Sigma} \big( (Q_{q_?}^i \sqcap S_a^{i,j,k}) \to S_a^{i+1,j,k} \big) \tag{49}$$

Figure 5: The TBox $\mathcal{T}_w'$.

use $B$ to denote the (shared) blank symbol of $\mathcal{M}$ and $\mathcal{M}'$. Lines (34) to (43) describe the behaviour of a Turing machine in the usual way: transitions follow the transition table, the computation starts in the initial configuration, etc. Line (45) ensures that the instance of NP can reach the (unique) instance of $\mathsf{Result} \sqcap R_i$ via the role $\mathsf{res}_i$, for all $i < m$. Lines (46) and (47) deal with transitions of $\mathcal{M}$ in the query state by looking up the result of the oracle call in the corresponding instance of $\mathsf{Result}$. Finally, Lines (48) and (49) ensure that the head position and tape symbols do not change when querying the oracle. The circumscription pattern for $\mathcal{T}_w'$ is simply $\mathsf{CP}' := (\emptyset, \emptyset, \emptyset, V)$, with $V$ the set of all predicates used in $\mathcal{T}_w'$.

The purpose of $\mathcal{T}_w''$ is to describe computations of the oracle Turing machine $\mathcal{M}'$. Note that we have to describe more than a single computation of $\mathcal{M}'$ (but at most polynomially many) since $\mathcal{M}$ may visit the state $q_?$ more than once. All these computations are represented in the NExp grid, where different computations are untangled by the use of different concept names for each computation. We use the counter $X_0, \ldots, X_{m'-1}$ to count configurations and the counter $Y_0, \ldots, Y_{m'-1}$ to count the tape cells of each configuration. Moreover, we use the following concept names:

- For all $a \in \Sigma$ and $i < m$, a concept name $S_a^i$. If $S_a^i$ is satisfied by some instance of NExp where $X_0, \ldots, X_{m'-1}$ has value $j$ and $Y_0, \ldots, Y_{m'-1}$ has value $k$, then the $i$-th computation of $\mathcal{M}'$ has, in its $j$-th step, symbol $a$ on the $k$-th cell.

- For all $q \in Q$ and $i < m$, a concept name $Q_q^i$. The purpose of this concept name is two-fold: first, it represents the current state of $\mathcal{M}'$ in the $i$-th computation. And second, it indicates the head position in the $i$-th computation.

- For all $a \in \Sigma$, $q \in Q$, $M \in \{L, R\}$ and all $i < m$, a concept name $A_{q,a,M}^i$ as a marker. The meaning of the marker $A_{q,a,M}^i$ is that, to reach the current configuration, $\mathcal{M}'$ has switched to $q$, written $a$, and moved its head in direction $M$. Additionally, the marker indicates the head position in the previous configuration.

- An additional concept name NH (for "nohead") that helps us to enforce that $\mathcal{M}'$ has only a single head position.

The details of $\mathcal{T}_w''$ are shown in Figure 6, where we require one copy of each line for every $i < m$. The purpose of Lines (50) and (51) is to regenerate the grid structure of NExp using the roles $r'$ und $u'$. This is necessary since the roles $r$ and $u$ are used in $\mathcal{T}_w$, and, to use Lemma 8, the TBoxes cannot share any role names. Lines (52) and (53) ensure that every instance of NExp reaches (only) the unique instance of NP via the role toNP, and (only) the unique instance of Result $\sqcap R_i$ via the role $\text{res}_i'$, for all $i < m$. Lines (54) to (64) describe the computation of $\mathcal{M}'$ in a straightforward way. More precisely, Lines (54) to (56) set up the initial configuration by reading the contents of $\mathcal{M}$'s oracle tape from the instance of NP. Lines (57) to (61) implement transitions, and Lines (62) to (64) enforce a unique label of the tape, a unique state, and a unique head position. Finally, Line (65) ensures that, if the $i$-th computation of $\mathcal{M}'$ is rejecting, then Rej is true in the instance of Result $\sqcap R_i$.

Note that $\mathcal{M}'$ is a non-deterministic machine and may have more than one computation. For storing Rej in Result $\sqcap R_i$, we need to know that *all* these computations are rejecting. To deal with this issue, Rej is minimized with all other predicates varying: if there exists an accepting computation of $\mathcal{M}'$ on the $i$-th input, then we can represent this computation in NExp and make Rej false in the instance of Result $\sqcap R_i$. Hence, Rej holds at Result $\sqcap R_i$ iff there exists no accepting computation. Note that we cannot fix the concept names $X_0, \ldots, X_{m'-1}, Y_0, \ldots, Y_{m'-1}$ while minimizing Rej since we would get an unbounded number of fixed concept names. This means that the elements of NExp can change their position during minimization, and with them the roles $r'$ and $u'$. This is not harmful since $\mathcal{T}_w$ and Lines (50) and (51) ensure that that the structure $(\text{NExp}^{\mathcal{I}}, (r')^{\mathcal{I}}, (u')^{\mathcal{I}})$ is always isomorphic to a grid, and the rest of $\mathcal{T}_w''$ ensures that the elements of NExp always encode computations of $\mathcal{M}'$. We thus use the circumscription pattern $\text{CP}'' := (\emptyset, \{\text{Rej}\}, \emptyset, V')$, where $V'$ contains all predicates used in $\mathcal{T}_w''$ except Rej.

The proof of the following lemma is left to the reader. In it's formulation, the union over all $Q_{q_{\text{acc}}}^i$ imposes that at least one state in the computation is accepting.

$$\text{NExp} \sqsubseteq (\neg(X = 2^{m'} - 1) \rightarrow \exists r'.\text{NExp}) \sqcap (\neg(Y = 2^{m'} - 1) \rightarrow \exists u'.\text{NExp}) \tag{50}$$

$$\text{NExp} \sqsubseteq \forall r'.(Y{=}Y) \sqcap \forall r'.(X{+}{+}) \sqcap \forall u'.(X{=}X) \sqcap \forall u'.(Y{+}{+}) \tag{51}$$

$$\text{NExp} \sqsubseteq \exists \text{res}'_i.(\text{Result} \sqcap R_i) \sqcap \forall \text{res}'_i.(\text{Result} \sqcap R_i) \tag{52}$$

$$\text{NExp} \sqsubseteq \exists \text{toNP.NP} \sqcap \forall \text{toNP.NP} \tag{53}$$

$$\text{NExp} \sqsubseteq \prod_{j<m} \prod_{a \in \Sigma} \Big( \big((X = 0) \sqcap (Y = j) \sqcap \forall \text{toNP}.S_a^{i,j,2}\big) \rightarrow S_a^i \Big) \tag{54}$$

$$\text{NExp} \sqsubseteq ((X = 0) \sqcap (Y \geq m)) \rightarrow S_B^i \tag{55}$$

$$\text{NExp} \sqsubseteq ((X = 0) \sqcap (Y = 0)) \rightarrow Q_{q_0'}^i \tag{56}$$

$$\text{NExp} \sqsubseteq \prod_{a \in \Sigma} \prod_{q \in Q'} \big( (S_a^i \sqcap Q_q^i) \rightarrow \bigsqcup_{(q,a,q',a',M) \in \Delta'} \forall r'.A_{q',a',M}^i \big) \tag{57}$$

$$\text{NExp} \sqsubseteq A_{q,a,R}^i \rightarrow (S_a^i \sqcap \forall u'.Q_q^i) \tag{58}$$

$$\text{NExp} \sqsubseteq A_{q,a,L}^i \rightarrow S_a^i \tag{59}$$

$$\text{NExp} \sqsubseteq \exists u'.A_{q,a,L}^i \rightarrow Q_q^i \tag{60}$$

$$\text{NExp} \sqsubseteq \neg \bigsqcup_{q \in Q'} Q_q^i \rightarrow \prod_{a \in \Sigma} (S_a^i \rightarrow \forall r'.S_a^i) \tag{61}$$

$$\text{NExp} \sqsubseteq \prod_{a,b \in \Sigma, a \neq b} \neg (S_a^i \sqcap S_b^i) \sqcap \prod_{q,q' \in Q', q \neq q'} \neg (Q_q^i \sqcap Q_{q'}^i) \tag{62}$$

$$\text{NExp} \sqsubseteq \big( \bigsqcup_{q \in Q'} Q_q^i \big) \rightarrow \forall u'.\text{NH} \tag{63}$$

$$\text{NExp} \sqsubseteq \text{NH} \rightarrow (\neg \bigsqcup_{q \in Q'} Q_q^i \sqcap \forall u'.\text{NH}) \tag{64}$$

$$\text{NExp} \sqsubseteq Q_{q_{rej}'}^i \rightarrow \forall \text{res}'_i.\text{Rej} \tag{65}$$

Figure 6: The TBox $\mathcal{T}_w''$.

**Lemma 17** $\mathcal{M}$ *accepts* $w$ *iff* $C_w := \text{NP} \sqcap \bigsqcup_{i<m} Q_{q_{acc}}^i$ *is simultaneously satisfiable w.r.t.* $\text{Circ}_{\text{CP}}(\mathcal{T}_w, \emptyset)$, $\text{Circ}_{\text{CP}'}(\mathcal{T}_w', \emptyset)$, *and* $\text{Circ}_{\text{CP}''}(\mathcal{T}_w'', \emptyset)$.

It remains to apply Lemmas 6, 4, and 8 to obtain the following result.

**Theorem 18** *In* $\mathcal{ALC}$, *satisfiability w.r.t. concept-circumscribed cKBs is* $\text{NP}^{\text{NExp}}$-*hard even if the TBox is acyclic, the ABox and preference relations are empty, there are no fixed predicates, and the number of minimized predicates is bounded by a constant.*

As already mentioned, we conjecture that the same result can be proved with empty TBoxes (but non-empty ABoxes). Corresponding lower bounds for subsumption and the instance problems follow from the reduction given in Section 2.

## 5. Circumscription with Fixed Roles

In the preceeding sections, we have analyzed the computational complexity of reasoning w.r.t. concept-circumscribed KBs and, in particular, established decidability. In the current section, we extend concept-circumscribed KBs to what we call role-fixing cKBs, which differ from the former by allowing role names to be fixed (but not minimized). Interestingly, the result of this seemingly harmless modification is that reasoning becomes highly undecidable. We start with defining the cKBs studied in this section.

**Definition 19** *A cKB* $\mathsf{Circ}_{\mathsf{CP}}(\mathcal{T}, \mathcal{A})$ *with* $\mathsf{CP} = (\prec, M, F, V)$ *is called* role-fixing *if $M$ contains no role names.* △

To pinpoint the exact complexity of reasoning in role-fixing cKBs, we present a reduction of the logical consequence problem in monadic second-order logic with binary relation symbols (over unrestricted structures, not over trees) to the instance problem w.r.t. role-fixing cKBs formulated in $\mathcal{ALC}$. It follows that the latter problem is harder than any problem definable in second-order arithmetic and thus outside the analytical hierarchy. Analogous results for satisfiability and subsumption follow by the reductions given in Section 2. Our reduction applies already to the case where TBox and preference relation are empty.

For a finite set $R$ of binary relation symbols, denote by $\mathrm{MSO}(R)$ the set of formulas constructed from a countably infinite set $P_1, P_2, \ldots$ of variables for sets, a countable infinite set $x_1, x_2, \ldots$ of individual variables, and the binary relation symbols in $R$, using Boolean connectives, first-order quantification, and monadic second-order quantification. It is not hard to see that reasoning with role-fixing cKBs corresponds to reasoning in a "tiny" fragment of $\mathrm{MSO}(R)$. More specifically, consider the standard translation of $\mathcal{ALC}$-concepts $C$ to FO-formulas (and thus $\mathrm{MSO}(R)$-formulas) $C^\sharp(x)$ with one free individual variable $x$ as e.g. given by Borgida (1996) and take a cKB $\mathsf{Circ}_{\mathsf{CP}}(\mathcal{T}, \mathcal{A})$ with $\mathsf{CP} = (\prec, M, F, V)$, $\prec = V = \emptyset$, $M = \{A\}$, and $F = \{r\}$. Translate $(\mathcal{T}, \mathcal{A})$ to the $\mathrm{MSO}(R)$-sentence

$$\varphi = \bigwedge_{C \sqsubseteq D \in \mathcal{T}} \forall x (C^\sharp(x) \to D^\sharp(x)) \wedge \bigwedge_{C(a) \in \mathcal{A}} C^\sharp(x_a) \wedge \bigwedge_{r(a,b) \in \mathcal{A}} r(x_a, x_b),$$

where the $x_a$ are individual variables.

Then an $\mathcal{ALC}$-concept $C$ is satisfiable w.r.t. $\mathsf{Circ}_{\mathsf{CP}}(\mathcal{T}, \mathcal{A})$ if, and only if, the $\mathrm{MSO}(R)$-formula

$$\varphi \wedge \exists x C^\sharp(x) \wedge \forall P (P \subset A^\sharp \to \neg\varphi[P/A^\sharp])$$

is satisfiable, where $P \subset A^\sharp$ stands for $\forall x \ (P(x) \to A^\sharp(x)) \wedge \exists x \ (A^\sharp(x) \wedge \neg P(x))$ and $\varphi[P/A^\sharp]$ denotes $\varphi$ with $A^\sharp$ replaced by $P$. This translation is easily extended to the case where an arbitrary number of concept names is minimized and an arbitrary number of concept and role names is fixed and varies.

When we prove that logical consequence in $\mathrm{MSO}(R)$ is reducible to the instance problem w.r.t. role-fixing cKBs, we thus establish the surprising result that reasoning in such sKBs does not only correspond to the above tiny fragment of $\mathrm{MSO}(R)$, but is as hard as *all of* $\mathrm{MSO}(R)$. Our reduction is indirect: instead of directly reducing logical consequence in $\mathrm{MSO}(R)$, we reduce the semantic consequence problem of modal logic and exploit Thomason's result that logical consequence in $\mathrm{MSO}(R)$ can be reduced to the latter problem, see the works by Thomason (1975b, 1975a) and the survey articles by Wolter et al. (2007) and Goldblatt (2003) for details.

We first define the semantic consequence problem of modal logic (in the framework of description logic) and present Thomason's result, starting with some notation.

Let $R$ be a finite set of role names. An $R$-*frame* is a structure $\mathfrak{F} = (\Delta^{\mathfrak{F}}, R^{\mathfrak{F}})$, where $\Delta^{\mathfrak{F}}$ is a non-empty domain and $r^{\mathfrak{F}} \subseteq \Delta^{\mathfrak{F}} \times \Delta^{\mathfrak{F}}$ for all $r \in R$. An interpretation $\mathcal{I} = (\Delta^{\mathcal{I}}, \cdot^{\mathcal{I}})$ is *based on* an $R$-frame $\mathfrak{F}$ iff $\Delta^{\mathfrak{F}} = \Delta^{\mathcal{I}}$ and $r^{\mathcal{I}} = r^{\mathfrak{F}}$ for all $r \in R$. We say that a concept $C$ is *valid* in $\mathfrak{F}$ and write $\mathfrak{F} \models C$ if $C^{\mathcal{I}} = \Delta^{\mathcal{I}}$ for every interpretation $\mathcal{I}$ based on $\mathfrak{F}$. The semantic consequence of modal logic is now defined as follows. Let $C$ and $D$ be $\mathcal{ALC}$ concepts with roles from $R$. Then $D$ *is a semantic consequence* of $C$, in symbols $C \Vdash D$, if for every $R$-frame $\mathfrak{F}$, from $\mathfrak{F} \models C$ it follows that $\mathfrak{F} \models D$. Note that since validity in an $R$-frame $\mathfrak{F}$ involves quantification over all possible interpretations of the symbols not contained in $R$, the relation $C \Vdash D$ is invariant over all uniform renamings of atomic concepts in $D$ (this will be used later on).

For simplicity, we consider only $\text{MSO}(r)$, monadic second-order logic with one binary relation symbol $r$. It is straighforward to extend our result to arbitrary finite sets $R$ of relation symbols. Given a set of role names $R$, an $\mathcal{ALC}$-concept is called an $\mathcal{ALC}_R$-concept if it uses no other role names other than those in $R$. The following theorem follows from the results by Thomason (1975b, 1975a).

**Theorem 20** *There exist an effective translation* $\sigma : \psi \mapsto \sigma(\psi)$ *of* $\text{MSO}(r)$ *sentences to* $\mathcal{ALC}_{\{r\}}$*-concepts and an* $\mathcal{ALC}_{\{r\}}$*-concept* $C_0$ *such that for all* $\text{MSO}(r)$ *sentences* $\varphi$ *and* $\psi$, *the following conditions are equivalent:*

- $\psi$ *is a logical consequence of* $\varphi$ *in* $\text{MSO}(r)$;

- $C_0 \sqcap \sigma(\varphi) \Vdash \sigma(\psi)$.

We can thus establish the reduction of $\text{MSO}(r)$ to the instance problem w.r.t. role-fixing cKBs by reducing instead the semantic consequence problem. In fact, such a reduction can be implemented in a very transparent way if we extend $\mathcal{ALC}$ with the *universal role*, whereas the reduction to $\mathcal{ALC}$ itself requires a number of rather technical intermediate steps. For this reason, we defer the $\mathcal{ALC}$ case to the appendix and give the proof with the universal role.

Let $u$ be a new role name, called the universal role. In every interpretation $\mathcal{I}$, $u$ has the fixed interpretation as $u^{\mathcal{I}} = \Delta^{\mathcal{I}} \times \Delta^{\mathcal{I}}$. Since the interpretation of $u$ is fixed anyway, we do not allow to use it in circumscription patterns.

Now suppose that $C$ and $D$ are $\mathcal{ALC}_{\{r\}}$-concepts. To establish the reduction, we construct a role-fixing cKB $\text{Circ}_{\text{CP}}(\emptyset, \{C_0(a)\})$ and concept $C_1$ such that $C \Vdash D$ if, and only if, $a$ is an instance of $C_1$ w.r.t. $\text{Circ}_{\text{CP}}(\emptyset, \{C_0(a)\})$. As noted above, we may assume that $C, D$ do not share any concept names (otherwise, simply replace the concept names in $D$ with fresh ones). Let $A$ be a concept name that does not occur in $C$ and $D$, let $\text{CP} = (\prec, M, \{r\}, V)$, where $\prec = \emptyset$, $M$ consists of $A$ and the set of concept names in $C$, and $V$ consists of the concept names in $D$. Set $\mathcal{A} = \{(\neg \forall u.C \sqcup \forall u. \bigsqcap_{B \in M} B)(a)\}$.

**Lemma 21** *The following conditions are equivalent:*

- $C \Vdash D$;

- $a$ is an instance of $(\neg \forall u.C) \sqcup D$ w.r.t. $\mathsf{Circ}_{\mathsf{CP}}(\emptyset, \mathcal{A})$.

**Proof.** To prove that Point 1 implies Point 2, assume that Point 2 does not hold. Let $\mathcal{I}$ be a model of $\mathsf{Circ}_{\mathsf{CP}}(\emptyset, \mathcal{A})$ such that $a^{\mathcal{I}} \in ((\forall u.C) \sqcap \neg D)^{\mathcal{I}}$. Let $\mathcal{I}$ be based on $\mathfrak{F}$. To prove that Point 1 does not hold, we show that $\mathfrak{F} \models C$ and $\mathfrak{F} \not\models D$. The latter is easy as it is witnessed by the interpretation $\mathcal{I}$. To show the former, let $\mathcal{J}$ be an interpretation based on $\mathfrak{F}$. We show that $C^{\mathcal{J}} = \Delta^{\mathcal{I}}$. First note that, since $\mathcal{I}$ is a model of $\mathcal{A}$ and $a^{\mathcal{I}} \in (\forall u.C)^{\mathcal{I}}$, we have $B^{\mathcal{I}} = \Delta^{\mathcal{I}}$, for all $B \in M$. We now distinguish two cases:

- $B^{\mathcal{J}} = \Delta^{\mathcal{I}}$, for all $B \in M$. In this case, all $B \in M$ have the same interpretation in $\mathcal{I}$ and $\mathcal{J}$. Thus, since all concept names in $C$ are in $M$ and $\mathcal{I}$ and $\mathcal{J}$ are based on the same frame, we obtain $C^{\mathcal{J}} = C^{\mathcal{I}} = \Delta^{\mathcal{I}}$.

- $B^{\mathcal{J}} \neq \Delta^{\mathcal{I}}$, for at least one $B \in M$. Then $\mathcal{J} <_{\mathsf{CP}} \mathcal{I}$. Assume that $C^{\mathcal{J}} \neq \Delta^{\mathcal{I}}$. Then $a^{\mathcal{I}} \in (\neg \forall u.C)^{\mathcal{J}}$ and $\mathcal{J}$ is a model of $\mathcal{A}$. Thus, we have derived a contradiction to the assumption that $\mathcal{I}$ is a model of $\mathsf{Circ}_{\mathsf{CP}}(\emptyset, \mathcal{A})$.

To prove that Point 2 implies Point 1, assume that Point 1 does not hold. Consider a frame $\mathfrak{F}$ such that $\mathfrak{F} \models C$, but $\mathfrak{F} \not\models D$. Let $\mathcal{I}$ be an interpretation based on $\mathfrak{F}$ such that $a^{\mathcal{I}} \in (\neg D)^{\mathcal{I}}$. We may also assume that $B^{\mathcal{I}} = \Delta^{\mathcal{I}}$ for all $B \in M$ (since no such $B$ occurs in $D$). Then $a^{\mathcal{I}} \in ((\forall u.C) \sqcap \neg D)^{\mathcal{I}}$ and $\mathcal{I}$ is a model of $\mathcal{A}$. It remains to show that there does not exist an $\mathcal{I}' <_{\mathsf{CP}} \mathcal{I}$ such that $\mathcal{I}'$ is a model of $\mathcal{A}$. This is straightforward: from $\mathfrak{F} \models C$, we obtain that there does not exist any $\mathcal{I}'$ such that $a^{\mathcal{I}'} \in (\neg \forall u.C)^{\mathcal{I}'}$. Moreover, there clearly does not exist an $\mathcal{I}'$ such that $B^{\mathcal{I}'} \subsetneq B^{\mathcal{I}}$ for some $B \in M$ and $a^{\mathcal{I}'} \in (\forall u. \bigsqcap_{B \in M} B)^{\mathcal{I}'}$. ❏

We have thus proved that the logical consequence problem of $\mathrm{MSO}(r)$ is effectively reducible to the instance problem w.r.t. role-fixing cKBs formulated in $\mathcal{ALC}$ extended with the universal role. In our reduction, the TBox and preference relation are empty. In the appendix, we show how the reduction above can be modified so as to prove the same result for $\mathcal{ALC}$ without the universal role.

**Theorem 22** *The logical consequence problem of $\mathrm{MSO}(r)$ is effectively reducible to the instance problem w.r.t. role-fixing cKBs formulated in $\mathcal{ALC}$. This even holds when the TBox and preference relation are empty.*

## 6. Circumscription with Minimized Roles

Unlike fixed concept names, fixed role names cannot be simulated using minimized role names. This is due to the fact that Boolean operators on roles are not available in standard DLs. Thus, Theorem 22 does not imply undecidability of reasoning in cKBs in which role names are allowed to be minimized, but not fixed. In this section, we investigate cKBs of this type. The formal definition is as follows.

**Definition 23** *A cKB $\mathsf{Circ}_{\mathsf{CP}}(\mathcal{T}, \mathcal{A})$ with $\mathsf{CP} = (\prec, M, F, V)$ is called* role-minimizing *if $F$ contains no role names.* △

We show that role-minimizing cKBs behave rather differently from concept-circumscribed KBs and from role-fixing cKBs. First, it turns out that reasoning with role-minimizing cKBs with empty TBox is $\mathrm{NExp}^{\mathrm{NP}}$-complete for $\mathcal{ALCQO}$, but undecidable for $\mathcal{ALCI}$. Thus, in contrast to the circumscribed KBs considered so far, we now observe a difference in complexity between $\mathcal{ALCQO}$ and $\mathcal{ALCI}$, and even between $\mathcal{ALC}$ and $\mathcal{ALCI}$. A second difference to the results obtained so far is that the $\mathrm{NExp}^{\mathrm{NP}}$-lower bound, which applies to cKBs formulated in $\mathcal{ALC}$ with an empty TBox, even holds for role-minimizing cKBs in which a single role is minimized and no other predicate is fixed or minimized. This result is of interest because it shows that complexity does not drop to $\mathrm{NP}^{\mathrm{NExp}}$ if the number of minimized predicates is constant. Finally, we show that, with non-empty TBoxes, reasoning with role-minimizing cKBs becomes undecidable already for $\mathcal{ALC}$.

## 6.1 Role-minimizing cKBs With Empty TBox in $\mathcal{ALCQO}$

We first prove the $\mathrm{NExp}^{\mathrm{NP}}$-completeness result discussed above for DLs without inverse roles. We start with the upper bound. To prove it, we first establish a bounded model property using a 'selective filtration'-style argument, see e.g. Blackburn et al. (2001). The difference to the bounded model property proof given above for concept-circumscribed KBs is that, here, we do not build a quotient model of a given model by identifying nodes using an equivalence relation, but construct a submodel of a given model by *selecting* relevant nodes. In contrast to forming quotient models, this technique works only with empty TBoxes since a TBox can force us to select infinitely many nodes. Similarly, the selection technique does not work for DLs with the inverse role because, as we shall see below, inverse roles can be used to simulate TBoxes.

Recall that the *role depth* $\mathrm{rd}(C)$ of a concept $C$ is defined as the nesting depth of the constructors $(\geqslant k\ r\ D)$ and $(\leqslant k\ r\ D)$ in $C$.

**Theorem 24** *In $\mathcal{ALCQO}$, satisfiability w.r.t. role-minimizing cKBs with empty TBox is in $\mathrm{NExp}^{\mathrm{NP}}$.*

**Proof.** Let $\mathsf{Circ}_{\mathsf{CP}}(\emptyset, \mathcal{A})$ be a role-minimizing cKB with $\mathsf{CP} = (\prec, M, F, V)$, and let $C_0$ be a concept that is satisfiable w.r.t. $\mathsf{Circ}_{\mathsf{CP}}(\emptyset, \mathcal{A})$. Let $m_0$ be the maximal parameter occurring in number restrictions of $\mathcal{A}$ or $C_0$. Set

$$n := \max(\{\mathrm{rd}(C_0)\} \cup \{\mathrm{rd}(C) \mid C(a) \in \mathcal{A}\}) \text{ and } m := ((m_0 + 1) \times (|\mathcal{A}| + |C_0|))^{n+1},$$

We show that there exists a model $\mathcal{J}$ of $\mathsf{Circ}_{\mathsf{CP}}(\emptyset, \mathcal{A})$ satisfying $C_0$ such that $\#\Delta^{\mathcal{J}} \leq m$. Let $\mathcal{I}$ be a model of $\mathsf{Circ}_{\mathsf{CP}}(\emptyset, \mathcal{A})$ such that there exists a $d_0 \in C_0^{\mathcal{I}}$. For each $d \in \Delta^{\mathcal{I}}$, fix a minimal set $D(d) \subseteq \Delta^{\mathcal{I}}$ such that,

- for every concept $(\geqslant k\ r\ C)$ which occurs in $\mathcal{A}$ or $C_0$ with $d \in (\geqslant k\ r\ C)^{\mathcal{I}}$ there exist at least $k$ distinct $e \in D(d)$ such that $(d, e) \in r^{\mathcal{I}}$ and $e \in C^{\mathcal{I}}$ and

- for every concept $(\leqslant k\ r\ C)$ which occurs in $\mathcal{A}$ or $C_0$ with $d \notin (\leqslant k\ r\ C)^{\mathcal{I}}$ there exist at least $k+1$ distinct $e \in D(d)$ such that $(d, e) \in r^{\mathcal{I}}$ and $e \in C^{\mathcal{I}}$.

Clearly, $\#D(d) \leq (m_0 + 1) \times (|C_0| + |\mathcal{A}|)$ for each $d \in \Delta^{\mathcal{I}}$. Next, define a set $D_0 \subseteq \Delta^{\mathcal{I}}$ by setting

$$D_0 := \{d_0\} \cup \{a^{\mathcal{I}} \mid a \in \mathsf{N_I} \text{ occurs in } \mathcal{A} \text{ or } C_0\}.$$

Define sets $D_i \subseteq \Delta^{\mathcal{I}}$, $1 \leq i \leq n$, inductively by

$$D_{i+1} := (\bigcup_{d \in D_i} D(d))$$

and set $\Delta_n := \bigcup_{0 \leq i \leq n} D_i$. Define an interpretation $\mathcal{I}'$ with domain $\Delta^{\mathcal{I}}$ as follows:

- $a^{\mathcal{I}'} = a^{\mathcal{I}}$, for all individual names $a$;

- for $r \in M \cup V$, $(d, e) \in r^{\mathcal{I}'}$ if $d \in \Delta_n \setminus D_n$, $e \in D(d)$, and $(d, e) \in r^{\mathcal{I}}$;

- for $A \in M \cup V$, $A^{\mathcal{I}'} = A^{\mathcal{I}} \cap \Delta_n$;

- for $A \in F$, $A^{\mathcal{I}'} = A^{\mathcal{I}}$.

A straightforward inductive argument shows that $\mathcal{I}'$ is a model of $\mathcal{A}$ such that $d_0 \in C_0^{\mathcal{I}'}$. Note that we did not change the interpretation of the $A \in F$. Moreover, we have $p^{\mathcal{I}'} \subseteq p^{\mathcal{I}}$ for every $p \in M$. Together with the fact that $\mathcal{I}'$ is a model of $\mathcal{A}$ and $\mathcal{I}' \not<_{\mathsf{CP}} \mathcal{I}$, we even get $p^{\mathcal{I}'} = p^{\mathcal{I}}$ for every $p \in M$. It follows that $\mathcal{I}'$ is a model for $\mathsf{Circ}_{\mathsf{CP}}(\emptyset, \mathcal{A})$ because $\mathcal{J} <_{\mathsf{CP}} \mathcal{I}'$ would imply $\mathcal{J} <_{\mathsf{CP}} \mathcal{I}$.

Note that $r^{\mathcal{I}'} \subseteq \Delta_n \times \Delta_n$, for every role $r$. Now define an interpretation $\mathcal{J}$ with domain $\Delta^{\mathcal{J}} = \Delta_n$ by putting

- $A^{\mathcal{J}} = A^{\mathcal{I}'} \cap \Delta_n$, for every concept name $A$;

- $r^{\mathcal{J}} = r^{\mathcal{I}'}$, for every role name $r$;

- $a^{\mathcal{J}} = a^{\mathcal{I}}$, for every individual name $a$ from $\mathcal{A}$ or $C_0$.

We still have that $\mathcal{J}$ is a model for $\mathcal{A}$ satisfying $C_0$. Moreover, any interpretation $\mathcal{J}' <_{\mathsf{CP}} \mathcal{J}$ satisfying $\mathcal{A}$ can be easily extended to an interpretation $\mathcal{I}'' <_{\mathsf{CP}} \mathcal{I}'$ satisfying $\mathcal{A}$. Hence, no such interpretation exists and $\mathcal{J}$ is a model for $\mathsf{Circ}_{\mathsf{CP}}(\emptyset, \mathcal{A})$. From $\#\Delta_n \leq m$ we derive $\#\Delta^{\mathcal{J}} \leq m$.

The proof of the $\mathrm{NExp}^{\mathrm{NP}}$-upper bound is now exactly the same as the proof of Theorem 10; it suffices to replace the bound $2^{4k}$ for the size of interpretations by the bound $m$. ❏

We now give a lower bound matching the upper bound of Theorem 24.

**Theorem 25** *In $\mathcal{ALC}$, satisfiability w.r.t. role-minimizing cKBs with empty TBox is $\mathrm{NExp}^{\mathrm{NP}}$-hard. This holds even if there is only one minimized role name and no fixed prediates*

**Proof.** By Theorem 15, in $\mathcal{ALC}$ it is $\mathrm{NExp}^{\mathrm{NP}}$-hard to decide whether a concept $C_0$ is satisfiable w.r.t. $\mathsf{Circ}_{\mathsf{CP}}(\emptyset, \mathcal{A})$, where $\mathsf{CP} = (\emptyset, M, \emptyset, V)$ and $M$ contains concept names only. Clearly, it is still $\mathrm{NExp}^{\mathrm{NP}}$-hard to decide whether there exists a common model of $C_0$ and $\mathsf{Circ}_{\mathsf{CP}}(\emptyset, \mathcal{A})$ of size at least $\#M$. Thus, it is sufficient to provide a polynomial reduction of this problem to the satisfiability problem w.r.t. cKBs in $\mathcal{ALC}$ with a single minimized role and all remaining predicates varying. Suppose $C_0$ and $\mathsf{Circ}_{\mathsf{CP}}(\emptyset, \mathcal{A})$ are given. Let $M = \{A_1, \ldots, A_k\}$ and take

- two fresh role names $r_0, r_1$;

- Boolean concepts $C_1, \ldots, C_k$ built using fresh concept names $B_1, \ldots, B_k$ such that every $C_i$, $i \leq k$, is satisfiable and every $C_i \sqcap C_j$, $i \neq j$, is unsatisfiable. One can take, for example, $C_i = B_1 \sqcap \cdots \sqcap B_{i-1} \sqcap \neg B_i \sqcap B_{i+1} \sqcap \cdots \sqcap B_k$, for $i \leq k$.

Let $\mathsf{CP}' = (\emptyset, \{r_0\}, \emptyset, V \cup \{B_1, \ldots, B_k, r_1\})$ and define $\mathcal{A}'$ and $C_0'$ as the result of replacing, in $\mathcal{A}$ and $C_0$, every occurrence of $A_i$ by $\exists r_0.C_i$, for $i \leq k$. Finally, set $\mathcal{A}^* = \mathcal{A}' \cup \{\exists r_1.C_i(a) \mid i \leq k\}$. We show the following:

$(*)$ $C_0$ is satisfiable w.r.t. $\mathsf{Circ}_{\mathsf{CP}}(\emptyset, \mathcal{A})$ in a model of size at least $\#M$ if, and only if, $C_0'$ is satisfiable w.r.t. $\mathsf{Circ}_{\mathsf{CP}'}(\emptyset, \mathcal{A}^*)$.

Let $\mathcal{I}$ be a model of $\mathsf{Circ}_{\mathsf{CP}}(\emptyset, \mathcal{A})$ and $C_0$ of size at least $\#M$. Define an interpretation $\mathcal{I}'$ with domain $\Delta^{\mathcal{I}}$ by extending $\mathcal{I}$ as follows: take mutually distinct $d_1, \ldots, d_k \in \Delta^{\mathcal{I}}$ and interpret $B_1, \ldots, B_k, r_0, r_1$, and $a$ in such a way that

- $C_i^{\mathcal{I}'} = \{d_i\}$, for $i \leq k$,

- $r_0^{\mathcal{I}'} = \{(d, d_i) \mid d \in A_i^{\mathcal{I}}, i \leq k\}$,

- $a^{\mathcal{I}} = d_1$,

- $r_1^{\mathcal{I}'} = \{(d_1, d_1), \ldots, (d_1, d_k)\}$.

It is readily checked that $\mathcal{I}'$ is a model of $C_0'$ and $\mathsf{Circ}_{\mathsf{CP}'}(\emptyset, \mathcal{A}^*)$.

Conversely, let $\mathcal{I}$ be a model of $\mathsf{Circ}_{\mathsf{CP}'}(\emptyset, \mathcal{A}^*)$ and $C_0'$. Define an interpretation $\mathcal{I}'$ by extending $\mathcal{I}$ with $A_i^{\mathcal{I}'} = (\exists r_0.C_i)^{\mathcal{I}}$, for $i \leq k$. It is readily checked that $\mathcal{I}'$ is a model of $C_0$ and $\mathsf{Circ}_{\mathsf{CP}}(\emptyset, \mathcal{A})$. ❏

## 6.2 Role-minimizing cKBs With Nonempty TBox

In the bounded model property proof above, it is important that the selection of nodes stops after $n$ iterations with the set $\Delta_n$, where $n$ is the maximum of the role depths of the concepts in the ABox and the concept $C$ we want to satisfy. Such a bound for the selection of nodes does not exist if the TBox is non-empty, and we now show that reasoning w.r.t. role-minimizing cKBs is indeed undecidable in this case. The proof is by a reduction of the $\mathbb{N} \times \mathbb{N}$-tiling problem (Berger, 1966).

**Definition 26** *A tiling problem is a quadruple triple $P = (T, H, V)$, where $T$ is a finite set of tile types and $H, V \subseteq T \times T$ are the horizontal and vertical matching conditions. A solution to $P$ is a mapping $\tau : \mathbb{N} \times \mathbb{N} \to T$ such that*

- $(\tau(i, j), \tau(i + 1, j)) \in H$ *for all $i, j \geq 0$;*

- $(\tau(i, j), \tau(i, j + 1)) \in V$ *for all $i, j \geq 0$.*

$\triangle$

Let $P = (T, H, V)$ be an instance of the tiling problem. We define a TBox $\mathcal{T}_P$ as follows:

$$\top \quad \sqsubseteq \quad \exists x.\top \sqcap \exists y.\top \tag{66}$$

$$\top \quad \sqsubseteq \quad \bigsqcup_{t \in T} \left( A_t \sqcap \bigsqcap_{t' \in T, t \neq t'} \neg A_{t'} \right) \tag{67}$$

$$\top \quad \sqsubseteq \quad \bigsqcap_{t \in T} \left( A_t \to \bigsqcup_{(t,t') \in H} \forall x.A_{t'} \right) \sqcap \bigsqcap_{t \in T} \left( A_t \to \bigsqcup_{(t,t') \in V} \forall y.A_{t'} \right) \tag{68}$$

$$\top \quad \sqsubseteq \quad N \sqcup (\exists x.\exists y.B \sqcap \exists y.\exists x.\neg B) \tag{69}$$

$$\neg N \quad \sqsubseteq \quad D \tag{70}$$

$$D \quad \sqsupseteq \quad \exists x.D \sqcup \exists y.D \tag{71}$$

$$D \quad \sqsubseteq \quad \forall x.D \sqcap \forall y.D \tag{72}$$

Let $\mathsf{CP} = (\emptyset, M, \emptyset, V)$ be the circumscription pattern in which $V = \{B, D\}$ and $M$ consists of the remaining concept and role names.

**Lemma 27** $\mathsf{Circ}_{\mathsf{CP}}(\mathcal{T}_P, \emptyset) \not\models D(a)$ iff $P$ has a solution.

**Proof.** Assume that $P$ has a solution $\tau$. Define an interpretation $\mathcal{I}$ as follows:

$$
\begin{aligned}
\Delta^{\mathcal{I}} &:= \mathbb{N} \times \mathbb{N} \\
x^{\mathcal{I}} &:= \{((i,j),(i+1,j)) \mid i,j \geq 0\} \\
y^{\mathcal{I}} &:= \{((i,j),(i,j+1)) \mid i,j \geq 0\} \\
A_t^{\mathcal{I}} &:= \{(i,j) \in \Delta^{\mathcal{I}} \mid \tau(i,j) = t\} \\
N^{\mathcal{I}} &:= \Delta^{\mathcal{I}} \\
B^{\mathcal{I}} &:= \{(i,j) \in \Delta^{\mathcal{I}} \mid i > 0 \text{ and } j > 0\} \\
D^{\mathcal{I}} &:= \emptyset \\
a^{\mathcal{I}} &:= (0,0)
\end{aligned}
$$

It is straightforward to verify that $\mathcal{I}$ is a model of $\mathcal{T}_P$. Additionally, we clearly have $a^{\mathcal{I}} \notin D^{\mathcal{I}}$. It thus remains to show that there is no model $\mathcal{J}$ of $\mathcal{T}_P$ with $\mathcal{J} <_{\mathsf{CP}} \mathcal{I}$. Assume there is such a $\mathcal{J}$. Since all concept and role names except $D$ and $B$ are minimized, it follows that

1. $x^{\mathcal{I}} = x^{\mathcal{J}}$ and $y^{\mathcal{I}} = y^{\mathcal{J}}$ because $\mathcal{J}$ is a model of (66);

2. $A_t^{\mathcal{I}} = A_t^{\mathcal{J}}$ for all $t \in T$ because of Point 1 and $\mathcal{J}$ is a model of (67);

3. $N^{\mathcal{I}} = N^{\mathcal{J}}$ because, no matter what $B^{\mathcal{J}}$ is, by Point 1 we have

$$(\exists x.\exists y.B \sqcap \exists y.\exists x.\neg B)^{\mathcal{J}} = \emptyset.$$

Thus and because $\mathcal{J}$ is a model of (69), $N^{\mathcal{I}} = N^{\mathcal{J}}$.

Thus, $\mathcal{I}$ and $\mathcal{J}$ differ at most in the interpretation of the concept names $D$ and $B$, which are varying. This is a contradiction to $\mathcal{J} <_{\mathsf{CP}} \mathcal{I}$.

Conversely, assume that $\mathsf{Circ}_{\mathsf{CP}}(\mathcal{T}_P, \emptyset) \not\models D(a)$, and let $\mathcal{I}$ be a model of $\mathsf{Circ}_{\mathsf{CP}}(\mathcal{T}_P, \emptyset)$ with $a^{\mathcal{I}} \notin D^{\mathcal{I}}$. By induction on $i + j$, we define a mapping $\pi$ that assigns to each $(i,j) \in \mathbb{N} \times \mathbb{N}$ an element $\pi(i,j) \in \Delta^{\mathcal{I}}$ such that for all $i, j \geq 0$, we have

1. $(\pi(i,j), \pi(i+1,j)) \in x^{\mathcal{I}}$;

2. $(\pi(i,j), \pi(i,j+1)) \in y^{\mathcal{I}}$;

To start, set $\pi(0,0) = a^{\mathcal{I}}$. In the $\ell$-th step, we do the following:

- Select a $d \in \Delta^{\mathcal{I}}$ such that $(\pi(0, \ell-1), d) \in y^{\mathcal{I}}$ and put $\pi(0, \ell) := d$. Such a $d$ exists since $\mathcal{I}$ is a model of (66).

- Select a $d \in \Delta^{\mathcal{I}}$ such that $(\pi(\ell-1, 0), d) \in x^{\mathcal{I}}$ and put $\pi(\ell, 0) := d$. Again, such a $d$ exists since $\mathcal{I}$ is a model of (66).

- Now let $i, j > 0$ such that $i + j = \ell$. Since $\mathcal{I}$ is a model of (66), there are $d, d' \in \Delta^{\mathcal{I}}$ such that $(\pi(i-1,j), d) \in x^{\mathcal{I}}$ and $(\pi(i, j-1), d') \in y^{\mathcal{I}}$. We show that $d = d'$, and then set $\pi(i,j) := d$.

  Assume to the contrary that $d \neq d'$. By (70)–(72) and since $a^{\mathcal{I}} \notin D^{\mathcal{I}}$, we have $\pi(i-1, j-1) \in N^{\mathcal{I}}$. Define a new interpretation $\mathcal{J}$ that is obtained from $\mathcal{I}$ by the following modifications:

  - $\pi(i-1, j-1)$ is removed from $N^{\mathcal{I}}$;
  - let $d' \in B^{\mathcal{J}}$ and $d \notin B^{\mathcal{J}}$;
  - let $D^{\mathcal{J}} = \Delta^{\mathcal{I}}$.

  Clearly, $\mathcal{J} <_{\mathsf{CP}} \mathcal{I}$. To obtain a contradiction against $\mathcal{I}$ being a model of $\mathsf{Circ}_{\mathsf{CP}}(\mathcal{T}_P, \emptyset)$, it thus remains to show that $\mathcal{J}$ is a model of $\mathcal{T}_P$. It suffices to consider (69) to (72), the only concept inclusions referring to $N$, $B$, and $D$. The axioms (70) to (72) hold because $D^{\mathcal{J}} = \Delta^{\mathcal{I}}$. To show (69), let $e \in \Delta^{\mathcal{I}}$. We show that $e \in C^{\mathcal{J}}$ for $C$ the concept on the right hand side of (69). Clearly, $e \in C^{\mathcal{J}}$ since $e \in C^{\mathcal{I}}$, if $e$ is not $\{x, y, x^-, y^-\}$-reachable from $a^{\mathcal{I}}$. We have $e \in N^{\mathcal{I}}$ if $e$ is $\{x, y, x^-, y^-\}$-reachable from $a^{\mathcal{I}}$, because otherwise we would have $a^{\mathcal{I}} \in D^{\mathcal{I}}$ by axioms (70) to (72). Thus, $e \in N^{\mathcal{J}} \subseteq C^{\mathcal{J}}$ if $e \neq \pi(i-1, j-1)$. Finally, $\pi(i-1, j-1) \in C^{\mathcal{J}}$, by definition of $B^{\mathcal{J}}$.

Now define a mapping $\tau : \mathbb{N} \times \mathbb{N} \to T$ by setting $\tau(i,j) := t$ if $\pi(i,j) \in A_t$. By (67), this mapping is well-defined. By (68), it satisfies the horizontal and vertical matching conditions. Thus, $P$ has a solution.

$\qquad\qquad\qquad\qquad\qquad\qquad\qquad\qquad\qquad\qquad\qquad\qquad\qquad\qquad\qquad\qquad$ ❏

Thus, we have shown the following result.

**Theorem 28** *In $\mathcal{ALC}$, satisfiability w.r.t role-minimizing cKBs is undecidable.*

## 6.3 Reasoning in Role-minimizing cKBs With Empty TBox in $\mathcal{ALCI}$

We prove undecidability of reasoning in role-minimizing cKBs with empty TBox in $\mathcal{ALCI}$. The proof uses the "spy-point" technique (Areces, Blackburn, & Marx, 1999); namely, we show that ABoxes can simulate TBox reasoning by employing inverse roles and a simulation of nominals by circumscription. Using this idea the proof is rather similar to the proof of Theorem 28.

Let $P$ be an instance of the tiling problem and consider the cKB $\mathsf{Circ}_{\mathsf{CP}}(\mathcal{T}_P, \emptyset)$ defined in the proof of Lemma 27. We simulate its TBox axioms $C \sqsubseteq C'$ by ABox assertions $((C \to C') \sqcap \forall r_0.(C \to C'))(a)$ and enforcing the role $r_0$ to connect all relevant points to $a$. This is achieved by forcing all relevant points in the domain to satisfy $\exists r_0^-.\{a\}$. Since we do not have nominals in the language we use a concept name $A$ instead of $\{a\}$ and ensure that it behaves like a nominal. We now present the details.

For the sake of readability, we write concept assertions $C(a)$ in the form $a : C$ and we set $\forall^1\{r\}.C = C \sqcap \forall r.C$. Let $A$, $B'$, and $N'$ be fresh concept names and $r_0$ be a fresh role name not occurring in $\mathcal{T}_P$. Then $\mathcal{A}_P$ consists of the assertions

$$a : \forall^1\{r_0\}.(C \to C'), \tag{73}$$

for $C \sqsubseteq C' \in \mathcal{T}_P$,

$$a : A, \quad a : \forall^1\{r_0\}. \bigsqcap_{s=x,y} \forall s.\exists r_0^-.A, \tag{74}$$

$$a : \forall^1\{r_0\}.\Big(N' \sqcup \big(((A \sqcap B') \sqcup \exists r_0^-.(A \sqcap B')) \sqcap \bigsqcup_{s=x,y} \exists s.\exists r_0^-.(A \sqcap \neg B')\big)\Big) \tag{75}$$

$$a : \forall^1\{r_0\}.(\neg N' \to D), \quad a : \exists r_0.D \to D \tag{76}$$

Now let $\mathsf{CP} = (\emptyset, M, \emptyset, \{D, B, B'\})$, where $M$ consists of all concept and role names distinct from $D$, $B$, and $B'$.

**Lemma 29** $\mathsf{Circ}_{\mathsf{CP}}(\emptyset, \mathcal{A}_P) \not\models D(a)$ *iff* $P$ *has a solution.*

**Proof.** Assume $P$ has a solution $\tau$. Take the interpretation $\mathcal{I}$ from the proof of Lemma 27 expanded by

$$A^{\mathcal{I}} = \{(0,0)\}, \quad N'^{\mathcal{I}} = \Delta^{\mathcal{I}}, \quad B'^{\mathcal{I}} = \emptyset, \quad r_0^{\mathcal{I}} = \{(a^{\mathcal{I}}, d) \mid d \in \Delta^{\mathcal{I}}\}.$$

We show that $\mathcal{I}$ is a model of $\mathsf{Circ}_{\mathsf{CP}}(\emptyset, \mathcal{A}_P)$. Clearly, $\mathcal{I}$ is a model of $\mathcal{A}_P$. Thus it remains to show that there is no model $\mathcal{J}$ of $\mathcal{A}_P$ with $\mathcal{J} <_{\mathsf{CP}} \mathcal{I}$. Assume there exists $\mathcal{J} <_{\mathsf{CP}} \mathcal{I}$ which is a model of $\mathcal{A}_P$. As $A$ is minimized and by (74), $A^{\mathcal{J}} = \{(0,0)\}$. It follows from axiom (66) encoded in (73) that $((0,0),(1,0)) \in x^{\mathcal{J}}$ and $((0,0),(0,1)) \in y^{\mathcal{J}}$. Now one can prove by induction on $\ell > 0$ using again the axiom (66) encoded in (73) and (74) that for all $(i,j)$ with $i + j = \ell$, $((0,0),(i,j)) \in r_0^{\mathcal{J}}$ and $((i,j),(i+1,j)) \in x^{\mathcal{J}}$, $((i,j),(i,j+1)) \in y^{\mathcal{J}}$. It follows that $x^{\mathcal{J}} = x^{\mathcal{I}}$ and $y^{\mathcal{J}} = y^{\mathcal{I}}$. Also observe that $N'^{\mathcal{J}} = \Delta^{\mathcal{I}}$ because, no matter how $B'$ is interpreted,

$$\Big(((A \sqcap B') \sqcup \exists r_0^-.(A \sqcap B')) \sqcap \bigsqcup_{s=x,y} \exists s.\exists r_0^-.(A \sqcap \neg B')\Big)^{\mathcal{J}} = \emptyset.$$

Now one can prove similarly to the proof of Lemma 27 that $\mathcal{J}$ can only differ from $\mathcal{I}$ in the interpretation of $B$, $B'$, and $D$, which is a contradiction.

Conversely, suppose $\mathcal{I}$ is a model of $\mathsf{Circ}_{\mathsf{CP}}(\emptyset, \mathcal{A}_P)$ with $a^{\mathcal{I}} \notin D^{\mathcal{I}}$. We first show that $(a^{\mathcal{I}}, d) \in r_0^{\mathcal{I}}$ whenever $d \neq a^{\mathcal{I}}$ and $d$ is $\{x, y\}$-reachable in $\mathcal{I}$ from $a^{\mathcal{I}}$ in a finite number of steps. Assume that this is not the case. Then there exist $d, d' \in \Delta^{\mathcal{I}}$ such that

- $d = a^{\mathcal{I}}$ or $(a^{\mathcal{I}}, d) \in r_0^{\mathcal{I}}$,

- $(d, d') \in x^{\mathcal{I}}$ or $(d, d') \in y^{\mathcal{I}}$,

- $(a, d') \notin r_0^{\mathcal{I}}$.

By (74), there exists $d'' \in \Delta^{\mathcal{I}}$ such that $a^{\mathcal{I}} \neq d''$, $(d'', d') \in r_0^{\mathcal{I}}$, and $d'' \in A^{\mathcal{I}}$. Observe that $d \in N'^{\mathcal{I}}$ by (76) and $a^{\mathcal{I}} \notin D^{\mathcal{I}}$. Define a new interpretation $\mathcal{J}$ by modifying $\mathcal{I}$ as follows:

- $d$ is removed from $N'^{\mathcal{I}}$;

- let $a^{\mathcal{I}} \in B'^{\mathcal{J}}$ and $d'' \notin B'^{\mathcal{J}}$;

- let $D^{\mathcal{J}} = \Delta^{\mathcal{I}}$.

Clearly $\mathcal{J} <_{\mathsf{CP}} \mathcal{I}$. To obtain a contradiction it is thus sufficient to show that $\mathcal{J}$ is a model of $\mathcal{A}_P$. Clearly, all assertion in $\mathcal{A}_P$ containing neither $D$, $B'$, nor $N'$ are satisfied in $\mathcal{J}$. For the remaining assertions except (75), it follows from $D^{\mathcal{J}} = \Delta^{\mathcal{I}}$ that they are satisfied in $\mathcal{J}$. Finally, for (75), observe that $N'^{\mathcal{I}} \supseteq \{a^{\mathcal{I}}\} \cup \{e \mid (a^{\mathcal{I}}, e) \in r_0^{\mathcal{I}}\}$ because $a \notin D^{\mathcal{I}}$ and assertions (76). Thus, by definition of $N'^{\mathcal{J}}$, we only have to consider the point $d$ removed from $N'^{\mathcal{I}}$. But $d \in \left( \left((A \sqcap B') \sqcup \exists r_0^-.(A \sqcap B')\right) \sqcap \bigsqcup_{s=x,y} \exists s. \exists r_0^-.(A \sqcap \neg B')\right)^{\mathcal{J}}$ by definition of $\mathcal{J}$.

Now one can use the assertions (73) to construct a solution $\tau$ of $P$ in the same way as in the proof of Lemma 27. ❏

We have thus proved the following result.

**Theorem 30** *In $\mathcal{ALCI}$, satisfiability w.r.t. role-minimizing cKBs with empty TBox is undecidable.*

# 7. Related Work

We have already pointed out in the introduction that circumscription is just one of several possible approaches to nonmonotonic DLs and that, in order to achieve decidability, each of these approaches has to adopt a suitable restriction on the expressive power of the DL component, on the non-monotonic features, or on the interaction of the DL and its non-monotonic features. In this section, we survey the existing approaches, discuss the adopted restrictions, and relate them to DLs with circumscription whenever possible. However, we point out that a full-fledged formal comparison of the different approaches is a serious research endeavor of its own and outside the scope of this paper. The main approaches to nonmonotonic DLs (excluding those relying on the integration of DLs and logic programming) are summarized in Table 1, where 'n.a.' stands for 'not analyzed' and the 'specificity' column states whether a formalism is equipped with a priority mechanism based on the the specificity (i.e., subsumption) of concepts.

We start with two early approaches based on circumscription. In the work by Brewka (1987), a frame system is given a nonmonotonic semantics based on circumscription. The focus is on showing the appropriateness of the proposed semantics, and the decidability and complexity of reasoning is not analyzed. Cadoli et al. (1990), apply circumscription to a DL in the same way as we do here. The authors consider only non-prioritized circumscription and apply it to a fragment of the description logic $\mathcal{ALE}$. Decidability of reasoning is shown by a reduction to propositional reasoning under the Extended Closed World Assumption (ECWA), which is in $\Pi_2^p$. To the best of our knowledge, this was the first effective reasoning method for a nonmonotonic description logic.

| Ref | DL | NM features | Complexity | Specificity |
|---|---|---|---|---|
| (Brewka, 1987) | frame lang. | Circ | n.a. | Y |
| (Cadoli, Donini, & Schaerf, 1990) | $< \mathcal{ALE}$ | Circ | in $\Sigma_2^p$ | N |
| (Padgham & Zhang, 1993) | $\mathcal{AL}$ with concrete domains | inheritance networks | n.a. | Y |
| (Straccia, 1993) | $\mathcal{ALC}$ | prioritized default logic | decidable | Y |
| (Baader & Hollunder, 1995a) | $\mathcal{ALCF}$ | default logic | decidable | N |
| (Baader & Hollunder, 1995b) | $\mathcal{ALC}$ | prioritized default logic | decidable | Y |
| (Lambrix, Shahmehri, & Wahlloef, 1998) | $\mathcal{ALQO}$+feature agreement | prioritized default logic | n.a. | Y |
| (Donini et al., 1997) | any decidable DL | MKNF with restrictions | depends on DL | N |
| (Donini et al., 2002) | $\mathcal{ALC}$ | MKNF with restrictions | in 3-ExpTime | N |
| (Giordano, Gliozzi, Olivetti, & Pozzato, 2008) | $\mathcal{ALC}$ | maximized typicality | in co-NExp$^{\text{NP}}$ | N |

Table 1: Some approaches to nonmonotonic DLs

In another early approach by Padgham and Zhang (1993), a nonmonotonic DL is obtained by an adaptation of the inheritance networks approach (Horty, 1994) where the underlying DL is essentially $\mathcal{AL}$ extended with concrete data values. The main contribution is the definition of the formalism and a discussion of applications, decidability and complexity are not analyzed.

A recent approach that is similar in spirit to circumscription has been taken by Giordano et al. (2008). The authors extend $\mathcal{ALC}$ with a modal operator $T$ representing typicality, and maximize $T$'s extension to achieve nonmonotonic inferences. Decidability is proved via a tableau algorithm that also establishes a co-NExp$^{\text{NP}}$ upper bound for subsumption. A lower bound is not given.

We now turn to approaches based on default logic (Reiter, 1980). The nonmonotonic DLs by Baader and Hollunder (1995a, 1995b), Straccia (1993), and Lambrix et al. (1998) share a common restriction: default rules can be applied to an individual only if it has a *name*, that is, if it is denoted by an individual constant that occurs explicitly in the knowledge base. This restriction is motivated by the observation that, when defaults are also applied to implicit (anonymous) individuals, then reasoning becomes undecidable (Baader &

Hollunder, 1995a). Since the models of DL knowledge bases usually include a large number of anonymous individuals due to existential quantification, this restriction introduces a strong asymmetry in the treatment of individuals.

Another line of nonmonotonic DLs (Donini et al., 1998, 1997, 2002) is based on first-order autoepistemic logics whose interpretation domains are restricted to a fixed and denumerable set of constants. The use of a unique domain resolves several issues related to quantification in modal logics (such as whether the Barcan formula should hold, and whether different worlds of the same Kripke structure should be allowed to have different domains). It also avoids the asymmetry of the approaches based on default logic because, by definition, all individuals have a name. The other side of the coin is that domains with finite or varying cardinality and knowledge bases that do not satisfy the unique name assumption can only be dealt with using rather technical encodings (such as an explicit axiomatization of the finite domain represented by a concept name $D$).

In the first paper mentioned above (Donini et al., 1998), decidability results apply only to monotonic knowledge bases[4] as the autoepistemic operator can be used in a nonmonotonic fashion only in queries. This restriction has been lifted in the subsequent publications. They make use of the logic of minimal knowledge and negation as failure (MKNF), which is equipped with two (auto)epistemic operators "K" and "A" (Donini et al., 1997, 2002). In the former paper (Donini et al., 1997), the underlying monotonic fragment can be any description logic with a decidable instance checking problem. The price payed for this generality is that decidability results apply only to so-called *simple* knowledge bases, where *quantifying-in* (that is, quantification across modal operators, as in $\forall R.\mathsf{K}\,C$) is not allowed. Nonetheless, such KBs are expressive enough to model default rules. A different restriction is explored by Donini et al. (2002). The underlying DL is restricted to $\mathcal{ALC}$ while limited forms of quantifying-in are allowed, in so-called *subjectively quantified* $\mathcal{ALCK}_{\mathcal{NF}}$ knowledge bases. Decidability results are obtained for the subclass of *simple subjectively quantified* knowledge bases, whose nonmonotonic part is restricted to inclusions of the form $\mathsf{K}\,C \sqsubseteq D$ such that $\top \sqsubseteq C$ is can *not* be inferred from the knowledge base. This restriction is incompatible with the traditional embedding of (priority-free) circumscription into autoepistemic logic, which is based on prerequisite-free default rules that would be equivalent to inclusions of the form $\mathsf{K}\top \sqsubseteq C$.

A recent line of research on integrating DLs and logic programming rules introduces further nonmonotonic extensions of DLs based on negation-as-failure. Some approaches (Eiter et al., 2004) use a loosely coupled integration of logic programs and DLs, where the interpretations of the DL component are not restricted while the logic program variables range only over the *named* DL individuals. Thus, these approaches are somewhat similar to the classical extensions of DLs based on defaults in that nonmonotonic inferences are limited to named individuals. A more recent approach (Motik & Rosati, 2007) is based on MKNF and shares with the MKNF-DLs discussed above pros and cons of adopting a fixed denumerable domain. If the complexity of reasoning in the underlying DL is $\mathcal{C} \not\subseteq \mathrm{NP}$, then the data complexity of entailment is bounded by $\mathcal{C}^{\mathcal{C}}$. Finally, we mention a 3-valued variant of this approach (Knorr, Alferes, & Hitzler, 2007) based on the well-founded semantics.

---

4. The autoepistemic operator can be used only in restricted contexts that suffice to encode so-called *procedural rules*, which are monotonic.

| | | $\mathcal{ALC}$ | $\mathcal{ALCQO}$ | $\mathcal{ALCI}$ | $\mathcal{ALCIO}$ |
|---|---|---|---|---|---|
| **Concept circ.** | $\#M \leq n, \#F \leq n$ | **NP$^{\mathbf{NExp}}$** | | | |
| | (unrestricted) | **NExp$^{\mathbf{NP}}$** even if $\prec = \emptyset$, and either TBox=$\emptyset$ or ABox=$\emptyset$ | | | |
| **Minim. roles** | TBox= $\emptyset$ | **NExp$^{\mathbf{NP}}$** even if $\#M \leq 1, \#F \leq 0$ | | **Undecidable** | |
| | TBox$\neq \emptyset$ | **Undecidable** | | | |
| **Fixed roles** | | **Highly undecidable**, even if TBox= $\emptyset$, $\prec = \emptyset$ | | | |

Table 2: Summary of complexity results for satisfiability w.r.t. cKBs

A common limitation of the nonmonotonic extensions of DLs based on MKNF is that they provide no support for specificity and priorities. In particular, defeasible inheritance is not mentioned in the expressiveness analysis for autoepistemic approaches (Donini et al., 1997, 2002) and it does not appear to be a goal of any MKNF-based approach. As pointed out in the introduction, it is well-known that, in the propositional case, nonmonotonic logics cannot modularly encode specificity-based priorities such as those needed by inheritance mechanisms (Horty, 1994).

## 8. Conclusions and Perspectives

We have shown that circumscription provides an elegant approach to defining nonmonotonic DLs, that the resulting formalisms have an appealing expressive power and are decidable if appropriate restrictions are adopted. The main such restriction, which leads to rather robust decidability results, is that only concept names are minimized and fixed whereas all role names vary. With empty TBoxes, decidability is retained if roles are allowed to be minimized, but not when they are fixed. The decidability and complexity results obtained in this paper are listed in more detail in Table 2. By the results of Section 3, all bounds with TBox $\neq \emptyset$ apply to both general and acyclic TBoxes.

We view this paper as a promising step towards establishing circumscribed DLs as a major family of nonmonotonic DLs to be used in practical applications. To reach this goal, some additional research topics have to be addressed, of which we mention two. First, the algorithms presented in this paper are based on massive non-deterministic guessing and thus do not admit an efficient implementation. Ideally, one would like to have algorithms which are well-behaved extensions of the tableau algorithms that underly state-of-the-art DL reasoners (Baader & Sattler, 2000). A crucial issue is whether the sophisticated optimization techniques implemented in such reasoners (tableaux caching, dependency-directed backtracking etc.; cf. Baader et al., 2003, Chap. 9) can be adapted to circumscribed DLs. Some first steps have been made by Grimm and Hitzler (2008). Second, it seems necessary to develop a design methodology for modelling defeasible inheritance. The examples given in this paper indicate that the main challenge is to find rules of thumb to determine which predicates should be fixed, varied, and minimized. It may be appealing to hide abnormality predicates behind explicit syntax for defeasible inclusions, and trade some generality for simplicity and usability.

Also from a theoretical perspective, our initial investigation leave open at least some interesting questions. For example, our current techniques are limited to circumscribed DLs that have the finite model property. It would be desirable to overcome this limitation and obtain decidability results for even more expressive DLs such as $\mathcal{SHIQ}$ or OWL. It is also possible to follow the opposite approach and consider circumscribed versions of inexpressive DLs such as those of the $\mathcal{EL}$ or DL-Lite family (Baader, Brandt, & Lutz, 2005a; Calvanese, Giacomo, Lembo, Lenzerini, & Rosati, 2007), which are currently very popular in a large number of applications. First steps have been made by Bonatti, Faella, and Sauro (2009), who investigated circumscribed versions of $\mathcal{EL}$ and DL-lite.

Finally, it is worth mentioning that our complexity results for circumscription can be used to prove complexity bounds for other, seemingly unrelated, reasoning problems in description logic. For example, certain reasoning services for conservative extensions and modularity in description logic and the satisfiability problem w.r.t. concept-circumscribed knowledge bases are mutually reducible to each other in polynomial time (Konev, Lutz, Walther, & Wolter, 2008). As there are not many problems that are known to be NExp$^{\text{NP}}$-complete, circumscription thus provides a new class of problems that can potentially be employed to give NExp$^{\text{NP}}$ lower bound proofs.

## Acknowledgments

The first author was partially supported by the network of excellence REWERSE, IST-2004-506779. The third author was partially supported by UK EPSRC grant no. GR/S63182/01.

## Appendix A. Missing Proofs in Section 3

**Lemma 5.** In $\mathcal{ALC}$, satisfiability w.r.t. (concept-)circumscribed KBs with empty TBox and without fixed roles can be polynomially reduced to satisfiability w.r.t. (concept-)circumscribed KBs with empty TBox and without fixed predicates.

**Proof.** In the proof of Lemma 4, we have used TBox axioms to state that some fresh concept names are interpreted as the complement of fixed concept names. In general, this cannot be done using ABox assertions only. Instead, we construct ABox assertions which force this to be the case only for objects which are relevant for the truth of the given ABox. Some care is required to do this using ABox assertions of polynomial size. The first part of the proof deals with this problem. The second part is then a straighforward modification of the proof of Lemma 4.

The first part of the proof consists of introducing some notation and proving a central technical claim. For $w = r_1 \cdots r_n \in \mathsf{N}_{\mathsf{R}}^*$, an interpretation $\mathcal{I}$, and $d, e \in \Delta^{\mathcal{I}}$, we write $(d, e) \in w^{\mathcal{I}}$ iff there are $d_0, \ldots, d_n \in \Delta^{\mathcal{I}}$ with $d = d_0$, $e = d_n$, and $(d_i, d_{i+1}) \in r_{i+1}^{\mathcal{I}}$ for all $i < n$.

Let $\mathcal{N}$ be a set of individual names and $\mathsf{Paths}$ a mapping from $\mathcal{N}$ to the powerset of $\mathsf{N}_{\mathsf{R}}^*$. An interpretation $\mathcal{I}$ is *well-behaved for the mapping* $\mathsf{Paths}$ if for every $d \in \Delta^{\mathcal{I}}$, there is an $a \in \mathcal{N}$ and a $w \in \mathsf{Paths}(a)$ such that $(a^{\mathcal{I}}, d) \in w^{\mathcal{I}}$. With each $\mathcal{ALC}$-concept $C$, we associate a set $\mathcal{P}(C)$ of pairs $(w, D)$ with $w \in \mathsf{N}_{\mathsf{R}}^*$ and $D$ a subconcept of $C$ as follows:

- if $C \in \mathsf{N}_{\mathsf{C}}$, then $\mathcal{P}(C) = \{(\varepsilon, C)\}$;

- if $C = \neg D$, then $\mathcal{P}(C) = \{(\varepsilon, C)\} \cup \mathcal{P}(D)$;

- if $C = D_1 \sqcap D_2$ or $C = D_1 \sqcup D_2$, then $\mathcal{P}(C) = \{(\varepsilon, C)\} \cup \mathcal{P}(D_1) \cup \mathcal{P}(D_2)$;

- if $C = \exists r.D$ or $C = \forall r.D$, then $\mathcal{P}(C) = \{(\varepsilon, C)\} \cup \{(rw, E) \mid (w, E) \in \mathcal{P}(D)\}$.

For a set of ABox assertions $\mathcal{S}$ and an individual name $a$, we use $\mathcal{P}(\mathcal{S}, a)$ to denote the set $\bigcup_{C(a) \in \mathcal{S}} \mathcal{P}(C)$. We write $\mathsf{Paths}(\mathcal{S}, a)$ for $\{w \mid \exists D : (w, D) \in \mathcal{P}(\mathcal{S}, a)\}$. We now formulate the announced claim.

Claim 1. Suppose $\mathsf{Circ}_{\mathsf{CP}}(\emptyset, \mathcal{A}) \not\models C_0(a_0)$, where $\mathsf{CP}$ does not contain fixed role names and $\mathcal{A}$ and $C_0(a)$ are formulated in $\mathcal{ALC}$. Let $\mathcal{S} = \mathcal{A} \cup \{C_0(a_0)\}$ and let $\mathcal{N}$ be the set of individual names in $\mathcal{S}$. Let $\mathcal{I}'$ be the restriction of $\mathcal{I}$ to those $d \in \Delta^{\mathcal{I}}$ such that, for some $a \in \mathcal{N}$ and some $w \in \mathsf{Paths}(\mathcal{S}, a)$, we have $(a^{\mathcal{I}}, d) \in w^{\mathcal{I}}$. Then $\mathcal{I}'$ is a model of $\mathsf{Circ}_{\mathsf{CP}}(\emptyset, \mathcal{A})$ and $\neg C_0(a_0)$ which is well-behaved for the mapping $\mathsf{Paths}(a) = \mathsf{Paths}(\mathcal{S}, a)$, $a \in \mathcal{N}$.

We now prove the claim. Let $\mathcal{I}$ be a model of $\mathsf{Circ}_{\mathsf{CP}}(\emptyset, \mathcal{A})$ with $a_0^{\mathcal{I}} \in (\neg C_0)^{\mathcal{I}}$. Note that for all $a \in \mathcal{N}$, we have $\varepsilon \in \mathsf{Paths}(\mathcal{S}, a)$ and thus $a^{\mathcal{I}} \in \Delta^{\mathcal{I}'}$. Clearly, $\mathcal{I}'$ is well-behaved for $\mathsf{Paths}$. One can prove by induction on $C$ that

$(*)$ for all $a \in \mathcal{N}$, $(w, C) \in \mathcal{P}(\mathcal{S}, a)$, and $d \in \Delta^{\mathcal{I}'}$ with $(a^{\mathcal{I}}, d) \in w^{\mathcal{I}}$, we have $d \in C^{\mathcal{I}}$ iff $d \in C^{\mathcal{I}'}$.

We show only the case $C = \exists r.D$, and leave the other cases to the reader. Let $d \in C^{\mathcal{I}}$. Then there is an $e \in D^{\mathcal{I}}$ with $(d, e) \in r^{\mathcal{I}}$. Since $(w, C) \in \mathcal{P}(\mathcal{S}, a)$, we have $(wr, D) \in \mathcal{P}(\mathcal{S}, a)$. Since $(a^{\mathcal{I}}, d) \in w^{\mathcal{I}}$, we have $(a^{\mathcal{I}}, e) \in (wr)^{\mathcal{I}}$. Thus, $e \in \Delta^{\mathcal{I}'}$ and the induction hypothesis yields $e \in D^{\mathcal{I}'}$. By definition of $\mathcal{I}'$ and the semantics, $d \in C^{\mathcal{I}'}$. Now let $d \in C^{\mathcal{I}'}$. Then there is an $e \in D^{\mathcal{I}'}$ with $(d, e) \in r^{\mathcal{I}}$. By definition of $\mathcal{I}'$, $(d, e) \in r^{\mathcal{I}}$. Since $(w, C) \in \mathcal{P}(\mathcal{S}, a)$, we have $(wr, D) \in \mathcal{P}(\mathcal{S}, a)$. Since $(a^{\mathcal{I}}, d) \in w^{\mathcal{I}}$, we have $(a^{\mathcal{I}}, e) \in (wr)^{\mathcal{I}}$. Thus, the induction hypothesis yields $e \in D^{\mathcal{I}}$.

Thus, $(*)$ is established. Using $(*)$ and the facts that $a_0^{\mathcal{I}} \in (\neg C_0)^{\mathcal{I}}$ and that $\mathcal{I}$ is a model of $\mathcal{A}$, it is not hard to verify that $a_0^{\mathcal{I}'} \in (\neg C_0)^{\mathcal{I}'}$ and that $\mathcal{I}'$ is a model of $\mathcal{A}$. We show that $\mathcal{I}'$ is also a model of $\mathsf{Circ}_{\mathsf{CP}}(\emptyset, \mathcal{A})$. Assume to the contrary that there is a model $\mathcal{J}'$ of $\mathcal{A}$ with $\mathcal{J}' <_{\mathsf{CP}} \mathcal{I}'$. Define an interpretation $\mathcal{J}$ as follows:

- $\Delta^{\mathcal{J}} = \Delta^{\mathcal{I}}$;

- $A^{\mathcal{J}} = A^{\mathcal{I}}$ for all $A \in F$;

- $A^{\mathcal{J}} = A^{\mathcal{J}'}$ for all $A \in V \cup M$;

- $r^{\mathcal{J}} = r^{\mathcal{J}'}$ for all $r \in \mathsf{N}_{\mathsf{C}}$;

- $b^{\mathcal{J}} = b^{\mathcal{I}}$ for all $b \in \mathsf{N}_{\mathsf{I}}$.

Using the assumption that $\mathsf{CP}$ does not contain fixed role names, it is not hard to verify that $\mathcal{J} <_{\mathsf{CP}} \mathcal{I}$. To obtain a contradiction to the fact that $\mathcal{I}$ is a model of $\mathsf{Circ}_{\mathsf{CP}}(\emptyset, \mathcal{A})$, it thus remains to show that $\mathcal{J}$ is a model of $\mathcal{A}$. To this end, we prove by induction on $C$ that

$(**)$ for all $a \in \mathcal{N}$, $(w, C) \in \mathcal{P}(\mathcal{S}, a)$, and $d \in \Delta^{\mathcal{J}'}$ with $(a^{\mathcal{J}}, d) \in w^{\mathcal{J}}$, we have $d \in C^{\mathcal{J}}$ iff $d \in C^{\mathcal{J}'}$.

Since the induction step is as in the proof of $(*)$, we only do the induction start. Thus, let $C = A \in \mathsf{N}_{\mathsf{C}}$. If $A \in V \cup M$, we are done by definition of $\mathcal{J}$. Now let $A \in F$. Since $\mathcal{I}'$ is a restriction of $\mathcal{I}$ and $A^{\mathcal{J}'} = A^{\mathcal{I}'}$, the definition of $\mathcal{J}$ yields $A^{\mathcal{J}} \cap \Delta^{\mathcal{J}'} = A^{\mathcal{J}'}$, as required. This finishes the proof of the claim.

To prove Lemma 5, we consider instance checking instead of satisfiability. Since we have provided polynomial reductions from satisfiability to (non)-instance checking and vice versa in Section 2, we nevertheless obtain the desired result. Let $\mathsf{Circ}_{\mathsf{CP}}(\emptyset, \mathcal{A})$ be a cKB with $\mathsf{CP} = (\prec, M, F, V)$ and $F = \{A_1, \ldots, A_k\}$. Take a concept assertion $C_0(a_0)$. Let $\mathcal{S} = \mathcal{A} \cup \{C_0(a_0)\}$ and $\mathcal{N}$ be the set of individual names in $\mathcal{S}$. Define

- $M' = M \cup \{A_1, \ldots, A_k, A'_1, \ldots, A'_k\}$, where the $A'_i$ are fresh concept names;

- $\mathcal{A}' = \mathcal{A} \cup \{\forall w.(A'_i \leftrightarrow \neg A_i)(a) \mid w \in \mathsf{Paths}(\mathcal{S}, a), a \in \mathcal{N}, i \leq k\}$;

- $\mathsf{CP}' = (\prec, M', \emptyset, V)$.

Then $\mathsf{Circ}_{\mathsf{CP}}(\emptyset, \mathcal{A}) \models C_0(a_0)$ iff $\mathsf{Circ}_{\mathsf{CP}'}(\emptyset, \mathcal{A}') \models C_0(a_0)$ follows immediately from Claim 1 and the fact that $\mathsf{Paths}(\mathcal{S}, a) = \mathsf{Paths}(\mathcal{S}', a)$, for $\mathcal{S}' = \mathcal{A}' \cup C_0(a_0)$. ❑

**Lemma 6** $C_0$ is satisfiable w.r.t. $\mathsf{Circ}_{\mathsf{CP}}(\mathcal{T}, \mathcal{A})$ iff $C_0 \sqcap B'$ is satisfiable w.r.t. $\mathsf{Circ}_{\mathsf{CP}'}(\mathcal{T}', \mathcal{A})$.

**Proof.** Suppose $\mathcal{I}$ is a model of $C_0$ and $\mathsf{Circ}_{\mathsf{CP}}(\mathcal{T}, \mathcal{A})$. Expand $\mathcal{I}$ to an interpretation $\mathcal{I}'$ by setting

$$A^{\mathcal{I}'} = B'^{\mathcal{I}'} = \Delta^{\mathcal{I}}, \quad B^{\mathcal{I}'} = A'^{\mathcal{I}'} = \emptyset, \quad u^{\mathcal{I}'} = \emptyset.$$

Clearly, $\mathcal{I}'$ is a model of $C_0 \sqcap B'$ and $(\mathcal{T}', \mathcal{A})$. We show that $\mathcal{I}'$ is a model of $\mathsf{Circ}_{\mathsf{CP}'}(\mathcal{T}', \mathcal{A})$. Assume to the contrary that there is a model $\mathcal{J}$ of $(\mathcal{T}', \mathcal{A})$ with $\mathcal{J} <_{\mathsf{CP}'} \mathcal{I}'$. Then $A'^{\mathcal{J}} = \emptyset$, $A^{\mathcal{J}} = \Delta^{\mathcal{I}}, B^{\mathcal{J}} = \emptyset$, and $B'^{\mathcal{J}} = \Delta^{\mathcal{I}}$. Since $u$ is varying and $\mathcal{J} <_{\mathsf{CP}'} \mathcal{I}'$, it is thus easy to show that $\mathcal{J} <_{\mathsf{CP}} \mathcal{I}$. This contradicts the fact that $\mathcal{I}$ is a model of $\mathsf{Circ}_{\mathsf{CP}}(\mathcal{T}, \mathcal{A})$.

Conversely, suppose $\mathcal{I}$ is a model of $C_0 \sqcap B'$ and $\mathsf{Circ}_{\mathsf{CP}'}(\mathcal{T}', \mathcal{A})$. We show that $\mathcal{I}$ is also a model of $C_0$ and $\mathsf{Circ}_{\mathsf{CP}}(\mathcal{T}, \mathcal{A})$. First observe that $A^{\mathcal{I}} = \Delta^{\mathcal{I}}$. For suppose that this is not the case. Define a new interpretation $\mathcal{J}$ in the same way as $\mathcal{I}$ except that $u^{\mathcal{J}} = \Delta^{\mathcal{I}} \times \Delta^{\mathcal{I}}$, $B^{\mathcal{J}} = \Delta^{\mathcal{I}}$, and $B'^{\mathcal{J}} = \emptyset$. Then $\mathcal{J} <_{\mathsf{CP}'} \mathcal{I}$ (since $B'^{\mathcal{I}} \neq \emptyset$) and $\mathcal{J}$ is a model of $(\mathcal{T}', \mathcal{A})$. Thus we have derived a contradiction. It follows that $C^{\mathcal{I}} = \Delta^{\mathcal{I}}$ and hence $\mathcal{I}$ is a model of $(\mathcal{T}, \mathcal{A})$ and $C_0$. It remains to show that there is no $\mathcal{J} <_{\mathsf{CP}} \mathcal{I}$ such that $\mathcal{J}$ is a model of $(\mathcal{T}, \mathcal{A})$. Assume such a $\mathcal{J}$ exists. Then $C^{\mathcal{J}} = \Delta^{\mathcal{I}}$. Define a model $\mathcal{J}'$ by expanding $\mathcal{J}$ as follows:

$$A^{\mathcal{J}'} = B'^{\mathcal{J}'} = \Delta^{\mathcal{I}}, \quad B^{\mathcal{J}'} = A'^{\mathcal{J}'} = \emptyset, \quad u^{\mathcal{J}'} = \emptyset.$$

Note that $A$, $B$, $A'$, $B'$, and $u$ are interpreted by $\mathcal{I}$ in the same way, then $\mathcal{J}' <_{\mathsf{CP}'} \mathcal{I}$. Moreover, $\mathcal{J}'$ is a model of $(\mathcal{T}', \mathcal{A})$. We have derived a contradiction. ❑

**Lemma 8** For all $\mathcal{L} \in \mathcal{DL}$, simultaneous satisfiability w.r.t. (concept) circumscribed KBs $\mathsf{Circ}_{\mathsf{CP}_1}(\mathcal{T}_1, \mathcal{A}_1), \ldots \mathsf{Circ}_{\mathsf{CP}_k}(\mathcal{T}_k, \mathcal{A}_k)$, such that $\mathsf{Circ}_{\mathsf{CP}_i}(\mathcal{T}_i, \mathcal{A}_i)$ and $\mathsf{Circ}_{\mathsf{CP}_j}(\mathcal{T}_j, \mathcal{A}_j)$ share no role names for $1 \leq i < j \leq k$, can be reduced to satisfiability w.r.t. single (concept) circumscribed KBs in polynomial time.

**Proof.** It suffices to reduce simultaneous satisfiability without shared role names to *the complement of instance checking* w.r.t. single cKBs. As a generalization is straightforward, we only give a proof for the case $k = 2$. Thus, let $\mathcal{L} \in \mathcal{DL}$ and $\mathsf{Circ}_{\mathsf{CP}_1}(\mathcal{T}_1, \mathcal{A}_1), \mathsf{Circ}_{\mathsf{CP}_2}(\mathcal{T}_2, \mathcal{A}_2)$ be cKBs formulated in $\mathcal{L}$ that share no role names, and $C_0$ an $\mathcal{L}$-concept. Moreover, let $A_0, \ldots, A_{n-1}$ be the concept names shared by the two cKBs, $\mathcal{R}$ the role names used in at least one of the two cKBs together with a fresh role name $r_0$, and $\mathsf{CP}_i = (\prec_i, M_i, F_i, V_i)$ for $i \in \{1, 2\}$. We obtain a new TBox $\mathcal{T}_2'$ from $\mathcal{T}_2$ by replacing each concept name $A_i$, $i < n$, with a new concept name $A_i'$. Let $\mathcal{A}_2'$ be obtained from $\mathcal{A}_2$ and $\mathsf{CP}_2' = (\prec_2', M_2', F_2', V_2')$ from $\mathsf{CP}_2$ in the same way. Define a TBox $\mathcal{T}^*$ as follows, where $P$ is a fresh concept name:

$$
\begin{aligned}
A_i \sqcap \neg A_i' &\sqsubseteq P \text{ for all } i < n \\
\neg A_i \sqcap A_i' &\sqsubseteq P \text{ for all } i < n \\
P &\sqsubseteq \forall r.P \text{ for all } r \in \mathcal{R} \\
\exists r.P &\sqsubseteq P \text{ for all } r \in \mathcal{R}
\end{aligned}
$$

Now set:

$$
\begin{aligned}
\mathcal{T} &:= \mathcal{T}_1 \cup \mathcal{T}_2' \cup \mathcal{T}^* \\
\mathcal{A} &:= \mathcal{A}_1 \cup \mathcal{A}_2' \cup \{r_0(b_1, b_2) \mid b_1, b_2 \text{ occur in } \mathcal{A}_1 \cup \mathcal{A}_2 \cup \mathcal{T}_1 \cup \mathcal{T}_2\} \\
\prec &:= \prec_1 \cup \prec_2' \\
M &:= M_1 \cup M_2' \\
F &:= F_1 \cup F_2' \\
V &:= V_1 \cup V_2' \cup \{P, r_0\} \\
\mathsf{CP} &:= (\prec, M, F, V)
\end{aligned}
$$

Let $a_0$ be an individual name from $\mathcal{A}_1$ (clearly, we may assume that $\mathcal{A}_1 \neq \emptyset$). It remains to show the following:

**Claim.** $C_0$ is simultaneously satisfiable w.r.t. $\mathsf{Circ}_{\mathsf{CP}_1}(\mathcal{T}_1, \mathcal{A}_1)$ and $\mathsf{Circ}_{\mathsf{CP}_2}(\mathcal{T}_2, \mathcal{A}_2)$ iff $a_0$ is not an instance of $\neg(\neg P \sqcap \exists r_0.C_0)$ w.r.t. $\mathsf{Circ}_{\mathsf{CP}}(\mathcal{T}, \mathcal{A})$.

"If". Assume that $a_0$ is not an instance of $\neg(\neg P \sqcap \exists r_0.C_0)$ w.r.t. $\mathsf{Circ}_{\mathsf{CP}}(\mathcal{T}, \mathcal{A})$. Then there is a model $\mathcal{I}$ of $\mathsf{Circ}_{\mathsf{CP}}(\mathcal{T}, \mathcal{A})$ with $a_0^{\mathcal{I}} \in (\neg P \sqcap \exists r_0.C_0)^{\mathcal{I}}$. We call $\mathcal{I}$ *connected* if the directed graph $G_{\mathcal{I}} = (\Delta^{\mathcal{I}}, \bigcup_{r \in \mathcal{R}} r^{\mathcal{I}} \cup (r^-)^{\mathcal{I}})$ is connected. A *connected component* $\mathcal{I}'$ of $\mathcal{I}$ is the restriction of $\mathcal{I}$ to domain $\Delta^{\mathcal{I}'}$ such that $\Delta^{\mathcal{I}'}$ is a (maximal) connected component in $G_{\mathcal{I}}$. We may assume without loss of generality that the chosen model $\mathcal{I}$ is connected: if it is not, then the use of the role $r_0$ ensures that there is a connected component $\mathcal{I}'$ of $\mathcal{I}$ that contains $b^{\mathcal{I}}$ for all individual names $b$ in $\mathcal{A}_1 \cup \mathcal{A}_2 \cup \mathcal{T}_1 \cup \mathcal{T}_2$. It is easy to see that $\mathcal{I}'$ is a model of $\mathsf{Circ}_{\mathsf{CP}}(\mathcal{T}, \mathcal{A})$ and $a_0^{\mathcal{I}'} \in (\neg P \sqcap \exists r_0.C_0)^{\mathcal{I}'}$, thus we may simply replace $\mathcal{I}$ by $\mathcal{I}'$.

We have to show that $C_0$ is simultaneously satisfiable with respect to $\mathsf{Circ}_{\mathsf{CP}_1}(\mathcal{T}_1, \mathcal{A}_1)$ and $\mathsf{Circ}_{\mathsf{CP}_2}(\mathcal{T}_1, \mathcal{A}_2)$. Clearly, $\mathcal{I}$ is a model of $C_0$. By construction of $\mathsf{Circ}_{\mathsf{CP}}(\mathcal{T}, \mathcal{A})$, it is a model of $\mathcal{T}_1$ and $\mathcal{A}_1$. To show that $\mathcal{I}$ is also a model of $\mathsf{Circ}_{\mathsf{CP}_1}(\mathcal{T}_1, \mathcal{A}_1)$, assume to the contrary that this is not the case. Then there exists a model $\mathcal{J}$ of $\mathcal{T}_1$ and $\mathcal{A}_1$ such that $\mathcal{J} <_{\mathsf{CP}_1} \mathcal{I}$. Define a model $\mathcal{J}'$ as follows:

- $\Delta^{\mathcal{J}'} = \Delta^{\mathcal{J}}$;

- all predicates used in $\mathcal{T}_1$ and $\mathcal{A}_1$ are interpreted as in $\mathcal{J}$;

- all predicates used in $\mathcal{T}_2'$ and $\mathcal{A}_2'$ are interpreted as in $\mathcal{I}$.

- $P^{\mathcal{J}'} := \begin{cases} \Delta^{\mathcal{I}} & \text{if } ((A_i \sqcap \neg A_i') \sqcup (\neg A_i \sqcap A_i'))^{\mathcal{J}} \neq \emptyset \text{ for some } i < n \\ \emptyset & \text{otherwise} \end{cases}$

- $r_0^{\mathcal{J}'} = \Delta^{\mathcal{J}'} \times \Delta^{\mathcal{J}'}$.

It is readily checked that $\mathcal{J}'$ is a model of $\mathcal{T}$ and $\mathcal{A}$, and that $\mathcal{J}' <_{\mathsf{CP}} \mathcal{I}$. Thus, we have derived a contradiction to the fact that $\mathcal{I}$ is a model of $\mathsf{Circ}_{\mathsf{CP}}(\mathcal{T}, \mathcal{A})$, and it follows that $\mathcal{I}$ is a model of $\mathsf{Circ}_{\mathsf{CP}}(\mathcal{T}_1, \mathcal{A}_1)$.

It remains to show that $\mathcal{I}$ is a model of $\mathsf{Circ}_{\mathsf{CP}}(\mathcal{T}_2, \mathcal{A}_2)$. Since $\mathcal{I}$ is connected, a model of $\mathcal{T}^*$, and satisfies $a_0^{\mathcal{I}} \notin P^{\mathcal{I}}$, we have that $A_i^{\mathcal{I}} = (A_i')^{\mathcal{I}}$ for all $i < n$. Therefore, $\mathcal{I}$ is also a model of $\mathcal{T}_2$ and $\mathcal{A}_2$. Analogously to the case of $\mathsf{Circ}_{\mathsf{CP}_1}(\mathcal{T}_1, \mathcal{A}_1)$, we can now show that $\mathcal{I}$ is a model of $\mathsf{Circ}_{\mathsf{CP}_2}(\mathcal{T}_2, \mathcal{A}_2)$.

"Only if". Assume that $C_0$ is simultaneously satisfiable w.r.t. the cKBs $\mathsf{Circ}_{\mathsf{CP}_1}(\mathcal{T}_1, \mathcal{A}_1)$ and $\mathsf{Circ}_{\mathsf{CP}_2}(\mathcal{T}_2, \mathcal{A}_2)$. Then there exists a model $\mathcal{I}$ of $C_0$ that is a model of $\mathsf{Circ}_{\mathsf{CP}_1}(\mathcal{T}_1, \mathcal{A}_1)$ and $\mathsf{Circ}_{\mathsf{CP}_2}(\mathcal{T}_2, \mathcal{A}_2)$. We modify $\mathcal{I}$ to a new model $\mathcal{I}'$ by setting

- $(A_i')^{\mathcal{I}'} := A_i^{\mathcal{I}}$ for all $i < n$;

- $P^{\mathcal{I}'} := \emptyset$;

- $r_0^{\mathcal{I}'} := \Delta^{\mathcal{I}} \times \Delta^{\mathcal{I}}$.

It is easy to see that $\mathcal{I}'$ is a model of $\mathcal{T}$ and $\mathcal{A}$, and that $a^{\mathcal{I}'} \in (\neg P \sqcap \exists r_0.C_0)^{\mathcal{I}'}$. To show that $a_0$ is not an instance of $\neg(\neg P \sqcap \exists r_0.C_0)$ w.r.t. $\mathsf{Circ}_{\mathsf{CP}}(\mathcal{T}, \mathcal{A})$, it thus remains to prove that $\mathcal{I}'$ is also model of $\mathsf{Circ}_{\mathsf{CP}}(\mathcal{T}, \mathcal{A})$. To do this, we first show the following:

(a) $\mathcal{I}'$ is a model of $\mathsf{Circ}_{\mathsf{CP}_1}(\mathcal{T}_1, \mathcal{A}_1)$. This is the case since any model $\mathcal{J}$ of $\mathcal{T}_1$ and $\mathcal{A}_1$ with $\mathcal{J} <_{\mathsf{CP}_1} \mathcal{I}'$ satisfies $\mathcal{J} <_{\mathsf{CP}_1} \mathcal{I}$. Thus, the existence of such a model contradicts the fact that $\mathcal{I}$ is a model of $\mathsf{Circ}_{\mathsf{CP}_1}(\mathcal{T}_1, \mathcal{A}_1)$.

(b) $\mathcal{I}'$ is a model of $\mathsf{Circ}_{\mathsf{CP}_2'}(\mathcal{T}_2', \mathcal{A}_2')$. Assume to the contrary that there is a model $\mathcal{J}$ of $\mathcal{T}_2'$ and $\mathcal{A}_2'$ with $\mathcal{J} <_{\mathsf{CP}_2'} \mathcal{I}'$. Convert $\mathcal{J}$ into an interpretation $\mathcal{J}^*$ by setting $A_i^{\mathcal{J}^*} := (A_i')^{\mathcal{J}}$ for all $i < n$. Then, $\mathcal{J}^*$ is a model of $\mathcal{T}_2$ and $\mathcal{A}_2$ and satisfies $\mathcal{J}^* <_{\mathsf{CP}_2} \mathcal{I}$. This is a contradiction to the fact that $\mathcal{I}$ is a model of $\mathsf{Circ}_{\mathsf{CP}_2}(\mathcal{T}_2, \mathcal{A}_2)$.

Now, assume to the contrary of what is to be shown that there is a model $\mathcal{J}'$ of $\mathcal{T}$ and $\mathcal{A}$ with $\mathcal{J}' <_{\mathsf{CP}} \mathcal{I}'$. By definition of $\mathsf{CP}$, $\mathcal{J}' <_{\mathsf{CP}} \mathcal{I}'$ implies that either $\mathcal{J}' <_{\mathsf{CP}_1} \mathcal{I}'$ or $\mathcal{J}' <_{\mathsf{CP}_2'} \mathcal{I}'$ hold. Since $\mathcal{J}'$ clearly satisfies $\mathcal{T}_1$, $\mathcal{A}_1$, $\mathcal{T}_2'$, and $\mathcal{A}_2'$, we obtain a contradiction to (a) and (b). ❏

## Appendix B. Missing Proofs in Section 4

**Lemma 14** $G$ is a yes-instance of co-3CERTCOL$_S$ iff $C_G$ is satisfiable w.r.t. $\mathsf{Circ}_{\mathsf{CP}_G}(\emptyset, \mathcal{A}_G)$, where $\mathsf{CP}_G = (\prec, M, F, V)$ with $\prec = \emptyset$, $M = \{\mathsf{Root}, \mathsf{Leaf}, \mathsf{Clash}\}$,

$$F = \{\mathsf{LeafFix}, \mathsf{Tr}, X_0, \dots, X_{n-1}, Y_0, \dots, Y_{n-1}, \},$$

and $V$ the set of all remaining predicates in $\mathcal{A}_G$.

**Proof.** "If". Suppose that $C_G$ is satisfiable w.r.t. $\mathsf{Circ}_{\mathsf{CP}_G}(\emptyset, \mathcal{A}_G)$ and let $\mathcal{I}$ be a model of $\mathsf{Circ}_{\mathsf{CP}_G}(\emptyset, \mathcal{A}_G)$ with $C_G^{\mathcal{I}} \neq \emptyset$. We have to show that $G$ is a yes-instance of co-CERT3COL$_S$. To start, we show that

(I) $a_0^{\mathcal{I}} \in C_G^{\mathcal{I}}$.

Assume to the contrary that this is not the case. Then there is a $d \in C_G^{\mathcal{I}}$ with $d \neq a_0^{\mathcal{I}}$. Since $d \in C_G^{\mathcal{I}}$, we have $d \in \mathsf{Root}^{\mathcal{I}}$. Let $\mathcal{J}$ be the interpretation obtained from $\mathcal{I}$ by setting $\mathsf{Root}^{\mathcal{J}} = \{a_0^{\mathcal{I}}\}$. By definition of $\mathcal{A}_G$, we have $a_0^{\mathcal{I}} \in \mathsf{Root}^{\mathcal{I}}$, and thus $\mathcal{J} \prec_{\mathsf{CP}_G} \mathcal{I}$. Moreover, it is easily seen that $\mathcal{J}$ is a model of $\mathcal{A}_G$. We have thus established a contradiction to the fact that $\mathcal{I}$ is a model of $\mathsf{Circ}_{\mathsf{CP}_G}(\emptyset, \mathcal{A}_G)$, which proves (I).

By Lines (1)-(4) (Fig. 2), there are elements $d_{i,w} \in \Delta^{\mathcal{I}}$ (the nodes of the tree) for $i \leq 2n$ and $w \in \{0,1\}^i$ such that

- $a_0^{\mathcal{I}} = d_{0,\varepsilon}$;

- $d_{i,w} \in X_j^{\mathcal{I}}$ iff the $j+1$-st bit of $w$ is 1, for all $i \leq 2n$ and $j \leq \min\{i, n-1\}$;

- $d_{i,w} \in Y_j^{\mathcal{I}}$ iff the $n+j+1$-st bit of $w$ is 1, for all $i, j$ with $n < i \leq 2n$ and $j \leq i - n$;

- $(d_{i,w}, d_{i+1,w0}), (d_{i,w}, d_{i+1,w1}) \in r^{\mathcal{I}}$, for all $i < 2n$;

- $d_{2n,w} \in \mathsf{Leaf}^{\mathcal{I}}$, for all $w \in \{0,1\}^{2n}$.

For $i, j < 2^n$, we use $\ell_{i,j}$ to denote the element $d_{2n,w}$ such that $w \in \{0,1\}^{2n}$ denotes the binary encoding of $i$ followed by that of $j$. We now show that

(II) $\mathsf{Leaf}^{\mathcal{I}} = \{\ell_{i,j} \mid i, j < 2^n\}$.

Assume to the contrary that this is not the case, i.e., that there is a $d \in \mathsf{Leaf}^{\mathcal{I}}$ such that $d \neq \ell_{i,j}$ for all $i, j < 2^n$. Let $i_d, j_d > 0$ be integers such that the truth values of $X_0, \ldots, X_{n-1}$ at $d$ encode $i_d$ and the truth values of $Y_0, \ldots, Y_{n-1}$ encode $j_d$. Starting from $\mathcal{I}$, we construct an interpretation $\mathcal{J}$ by setting

$$
\begin{aligned}
\mathsf{Leaf}^{\mathcal{J}} &:= \mathsf{Leaf}^{\mathcal{I}} \setminus \{d\} \\
r^{\mathcal{J}} &:= (r^{\mathcal{I}} \setminus (\Delta^{\mathcal{I}} \times d)) \cup \{(e, \ell_{i_d, j_d}) \mid (e, d) \in r^{\mathcal{I}}\}
\end{aligned}
$$

Further modify $\mathcal{J}$ into $\mathcal{J}'$ by setting

$$
\begin{aligned}
P^{\mathcal{J}'} := \quad &(\exists r^{2n}.\exists \mathsf{var1}.\neg \mathsf{Leaf})^{\mathcal{J}} \cup \\
&(\exists r^{2n}.\exists \mathsf{var2}.\neg \mathsf{Leaf})^{\mathcal{J}} \cup \\
&(\exists r^{2n}.\exists \mathsf{col1}.\neg \mathsf{Leaf})^{\mathcal{J}} \cup \\
&(\exists r^{2n}.\exists \mathsf{col2}.\neg \mathsf{Leaf})^{\mathcal{J}}
\end{aligned}
$$

By going through Lines (1) to (26), it is straightforward to check that $\mathcal{J}'$ is a model of $\mathcal{A}_G$. Moreover, we clearly have $\mathcal{J}' <_{\mathsf{CP}_G} \mathcal{I}$, and thus obtain a contradiction to the fact that $\mathcal{I}$ is a model of $\mathsf{Circ}_{\mathsf{CP}_G}(\emptyset, \mathcal{A}_G)$. We have thus shown (II).

The following is an easy consequence of (I), the fact that $\neg P$ is a conjunct of $C_G$, and Lines (8), (9), (20), and (21):

(III) $(\ell_{i,j}, d) \in r^{\mathcal{I}}$ implies $d \in \mathsf{Leaf}^{\mathcal{I}}$ for all $i, j < 2^n$, $d \in \Delta^{\mathcal{I}}$, and $r \in \{\mathsf{var1}, \mathsf{var2}, \mathsf{col1}, \mathsf{col2}\}$.

Now suppose to the contrary of what we aim to prove that $G$ is not a yes-instance. Then, for all truth assignments $t$, the subgraph $t(G)$ is 3-colorable. In particular, this holds for the assignment $t$ defined by setting, for all $i, j < 2^n$,

$$t(V_{ij}) := 1 \text{ iff } \ell_{i,j} \in \mathsf{Tr}^{\mathcal{I}}.$$

Let $c : \{0, \ldots, 2^n - 1\} \rightarrow \{R, G, B\}$ be a 3-coloring of $t(G)$ and construct an interpretation $\mathcal{J}$ by starting from $\mathcal{I}$ and applying the following modifications:

$$
\begin{aligned}
C^{\mathcal{J}} &= \{\ell_{i,0} \mid i < 2^n, c(i) = C\} \text{ for all } C \in \{R, G, B\} \\
\mathsf{Clash}^{\mathcal{J}} &= \emptyset.
\end{aligned}
$$

Clearly, $\mathcal{J} <_{\mathsf{CP}_G} \mathcal{I}$: the minimized predicate $\mathsf{Clash}$ is empty in $\mathcal{J}$, but non-empty in $\mathcal{I}$ since $C_G^{\mathcal{I}}$ is non-empty. To obtain a contradiction, it thus suffices to show that $\mathcal{J}$ is a model of $\mathcal{A}_G$.

Since $\mathcal{I}$ and $\mathcal{J}$ agree on all predicates but $R, G, B$, and $\mathsf{Clash}$, all inclusions that do not mention these concepts are satisfied in $\mathcal{J}$. These are Lines (1) to (21). Lines (22) and (23) are satisfied by construction of $\mathcal{J}$ and since, due to Line (5) and (II), $(a_0^{\mathcal{J}}, d) \in (r^{2n})^{\mathcal{J}}$ implies $d = \ell_{i,j}$ for some $i, j < 2^n$. It thus remains to consider Lines (24) to (26). We first show that

(IV) for all $i, j < 2^n$, $(i, j)$ is an edge of $t(G)$ iff $\ell_{i,j} \notin \mathsf{Elim}^{\mathcal{J}}$.

Let $i, j < 2^n$ and let the potential edge $(i, j)$ be labeled with $V_{k_1, k_2} \vee V_{k_3, k_4}$ (since the circuits $c_S^{(i)}$ and $c_j^{(i)}$ deliver an output for any input, we can assume that also potential, but non-existing edges have a label). By (II) and (III) together with Lines (8) and (9), $\mathsf{var1}^{\mathcal{J}} = \mathsf{var1}^{\mathcal{I}} = \{\ell_{k_1, k_2}\}$ and $\mathsf{var2}^{\mathcal{J}} = \mathsf{var2}^{\mathcal{I}} = \{\ell_{k_3, k_4}\}$. Thus, the definition of $t$ together with Lines (12) to (16) yields that $(i, j)$ is an edge of $t(G)$ iff $\ell_{i,j} \notin \mathsf{Elim}^{\mathcal{I}}$. To prove (IV), it remains to note that $\mathcal{I}$ and $\mathcal{J}$ interpret the concept name $\mathsf{Elim}$ in the same way.

Now, we prove that (24) to (26) are satisfied in $\mathcal{J}$. Let $(a_0^{\mathcal{J}}, d) \in (r^{2n})^{\mathcal{J}}$. By Line 5 and (II), $d = \ell_{i,j}$ for some $i, j < 2^n$. Let $\ell_{i,j} \notin \mathsf{Elim}^{\mathcal{J}}$. By (IV) and since $c$ is a 3-coloring of $t(G)$, we get $c(i) \neq c(j)$. Thus, by construction of $\mathcal{J}$, $\ell_{i,0} \notin C^{\mathcal{J}}$ or $\ell_{j,0} \notin C^{\mathcal{J}}$ for all $C \in \{R, G, B\}$. By (II) and (III) together with Lines (18) and (19), $\mathsf{col1}^{\mathcal{J}} = \mathsf{col1}^{\mathcal{I}} = \{\ell_{i,0}\}$ and $\mathsf{col2}^{\mathcal{J}} = \mathsf{col2}^{\mathcal{I}} = \{\ell_{j,0}\}$. Therefore, $\ell_{i,j} \notin (\exists \mathsf{col1}.C \sqcap \exists \mathsf{col2}.C)^{\mathcal{J}}$ for all $C \in \{R, G, B\}$. Since this holds for any $\ell_{i,j} \notin \mathsf{Elim}^{\mathcal{J}}$, the preconditions of the implications in Lines (24) to (26) are never satisfied. Thus, the implications are satisfied.

"Only if". Suppose that $G$ is a yes-instance and let $t$ be a truth assignment such that $t(G)$ is not 3-colorable. Let $c : \{0, \ldots, 2^n - 1\} \rightarrow \{R, G, B\}$ be a color assignment that minimizes (w.r.t. set inclusion) the set $\{(i, j) \mid (i, j) \text{ an edge in } G \text{ with } c(i) = c(j)\}$. Define an interpretation $\mathcal{I}$ as follows (here and in the following, we do not distinguish between a number and its binary encoding as a string):

$$\begin{aligned}
\Delta^{\mathcal{I}} &= \{d_{i,w} \mid i \leq 2^n, w \in \{0,1\}^i\} \\
\mathsf{Root}^{\mathcal{I}} &= \{d_{0,\varepsilon}\} \\
\mathsf{Leaf}^{\mathcal{I}} &= \{d_{2n,w} \mid w \in \{0,1\}^{2n}\} \\
\mathsf{LeafFix}^{\mathcal{I}} &= \{d_{2n,w} \mid w \in \{0,1\}^{2n}\} \\
X_j^{\mathcal{I}} &= \{d_{i,w} \mid \text{ the } j+1\text{-st bit of } w \text{ is } 1,\ i \leq 2n,\text{ and } j \leq \min\{i, n-1\}\}; \\
Y_j^{\mathcal{I}} &= \{d_{i,w} \mid \text{ the } n+j+1\text{-st bit of } w \text{ is } 1,\ n < i \leq 2n,\text{ and } j \leq i-n\}; \\
\mathsf{Tr}^{\mathcal{I}} &= \{d_{2n,ij} \mid t(V_{ij}) = 1\} \\
\mathsf{Tr}_\ell^{\mathcal{I}} &= \{d_{2n,ij} \mid t(V_{ij}) \leftrightarrow c_S^{(\ell)}(i,j)\} \quad (\ell = 1, 2) \\
\mathsf{Elim}^{\mathcal{I}} &= \{d_{2n,ij} \mid (i,j) \text{ is not an edge of } t(G)\} \\
C^{\mathcal{I}} &= \{d_{2n,i0} \mid c(i) = C\} \quad (C = R, G, B) \\
\mathsf{Clash}^{\mathcal{I}} &= \{d_{2n,ij} \mid (i,j) \text{ is an edge of } t(G) \text{ and } c(i) = c(j)\} \\
r^{\mathcal{I}} &= \{(d_{i,w}, d_{i+1,w0}), (d_{i,w}, d_{i+1,w1}) \mid i < 2n\} \\
\mathsf{var1}^{\mathcal{I}} &= \{(d_{2n,ij}, d_{2n,kl}) \mid \text{the first variable in the label of } (i,j) \text{ is } V_{kl}\} \\
\mathsf{var2}^{\mathcal{I}} &= \{(d_{2n,ij}, d_{2n,kl}) \mid \text{the second variable in the label of } (i,j) \text{ is } V_{kl}\} \\
\mathsf{col1}^{\mathcal{I}} &= \{(d_{2n,ij}, d_{2n,i0}) \mid i < 2^n\} \\
\mathsf{col2}^{\mathcal{I}} &= \{(d_{2n,ij}, d_{2n,j0}) \mid i < 2^n\} \\
P^{\mathcal{I}} &= \emptyset \\
a_0^{\mathcal{I}} &= d_{0,\varepsilon}.
\end{aligned}$$

For each Boolean circuit $c$, the corresponding output concept name $\mathsf{Out}_c$ is interpreted as $\mathsf{Out}_c^{\mathcal{I}} := \{d_{2n,ij} \mid i, j < 2n \text{ and } c(i,j) \text{ is } \textit{true}\}$.

To show that $C_G$ is satisfiable w.r.t. $\mathsf{Circ}_{\mathsf{CP}_G}(\emptyset, \mathcal{A}_G)$, it suffices to show that $a_0^{\mathcal{I}} \in C_G^{\mathcal{I}}$ and $\mathcal{I}$ is a model of $\mathsf{Circ}_{\mathsf{CP}_G}(\emptyset, \mathcal{A}_G)$. The former is easy: recall that $C_G = \mathsf{Root} \sqcap \neg P \sqcap \exists r^{2n}.\mathsf{Clash}$. By definition of $\mathcal{I}$, $a_0^{\mathcal{I}} \in (\mathsf{Root} \sqcap \neg P)^{\mathcal{I}}$. Since $c$ is not a 3-coloring, $a_0^{\mathcal{I}} \in (\exists r^{2n}.\mathsf{Clash})^{\mathcal{I}}$. It thus remains to show that $\mathcal{I}$ is a model of $\mathsf{Circ}_{\mathsf{CP}_G}(\emptyset, \mathcal{A}_G)$. Since it is easy to verify that $\mathcal{I}$ is a model of $\mathcal{A}_G$, this boils down to showing that there is no model $\mathcal{J}$ of $\mathcal{A}_G$ with $\mathcal{J} <_{\mathsf{CP}_G} \mathcal{I}$.

Assume to the contrary that there is such a $\mathcal{J}$. By Lines (1)–(5) and since $\mathcal{J}$ is a model of $\mathcal{A}_G$, we have $\#\mathsf{Leaf}^{\mathcal{J}} \geq 2^{2n}$. Since $\#\mathsf{Leaf}^{\mathcal{I}} = 2^{2n}$ and $\mathcal{J} <_{\mathsf{CP}_G} \mathcal{I}$, we get $\mathsf{Leaf}^{\mathcal{J}} = \mathsf{Leaf}^{\mathcal{I}}$. For similar but simpler reasons, $\mathsf{Root}^{\mathcal{J}} = \mathsf{Root}^{\mathcal{I}}$. Thus, $\mathcal{J} <_{\mathsf{CP}_G} \mathcal{I}$ implies $\mathsf{Clash}^{\mathcal{J}} \subsetneq \mathsf{Clash}^{\mathcal{I}}$. By Lines (1) to (5) and since $\mathcal{I}$ and $\mathcal{J}$ are models of $\mathcal{A}_G$ with $\mathsf{Leaf}^{\mathcal{J}} = \mathsf{Leaf}^{\mathcal{I}}$ and $\#\mathsf{Leaf}^{\mathcal{I}} = \#\mathsf{Leaf}^{\mathcal{J}} = 2^{2n}$, we have

(I) $\{d \in \Delta^{\mathcal{I}} \mid (a_0^{\mathcal{I}}, d) \in (r^{2n})^{\mathcal{I}}\} = \{d \in \Delta^{\mathcal{J}} \mid (a_0^{\mathcal{J}}, d) \in (r^{2n})^{\mathcal{J}}\} = \mathsf{Leaf}^{\mathcal{I}} = \mathsf{Leaf}^{\mathcal{J}}$

Thus, Lines (24) to (26) and the fact that $\mathcal{J}$ is a model of $\mathcal{A}_G$ ensure that

(II) $\bigcup_{C \in \{R,G,B\}} (\mathsf{Leaf} \sqcap \neg\mathsf{Elim} \sqcap \exists\mathsf{col1}.C \sqcap \exists\mathsf{col2}.C)^{\mathcal{J}} \subseteq \mathsf{Clash}^{\mathcal{J}}$.

Define a coloring $c'$ by setting

$$c'(i) = C \text{ iff } d_{2n,i0} \in C^{\mathcal{J}} \quad (C = R, G, B).$$

Suppose that

(III) (a) $\mathsf{Elim}^{\mathcal{I}} \cap \mathsf{Leaf}^{\mathcal{I}} = \mathsf{Elim}^{\mathcal{J}} \cap \mathsf{Leaf}^{\mathcal{I}}$,
    (b) $\mathsf{col1}^{\mathcal{I}} \cap (\mathsf{Leaf}^{\mathcal{I}} \times \mathsf{Leaf}^{\mathcal{I}}) = \mathsf{col1}^{\mathcal{J}} \cap (\mathsf{Leaf}^{\mathcal{J}} \times \mathsf{Leaf}^{\mathcal{J}})$, and
    (c) $\mathsf{col2}^{\mathcal{I}} \cap (\mathsf{Leaf}^{\mathcal{I}} \times \mathsf{Leaf}^{\mathcal{I}}) = \mathsf{col2}^{\mathcal{J}} \cap (\mathsf{Leaf}^{\mathcal{J}} \times \mathsf{Leaf}^{\mathcal{J}})$.

Then (II) guarantees that if $(i,j)$ is an edge in $G$ with $c'(i) = c'(j)$, then $d_{2n,ij} \in \mathsf{Clash}^{\mathcal{J}}$. Since $\mathsf{Clash}^{\mathcal{J}} \subsetneq \mathsf{Clash}^{\mathcal{I}}$, we get that

1. if $c'(i) = c'(j)$, then $c(i) = c(j)$;

2. there are $i, j$ with $c'(i) \neq c'(j)$, but $c(i) = c(j)$.

This contradicts our initial minimality assumption on the coloring $c$.

It thus remains to prove (III). We start with (a). Assume that

(d) $\mathsf{var1}^{\mathcal{I}} \cap (\mathsf{Leaf}^{\mathcal{I}} \times \mathsf{Leaf}^{\mathcal{I}}) = \mathsf{var1}^{\mathcal{J}} \cap (\mathsf{Leaf}^{\mathcal{J}} \times \mathsf{Leaf}^{\mathcal{J}})$ and

(e) $\mathsf{var2}^{\mathcal{I}} \cap (\mathsf{Leaf}^{\mathcal{I}} \times \mathsf{Leaf}^{\mathcal{I}}) = \mathsf{var2}^{\mathcal{J}} \cap (\mathsf{Leaf}^{\mathcal{J}} \times \mathsf{Leaf}^{\mathcal{J}})$.

Then, Lines (12) to (16) together with (I) and the fact that $\mathsf{Tr}^{\mathcal{I}} = \mathsf{Tr}^{\mathcal{J}}$ implies (a). It thus remains to prove (b) to (e). We concentrate on (b) as the other cases are analogous. Take $(d,d') \in \mathsf{col1}^{\mathcal{I}} \cap (\mathsf{Leaf}^{\mathcal{I}} \times \mathsf{Leaf}^{\mathcal{I}})$. Then $d = d_{2n,ij}$ and $d' = d_{2n,i'j'}$ for some $i, j, i', j' < 2^n$. By Line (18), $i' = i$ and $j' = 0$. By (I) and Lines (17) and (18) and since $\mathcal{I}$ and $\mathcal{J}$ agree on the interpretation of $X_0, \ldots, X_{n-1}, Y_0, \ldots, Y_{n-1}$, there is an $e \in \mathsf{LeafFix}^{\mathcal{J}}$ such that $(d_{2n,ij}, e) \in \mathsf{col1}^{\mathcal{J}}$, the value encoded by $X_0, \ldots, X_{n-1}$ at $e$ in $\mathcal{J}$ is $i$, and the value encoded by $Y_0, \ldots, Y_{n-1}$ at $e$ in $\mathcal{J}$ is 0. Since $\mathsf{LeafFix}^{\mathcal{I}} = \mathsf{LeafFix}^{\mathcal{J}}$, we have $\mathsf{LeafFix}^{\mathcal{J}} = \mathsf{Leaf}^{\mathcal{J}}$. However, there is only a single element of $\mathsf{Leaf}^{\mathcal{J}}$ where $X_0, \ldots, X_{n-1}$ encodes $i$ and $Y_0, \ldots, Y_{n-1}$ encodes 0 and this is $d_{2n,i0} = d'$. The converse direction is analogous.   □

**Corollary 16** In $\mathcal{ALC}$, satisfiability w.r.t. concept-circumscribed KBs is $\mathrm{NExp}^{\mathrm{NP}}$-hard even if the TBox is acyclic, the ABox and preference relations are empty, and there are no fixed predicates.

**Proof.** The ABox $\mathcal{A}_G$ from the reduction given in Section 4.2.1 is of the form $\{C_0(a_0)\}$ and the circumscription pattern $\mathsf{CP}_G$ has an empty preference relation. It thus suffices to show that there is a polynomial reduction of satisfiability w.r.t. concept-circumscribed KBs of this form to satisfiability w.r.t. concept-circumscribed KBs with an acyclic TBox, empty ABox and preference relation, and no fixed predicates.

Let $\mathsf{Circ}_{\mathsf{CP}}(\emptyset, \mathcal{A})$ be a concept-circumscribed KB with $\mathcal{A} = \{C_0(a_0)\}$ and $\mathsf{CP} = (\prec, M, V, F)$ with $\prec = \emptyset$, and let $C$ be an $\mathcal{ALC}$ concept. Define $\mathcal{T} = \{A \sqsubseteq \exists u.C_0\}$, where $A$ is a concept name that does not occur in $\mathcal{A}$ and $C$ and $u$ is a role name that does not occur in $\mathcal{A}$ and $C$. Also define $\mathsf{CP}' = (\prec, M, V \cup \{u\}, F \cup \{A\})$. Then $C$ is satisfiable w.r.t. $\mathsf{Circ}_{\mathsf{CP}}(\emptyset, \mathcal{A})$ iff $A \sqcap C$ is satisfiable w.r.t. $\mathsf{Circ}_{\mathsf{CP}'}(\mathcal{T}, \emptyset)$:

"If". If $A \sqcap C$ is satisfiable w.r.t. $\mathsf{Circ}_{\mathsf{CP}'}(\mathcal{T}, \emptyset)$, then there is a model $\mathcal{I}$ of $\mathsf{Circ}_{\mathsf{CP}'}(\mathcal{T}, \emptyset)$ and a $d_0 \in (A \sqcap C)^{\mathcal{I}}$. Thus, there is an $e_0 \in C_0^{\mathcal{I}}$. Modify $\mathcal{I}$ to obtain a new interpretation $\mathcal{J}$ by setting $a_0^{\mathcal{J}} = e_0$. Clearly, $\mathcal{J}$ is a model of $\mathcal{A}$ and $C$. To show that it is also a model of $\mathsf{Circ}_{\mathsf{CP}}(\emptyset, \mathcal{A})$, assume to the contrary that there is a model $\mathcal{J}'$ of $\mathcal{A}$ with $\mathcal{J}' <_{\mathsf{CP}} \mathcal{J}$. Modify $\mathcal{J}'$ into an interpretation $\mathcal{I}'$ by setting $A^{\mathcal{I}'} = A^{\mathcal{I}}$ and $u^{\mathcal{I}'} = \Delta^{\mathcal{I}} \times \Delta^{\mathcal{I}}$. It is readily checked that $\mathcal{I}'$ is a model of $\mathcal{T}$ and $\mathcal{I}' <_{\mathsf{CP}'} \mathcal{I}$, thus we have obtained a contradiction to the fact that $\mathcal{I}$ is a model of $\mathsf{Circ}_{\mathsf{CP}'}(\mathcal{T}, \emptyset)$.

"Only if". If $C$ is satisfiable w.r.t. $\mathsf{Circ}_{\mathsf{CP}}(\emptyset, \mathcal{A})$, then there is a model $\mathcal{I}$ of $\mathsf{Circ}_{\mathsf{CP}}(\emptyset, \mathcal{A})$ and a $d_0 \in C^{\mathcal{I}}$. Let $\mathcal{J}$ be defined as $\mathcal{I}$, except that $A^{\mathcal{J}} = \{d_0\}$ and $u^{\mathcal{J}} = \Delta^{\mathcal{I}} \times \Delta^{\mathcal{I}}$. Clearly, $\mathcal{J}$ is a model of $\mathcal{T}$ and $A \sqcap C$. To show that $\mathcal{J}$ is also a model of $\mathsf{Circ}_{\mathsf{CP}}(\mathcal{T}, \emptyset)$, assume to the contrary that there is a model $\mathcal{J}'$ of $\mathcal{T}$ with $\mathcal{J}' <_{\mathsf{CP}'} \mathcal{J}$. Since $A$ is fixed in $\mathsf{CP}'$, $d_0 \in A^{\mathcal{J}'}$, thus $d_0 \in (\exists u.C_0)^{\mathcal{J}'}$ and there is an $e_0 \in C_0^{\mathcal{J}'}$. Modify $\mathcal{J}'$ into a new interpretation $\mathcal{I}'$ by setting $a_0^{\mathcal{I}'} = e_0$. It is readily checked that $\mathcal{I}'$ is a model of $\mathcal{A}$ and $\mathcal{I}' <_{\mathsf{CP}} \mathcal{I}$, thus we have obtained a contradiction to the fact that $\mathcal{I}$ is a model of $\mathsf{Circ}_{\mathsf{CP}'}(\emptyset, \mathcal{A})$. To get rid of fixed predicates, it suffices to apply Lemma 5. ❏

## Appendix C. Missing Proofs in Section 5

We show that the semantic consequence problem can be reduced to instance checking w.r.t. role-fixing cKBs in $\mathcal{ALC}$. We have already proved that for $\mathcal{ALC}$ extended with the universal role. In fact, the only remaining problem is to 'approximate' concepts $\forall u.C$ using $\mathcal{ALC}$ concepts that state that the extension of $C$ contains all points within a certain, sufficiently large neighbourhood.

To construct such an approximation, we first introduce a local version of the notion of frame validity of concepts. A *pointed $R$-frame* is a pair $(\mathfrak{F}, d)$ such that $d \in \Delta^{\mathfrak{F}}$ and $\mathfrak{F}$ is an $R$-frame. A concept $C$ is *valid* in a pointed $R$-frame $(\mathfrak{F}, d)$, in symbols $(\mathfrak{F}, d) \models C$, iff $d \in C^{\mathcal{I}}$ for every interpretation $\mathcal{I}$ based on $\mathfrak{F}$. For $m \in \mathbb{N}$ and $R$ a finite set of role names, $\forall^m R.C$ denotes $C$ if $m = 0$, and

$$\forall^{m-1} R.C \sqcap \bigsqcap_{r \in R} \forall r. \forall^{m-1} R.C$$

if $m > 0$. In what follows, we will use concepts of the form $\forall^m R.C$ as approximations of $\forall u.C$. We remind the reader of some correspondence results of modal logic. Let $\mathsf{trans}_A = \neg \forall s.A \sqcup \forall s. \forall s.A$ and $\mathsf{cont}_A = \neg \forall s.A \sqcup \forall r.A$. Then it is well known (Blackburn & van Benthem, 2007) and easy to prove that for every $\{r, s\}$-frame $\mathfrak{F}$, the following holds:

- $\mathsf{trans}_A$ is valid in $\mathfrak{F}$ if, and only if, $s^{\mathfrak{F}}$ is transitive;

- $\mathsf{cont}_A$ is valid in $\mathfrak{F}$ if, and only if, $r^{\mathfrak{F}} \subseteq s^{\mathfrak{F}}$.

Say that $d' \in \Delta^{\mathfrak{F}}$ is $\{r, s\}$-*reachable* from $d$ in $\mathfrak{F}$ if $(d, d') \in (r^{\mathfrak{F}} \cup s^{\mathfrak{F}})^*$ and call $d \in \Delta^{\mathfrak{F}}$ a *root* of $\mathfrak{F}$ if every $d' \in \Delta^{\mathfrak{F}}$ is $\{r, s\}$-reachable from $d$ in $\mathfrak{F}$. If $\mathfrak{F}$ is an $\{r, s\}$-frame with root $d$, then the following conditions are equivalent:

- $\forall^1 \{r, s\}.(\mathsf{trans}_A \sqcap \mathsf{cont}_A)$ is valid in $(\mathfrak{F}, d)$;

- $s^{\mathfrak{F}}$ is transitive and $r^{\mathfrak{F}} \subseteq s^{\mathfrak{F}}$.

These observations are used in the proof of Lemma 31 below. As before, we sometimes write concept assertions $C(a)$ in the form $a : C$. Recall that the role depth $\mathsf{rd}(C)$ of a concept $C$ is defined as the nesting depth of the constructors $\exists r.D$ and $\forall r.D$, $r \in R$, in $C$.

**Lemma 31** *Let $C$ and $D$ be $\mathcal{ALC}_{\{r\}}$-concepts not sharing any concept names and let $A$ be a fresh concept name. Let $\mathsf{CP} = (\emptyset, M, \{r, s\}, V)$ be a circumscription pattern, where $M$ consists of $A$ and the concept names in $C$ and $V$ consists of the concept names in $D$. Let $a$ be an individual name. Then the following conditions are equivalent:*

1. *for all $\{r, s\}$-frames $\mathfrak{F}$ with $r^{\mathfrak{F}} \subseteq s^{\mathfrak{F}}$ and $s^{\mathfrak{F}}$ transitive: if $\mathfrak{F} \models C$, then $\mathfrak{F} \models D$;*

2. *for all pointed $\{r, s\}$-frames $(\mathfrak{F}, d)$:*

$$(\mathfrak{F}, d) \models \forall^1 \{r, s\}.(\mathsf{trans}_A \sqcap \mathsf{cont}_A \sqcap C) \quad \Rightarrow \quad (\mathfrak{F}, d) \models D$$

3. *$a$ is an instance of $\neg \forall^1 \{r, s\}.(\mathsf{trans}_A \sqcap \mathsf{cont}_A \sqcap C) \sqcup D$ w.r.t. $\mathsf{Circ}_{\mathsf{CP}}(\emptyset, \mathcal{A})$, where*

$$\mathcal{A} = \{a : (\neg \forall^1 \{r, s\}.(\mathsf{trans}_A \sqcap \mathsf{cont}_A \sqcap C) \sqcup \forall^{1 + \max \{2, \mathsf{rd}(C)\}} \{r, s\}. \bigsqcap_{B \in M} B)\}.$$

**Proof.** Point 1 implies Point 2. Suppose Point 2 does not hold. Let $(\mathfrak{F}, d)$ be a pointed $\{r, s\}$-frame such that $(\mathfrak{F}, d) \models \forall^1 \{r, s\}.(\mathsf{trans}_A \sqcap \mathsf{cont}_A \sqcap C)$ and $(\mathfrak{F}, d) \not\models D$. We may assume that $d$ is a root of $\mathfrak{F}$. From $(\mathfrak{F}, d) \models \forall^1 \{r, s\}.(\mathsf{trans}_A \sqcap \mathsf{cont}_A)$ we obtain $r^{\mathfrak{F}} \subseteq s^{\mathfrak{F}}$ and $s^{\mathfrak{F}}$ transitive. Therefore, from $(\mathfrak{F}, d) \models \forall^1 \{s\}.C$ we obtain $\mathfrak{F} \models C$. It follows that $\mathfrak{F}$ is a frame refuting Point 1.

In what follows we use, for every $\mathcal{I}$ and $d \in \Delta^{\mathcal{I}}$, $d^{\uparrow, \mathcal{I}}$ to denote the set of all $e \in \Delta^{\mathcal{I}}$ which are $\{r, s\}$-reachable from $d$ in at most $1 + \max \{2, \mathsf{rd}(C)\}$ steps.

Point 2 implies Point 3. Suppose Point 3 does not hold. Let $\mathcal{I}$ be a model of $\mathsf{Circ}_{\mathsf{CP}}(\emptyset, \mathcal{A})$ such that

$$a^{\mathcal{I}} \in (\forall^1 \{r, s\}.(\mathsf{trans}_A \sqcap \mathsf{cont}_A \sqcap C) \sqcap \neg D)^{\mathcal{I}}. \tag{77}$$

Let $\mathcal{I}$ be based on $\mathfrak{F}$ and set $d := a^{\mathcal{I}}$. We show $(\mathfrak{F}, d) \models \forall^1 \{r, s\}.(\mathsf{trans}_A \sqcap \mathsf{cont}_A \sqcap C)$ and $(\mathfrak{F}, d) \not\models D$. The latter is easy as it is witnessed by the interpretation $\mathcal{I}$. To show the former, let $\mathcal{J}$ be an interpretation based on $\mathfrak{F}$. We show that $d \in (\forall^1 \{r, s\}.(\mathsf{trans}_A \sqcap \mathsf{cont}_A \sqcap C))^{\mathcal{J}}$. By (77) and since $\mathcal{I}$ is a model of $\mathsf{Circ}_{\mathsf{CP}}(\emptyset, \mathcal{A})$, $a^{\mathcal{I}} \in (\forall^{1 + \max \{2, \mathsf{rd}(C)\}} \{r, s\}. \bigsqcap_{B \in M} B)^{\mathcal{I}}$. It follows immediately that

$$B^{\mathcal{I}} = d^{\uparrow, \mathcal{I}}, \tag{78}$$

for all $B \in M$. We now distinguish two cases:

- $B^{\mathcal{J}} \supseteq d^{\uparrow, \mathcal{I}}$, for all $B \in M$. Since $\mathcal{I}$ and $\mathcal{J}$ are based on the same frame and all concept names in $\mathsf{cont}_A$, $\mathsf{trans}_A$, and $C$ are in $M$, the truth of $\forall^1 \{r, s\}.(\mathsf{trans}_A \sqcap \mathsf{cont}_A \sqcap C)$ at $d$ depends on the truth value of concept names from $M$ in $d^{\uparrow, \mathcal{I}}$ only. From (78) we obtain $B^{\mathcal{I}} \cap d^{\uparrow, \mathcal{I}} = B^{\mathcal{J}} \cap d^{\uparrow, \mathcal{I}}$, for all $B \in M$. Hence, by (77), $d \in (\forall^1 \{r, s\}.(\mathsf{trans}_A \sqcap \mathsf{cont}_A \sqcap C))^{\mathcal{J}}$, as required.

- $B^{\mathcal{J}} \not\supseteq d^{\uparrow, \mathcal{I}}$, for at least one $B \in M$.

  Let $\mathcal{J}'$ be the modification of $\mathcal{J}$ where $B^{\mathcal{J}'} = B^{\mathcal{J}} \cap d^{\uparrow, \mathcal{I}}$, for $B \in M$. By (78), $\mathcal{J}' <_{\mathsf{CP}} \mathcal{I}$. If $d \in (\neg \forall^1 \{r, s\}.(\mathsf{trans}_A \sqcap \mathsf{cont}_A \sqcap C))^{\mathcal{J}'}$, then $\mathcal{J}'$ is a model of $\mathcal{A}$ and we have a contradiction to the fact that $\mathcal{I}$ is a model of $\mathsf{Circ}_{\mathsf{CP}}(\emptyset, \mathcal{A})$. Thus, $d \in (\forall^1 \{r, s\}.(\mathsf{trans}_A \sqcap \mathsf{cont}_A \sqcap C))^{\mathcal{J}'}$. Since, again, the truth of $\forall^1 \{r, s\}.(\mathsf{trans}_A \sqcap \mathsf{cont}_A \sqcap C)$ at $d$ depends on the truth value of $B \in M$ on $d^{\uparrow, \mathcal{I}}$ only, we have $d \in (\forall^1 \{r, s\}.(\mathsf{trans}_A \sqcap \mathsf{cont}_A \sqcap C))^{\mathcal{J}}$, as required.

Point 3 implies Point 1. Suppose Point 1 does not hold. Consider a frame $\mathfrak{F}$ such that $s^{\mathfrak{F}}$ is transitive, $r^{\mathfrak{F}} \subseteq s^{\mathfrak{F}}$, $\mathfrak{F} \models C$, and $\mathfrak{F} \not\models D$. It follows that $\mathfrak{F} \models \text{trans}_A \sqcap \text{cont}_A$. Let $\mathcal{I}$ be an interpretation based on $\mathfrak{F}$ such that $d \in (\neg D)^{\mathcal{I}}$. We may assume that $d$ is a root of $\mathfrak{F}$. We may also assume that $B^{\mathcal{I}} = d^{\uparrow,\mathcal{I}}$ for all $B \in M$ (since no such $B$ occurs in $D$) and $a^{\mathcal{I}} = d$. Then $a^{\mathcal{I}} \in (\forall^1\{r,s\}.(\text{trans}_A \sqcap \text{cont}_A \sqcap C) \sqcap \neg D))^{\mathcal{I}}$ and $\mathcal{I}$ is a model of $\mathcal{A}$. It remains to show that there does not exist an $\mathcal{I}' <_{\text{CP}} \mathcal{I}$ such that

$$a^{\mathcal{I}'} \in (\neg \forall^1\{r,s\}.(\text{trans}_A \sqcap \text{cont}_A \sqcap C) \sqcup \forall^{1+\max\{2,\text{rd}(C)\}}\{r,s\}. \bigsqcap_{B \in M} B)^{\mathcal{I}'}.$$

This is straightforward: from $(\mathfrak{F},d) \models \forall^1\{r,s\}.C$, we obtain that there does not exist any $\mathcal{I}'$ such that $d \in (\neg\forall^1\{r,s\}.C)^{\mathcal{I}'}$ and clearly there does not exist any $B \in M$ with $B^{\mathcal{I}'} \subset B^{\mathcal{I}}$ such that $d \in (\forall^{1+\max\{2,\text{rd}(C)\}}\{r,s\}.B)^{\mathcal{I}'}$. ❏

We are in a position now to prove the reduction for $\mathcal{ALC}$.

**Theorem 22** The logical consequence problem of $\text{MSO}(r)$ is effectively reducible to the instance problem w.r.t. role-fixing cKBs formulated in $\mathcal{ALC}$. This even holds when the TBox and preference relation are empty.

**Proof.** By Theorem 20, the logical consequence problem of $\text{MSO}(r)$ is effectively reducible to the modal consequence problem for $\mathcal{ALC}_{\{r\}}$-concepts. Hence, it suffices to reduce the modal consequence problem for $\mathcal{ALC}_{\{r\}}$-concepts. Let $\mathcal{ALC}_{\{r\}}$-concepts $C$ and $D$ be given. We may assume that $C$ and $D$ have no concept names in common (if they have, then replace every concept name $B$ in $D$ by a new concept name $B'$ and denote the resulting concept by $D'$; as noted above, $C \Vdash D$ iff $C \Vdash D'$.) Let $\text{CP} = (\prec, M, \{s, r\}, V)$ where $\prec = \emptyset$, $M$ consists of $A$ and all concept names in $C$, and $V$ consists of all concept names in $D$. Let

$$\mathcal{A} = \{a : (\neg\forall^1\{r,s\}.(\text{trans}_A \sqcap \text{cont}_A \sqcap C) \sqcup \forall^{1+\max\{2,\text{rd}(C)\}}\{r,s\}. \bigsqcap_{B \in M} B)\}$$

and $C_0 = \neg\forall^1\{r,s\}.(\text{trans}_A \sqcap \text{cont}_A \sqcap C) \sqcup D$. By the equivalence of Point 1 and Point 3 in Lemma 31, $\text{Circ}_{\text{CP}}(\emptyset, \mathcal{A}) \models C_0(a)$ if, and only if, for all frames $\mathfrak{F}$ with $r^{\mathfrak{F}} \subseteq s^{\mathfrak{F}}$ and $s^{\mathfrak{F}}$ transitive, from $\mathfrak{F} \models C$ follows $\mathfrak{F} \models D$. As $C$ and $D$ do not contain $s$,

$$\text{Circ}_{\text{CP}}(\emptyset, \mathcal{A}) \models C_0(a) \quad \Leftrightarrow \quad C \Vdash D.$$

❏